\documentclass[twocolumn,tighten,times]{aastex631}

\usepackage{gensymb}

\newcommand{\msun}{\ensuremath{M_\sun}}

\newcommand{\nue}{\ensuremath{\nu_{e}}}
\newcommand{\nuebar}{\ensuremath{\bar \nu_e}}

\newcommand{\numt}{\ensuremath{\nu_{\mu\tau}}}
\newcommand{\numtbar}{\ensuremath{\bar \nu_{\mu\tau}}}

\newcommand{\numubar}{\ensuremath{\bar \nu_{\mu}}}
\newcommand{\nutaubar}{\ensuremath{\bar \nu_{\tau}}}

\newcommand{\Ye}{\ensuremath{Y_{\rm e}}}

\newcommand{\kbbar}{\ensuremath{k_{\rm B} \; {\rm baryon}^{-1}}}

\newcommand{\MHe}{\ensuremath{M_{\rm He}}}

\newcommand{\ccsne}{{\sc CCSN}e}

\newcommand{\isotope}[2]{\ensuremath{\mathrm {^{#2}#1}}}
\newcommand{\gcc}{\ensuremath{{\mbox{g~cm}}^{-3}}}

\newcommand{\Bethes}{\ensuremath{{\mbox{B~s}}^{-1}}}

\newcommand{\chimera}{{\sc Chimera}}

\newcommand{\UTphys}{Department of Physics and Astronomy, University of Tennessee, Knoxville, TN 37996-1200, USA}
\newcommand{\ORNLphys}{Physics Division, Oak Ridge National Laboratory, P.O. Box 2008, Oak Ridge, TN 37831-6354, USA}
\newcommand{\NCCS}{National Center for Computational Sciences, Oak Ridge National Laboratory, P.O. Box 2008, Oak Ridge, TN 37831-6164, USA}
\newcommand{\FAU}{Department of Physics, Florida Atlantic University, 777 Glades Road, Boca Raton, FL 33431-0991, USA}
\newcommand{\OSU}{Department of Astronomy and Center of Cosmology and AstroParticle Physics, The Ohio State University, Columbus, OH 43210, USA}

\shorttitle{2D Core-Collapse Simulations of Two 15.8-\msun\ stars}
\shortauthors{Bruenn et al.}

\begin{document}

\title{Comparison of the Core-Collapse Evolution of Two Nearly Equal Mass Progenitors}

\author[0000-0003-0999-5297]{Stephen W. Bruenn}
\affiliation{\FAU}

\author[0000-0001-8235-5910]{Andre Sieverding}
\affiliation{\ORNLphys}

\author[0000-0002-5231-0532]{Eric J. Lentz}
\affiliation{\UTphys}
\affiliation{\ORNLphys}

\author[0000-0002-1728-1561]{Tuguldur Sukhbold}
\affiliation{\OSU}

\author[0000-0002-9481-9126]{W. Raphael Hix}
\affiliation{\ORNLphys}
\affiliation{\UTphys}

\author[0000-0002-1530-7297]{Leah N. Huk}
\affiliation{\NCCS}

\author[0000-0003-3023-7140]{J. Austin Harris}
\affiliation{\NCCS}

\author[0000-0002-5358-5415]{O. E. Bronson Messer}
\affiliation{\NCCS}
\affiliation{\ORNLphys}
\affiliation{\UTphys}

\author[0000-0001-9816-9741]{Anthony Mezzacappa}
\affiliation{\UTphys}

\correspondingauthor{Stephen W. Bruenn}
\email{bruenn@fau.edu}

\begin{abstract}
We compare the core-collapse evolution of a pair of 15.8~\msun\ stars with significantly different internal structures, a consequence of the bimodal variability exhibited by massive stars during their late evolutionary stages. 
The 15.78 and 15.79~\msun\ progenitors have core masses (masses interior to an entropy of 4~\kbbar) of 1.47 and 1.78~\msun\ and compactness parameters $\xi_{1.75}$ of 0.302 and 0.604, respectively.
The core collapse simulations are carried out in 2D to nearly 3~s post-bounce and show substantial differences in the times of shock revival and explosion energies.
The 15.78~\msun\ model begins exploding promptly at 120 ms post-bounce when a strong density decrement at the Si--Si/O shell interface, not present in the 15.79~\msun\ progenitor, encounters the stalled shock.
The 15.79~\msun\ model takes 100 ms longer to explode but ultimately produces a more powerful explosion. 
The larger mass accretion rate of the 15.79~\msun\ model during the first 0.8~s post-bounce time results in larger \nue/\nuebar\ luminosities and rms energies.  
Additionally, the \nue/\nuebar\ luminosities and rms energies arising from the inner core are also larger in the 15.79~\msun\ model throughout due to the larger negative temperature gradient of this core, a consequence of greater adiabatic compression.
Larger \nue/\nuebar\ luminosities and rms energies in the 15.79~\msun\ model, along with its flatter and higher density heating region, result in more neutrino energy deposition behind the shock and more and higher enthalpy matter ejected.
Using a limited nuclear reaction network, we find the ejected \isotope{Ni}{56} mass of the more energetic 15.79~\msun\ model is more than double that of the less energetic 15.78~\msun\ model.
Most of the ejecta in both models is moderately proton-rich, though counterintuitively the highest electron fraction ($Y_e=0.61$) ejecta in either model is in the less energetic 15.78~\msun\ model while the lowest electron fraction ($Y_e=0.45$) ejecta in either model is in the 15.79~\msun\ model.
\end{abstract}


\section{Introduction}

The utility of models of supernovae and other stellar phenomena are the insight they provide into parts of the star obscured from our view either by the star's outer precincts or the mists of time.   Verification that such models can be trusted comes from comparison to observations of the supernova or other event.    For core-collapse supernovae, observations have long provided a set of gross properties of the supernova, including the (kinetic) energy of the explosion, the total mass of the ejecta, the mass of \isotope{Ni}{56} in the ejecta, etc. \citep[see, e.g.][]{Hamu03}, although with uncertainties of several 10\%. 
These large uncertainties make the data more useful for revealing general trends, like the correlation between increasing ejecta (and presumably stellar) mass, \isotope{Ni}{56} mass and the explosion energy, than for detailed constraint of supernova models.  In recent years, the archive of galactic imagery from Hubble Space Telescope has opened another opportunity for determining the mass of the star that subsequently exploded.  Such information is critical to answer fundamental questions like which stars are responsible for supernovae of various types and what are the initial mass functions of neutron stars and black holes.  The mass determination comes from a comparison between observational placement on the H-R diagram and evolutionary tracks from stellar evolution models \citep[see ][]{Smar09}. 
The much smaller uncertainty from these analyzes allows useful comparison between models and observations \citep[see, e.g.,][]{BrLeHi16}, however there are systematic uncertainties which need to be better understood, for example, use of a different stellar evolution code shifts the stellar masses by 1--3~\msun \citep{Smar15}.

Addressing the same fundamental questions has also prompted several numerical studies of the progenitor dependence on the evolution of core-collapse supernovae \ccsne\, with various numerical detail and comprising various numbers of progenitors \citep{OCOt11, UgJaMa12, OCOt13, {BrLeHi16}, SuHaJa16, OtRodSS18, BuRaVa19, BuRaVa20, VaLaRe21, WaVaBu22, BoRoLi22}. 
Most crudely, the question is, for a given progenitor mass and structure, does an explosion occur or does collapse to a black hole prevent the explosion from becoming sufficiently powered?
A more refined mapping might indicate, on average, the explosion energy, nucleosynthesis yields, and neutrino and gravitational wave signatures as functions of progenitor mass.
What has become clear from the computational investigations thus far is that the core-collapse evolution, even at the most basic level of whether or not an explosion ensues, is a non-monotonic function of the progenitor mass because it is highly dependent on the core structure at the time of collapse.
One challenge to these analysis is the unfortunate lack, at the present time, of convergence in the outcomes of different numerical simulations for a given progenitor.
For example, note the differences in the \ccsne\ simulation outcomes initiated from four 12--25~\msun \citet{WoHe07} progenitors by \citet{BrLeHi16}; \citet{DoBuZh15}; and \citet{SuHaJa16}.
It should be noted, however, that even large differences (e.g., explosions versus duds) in the outcomes of \ccsne\ simulations may indicate an extreme sensitivity of the outcomes to the details of the simulations, particularly if the models are close to the explosion criteria, rather than large differences between the simulation codes themselves.

\citet{SuWoHe18} have further complicated the potential mapping between stellar mass and supernova outcome by showing that shell-burning behaviors within the carbon core can generate stochastic variations in the final structure of the core for solar metallicity progenitors of 14--19~\msun.
The goal of this paper is to compare the core collapse and subsequent evolution stemming from two progenitors of almost equal masses (difference in mass $< 0.1$\%) but different internal structures.
A similar but very brief investigation of the consequences of internal structure differences for supernova explosions was presented by \citet{SuYaTa16}, who evolved five 15-\msun\ progenitors provided by different stellar evolution groups.
Because these differences in the \citet{SuWoHe18} progenitors develop very late in the stars' lives, these models would be visually indistinguishable (surface radii vary by less than 0.2\%, surface temperature and density by less than 0.05\%). 
This isolates the effects of the internal structure from those of stellar mass on the core collapse outcomes, providing a direct indicator of the sensitivity of the core-collapse and subsequent evolution to the structure of a given mass progenitor.
The variation of explosion outcomes for visually indistinguishable stars is also a measure of the physical scatter in these quantities we should expect for explosions coming from seemingly identical stars.

In Section~\ref{sec:preSN} we examine the pre-supernova evolution of two (15.78 and 15.79~\msun) updated \citet{SuWoHe18} progenitors.
Section~\ref{sec:methods} lays out the computational set up of our supernova simulations from these progenitors.
The basic properties of the core collapse and shock revival are presented in Section~\ref{sec:resullts} and the dynamics of the explosions are presented in Section~\ref{sec:expl}.
A comparison of the basic nucleosynthesis of the two models is presented in Section~\ref{sec:nucleo}, and a summary of the results of this paper are given is Section~\ref{sec:summary}.

\section{Pre-supernova evolution}
\label{sec:preSN}

\begin{figure}
	\fig{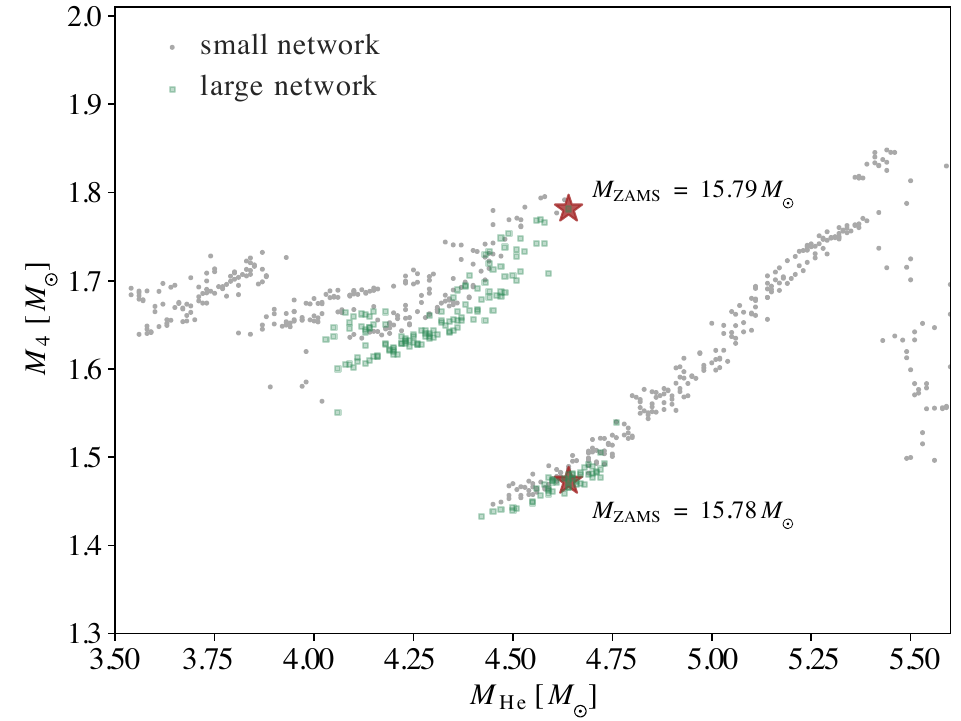}{\columnwidth}{}
	\caption{\label{fig:m4he}
	Mass with entropy $s < 4$ ~\kbbar ($M_4$) plotted versus \isotope{He}{4} core mass (\MHe) for the small network models of  \citet{SuWoHe18} (gray circles) and the large network models of Suhkbold (unpublished; green squares). The two models used in this paper are marked with stars.
	}
\end{figure}

\begin{figure}
        \fig{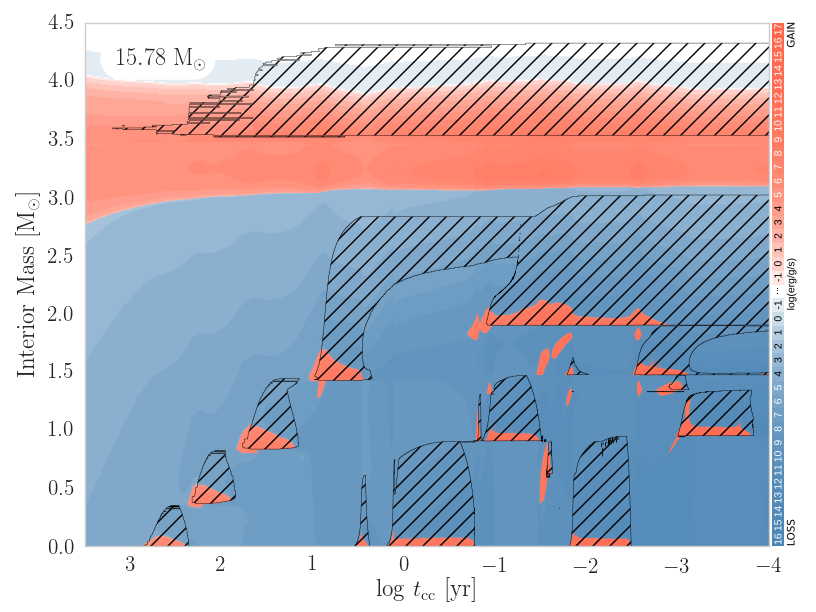}{\columnwidth}{(a)}
        	\fig{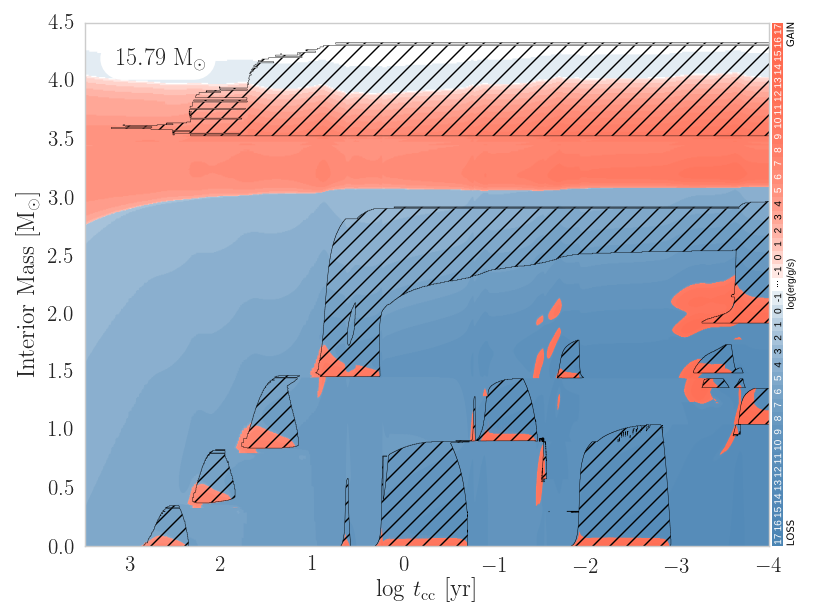}{\columnwidth}{(b)}
	\caption{\label{fig:preSNconv}
	 Carbon-core convective histories of (a) 15.78~\msun\ and (b) 15.79~\msun\ pre-supernova models.
		}
\end{figure}

Inside the carbon core of massive stars there are episodes of core carbon, neon, oxygen, and silicon burning, each followed by shell burning episodes.
These burning regions can influence concurrent and ensuing episodes, making for stochasticity in the evolution of massive star cores.
\citet{SuWoHe18} identified some of the variability as being bimodal and related to specific patterns of shell interactions.
A useful quantity to characterize the core of a massive star is the mass of the region where the entropy $s < 4$ ~\kbbar, which as often been used to 
estimate the remnant mass, or mass cut, for parameterized explosions in spherical symmetry \citep{WoHe07} and for a criterion to estimate the likelihood of a successful, neutrino driven explosion \citep{ErJaWo16}.
The multivalued, branched nature of $M_4$ for helium core masses, \MHe, in the range of 4.4--4.6~\msun\ can be clearly seen in Figure~\ref{fig:m4he}.
In both branches, $M_4$, which is typically tied to the mass coordinate of maximum oxygen burning at collapse, increases with \MHe, while a clear gap remains.
This is true both of the small network models from \citet{SuWoHe18} (gray) and the unpublished models using a larger network, shown in green.
The unpublished `large network' models incorporate an adaptive nuclear network of up to 300 species as needed into the evolution of the stellar structure equations instead of the 19-species network supplemented by the quasi-equilibrium network for silicon burning for the `small network' runs as detailed in Section~2.2 of \citet{SuWoHe18}, but are otherwise identical in numerical configuration.
The two models from which we will compute supernovae are on different branches despite differing in initial mass by only 0.01~\msun.
The upper branch at $\MHe = 4.5$~\msun\ contains the 15.79~\msun\ model near its upper mass limit with $M_4 = 1.78$~\msun\ and the lower branch contains the 15.78~\msun\ model with  $M_4 = 1.47$~\msun.
\begin{figure}
	\fig{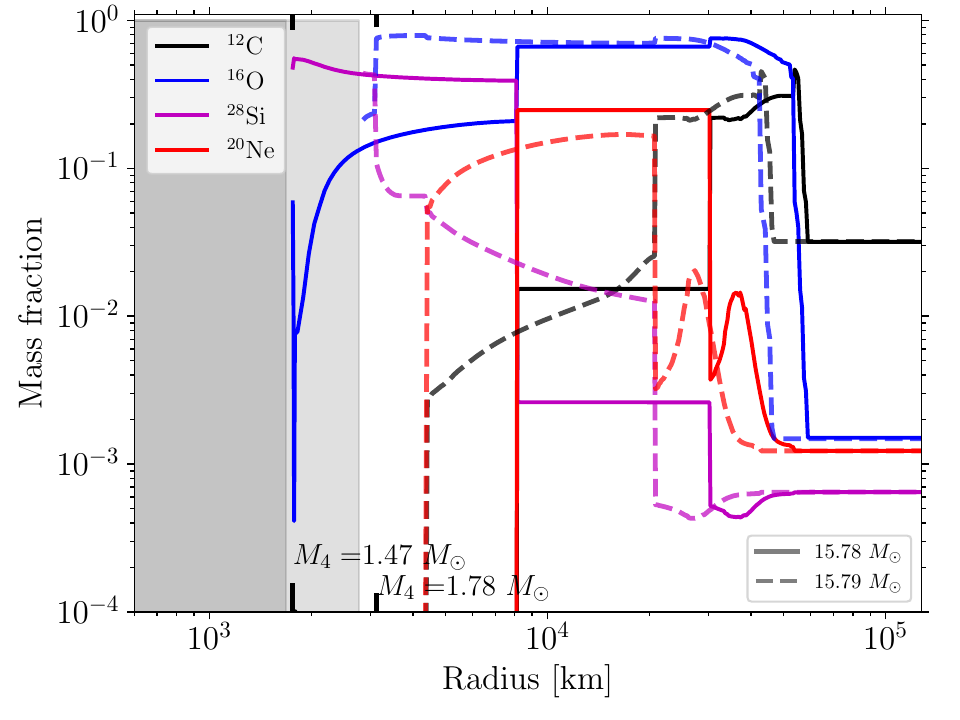}{\columnwidth}{}
	\caption{\label{fig:mf_profiles}Mass fraction profiles of the two progenitors. The thick tickmarks at  1.47~\msun\ and 1.78~\msun\ indicate $M_4$ for the 15.78~\msun\ and the 15.79~\msun\ model respectively. The composition of the Fe-core in nuclear statistical equilibrium (NSE) is not shown.}
\end{figure}

\begin{figure}
	\fig{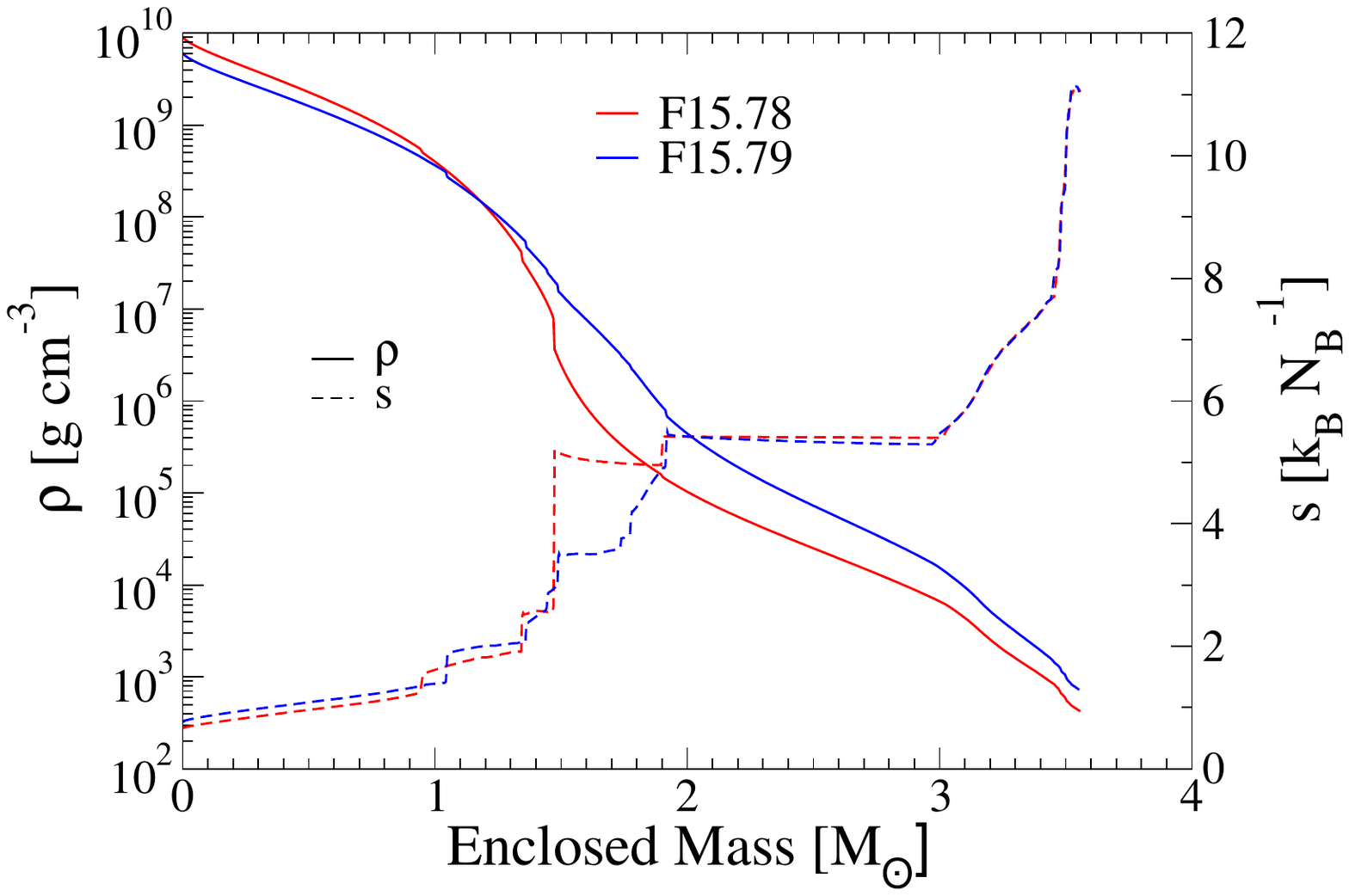}{\columnwidth}{}
	\caption{\label{fig:init_den_s}
	The density and entropy profiles of the progenitors.
	}
\end{figure}

Figure~\ref{fig:preSNconv} shows the convective histories (Kippenhahn diagrams) of the two models, where hatched areas indicate convective regions, red color indicates heating by nuclear burning and blue color cooling by neutrino emission.  
About 1000 years before collapse, both models ignite convective core C-burning followed by three C-burning shells moving outward in mass. While the third C-burning shell is active, Ne burning takes place in the core and the end of the shell burning episode is coincident with the start of core Oxygen burning at about one year before collapse. 
Both models, and the two $M_4$ branches to which they belong, are characterized by the suppression of core oxygen burning until the third carbon-burning shell is complete due to the ignition of that shell within the degenerate effective Chandrasekhar mass.  This is unlike the branch for $\MHe < 3.7$~\msun\ (see Figure~\ref{fig:m4he}), where the C-burning shell ignition occurs outside the Chandrasekhar mass and does not suppress core oxygen burning.
After core O burning, the evolution of the two models we consider here diverges.
In the 15.78~\msun\ model, a fourth convective C-burning shell ignites about 0.1~years before collapse, encompassing the region between 2~\msun\ and 3~\msun\ at the same time as the first O-shell ignites. 
This leads to a flat $^{12}$C profile in this region (see Figure~\ref{fig:mf_profiles}).
This C-shell burning episode slows the contraction of the core and suppresses the 2nd O-burning shell and leads to an initial Si-core that is smaller than in the 15.79~\msun\ model, in which the fourth C-burning shell only ignites about 8~hours before collapse and leaves behind gradients in the $^{12}$C mass fraction profile. 
In both models, the final O-shell burning episode before collapse then takes place on top of this Si-core. The difference in the Si-core sizes at the time of Si-core burning is also highlighted in Figure 13 of \citet{SuWoHe18} and it is closely connected to the entropy jump that defines the value of $M_4$.
The suppressed second oxygen-burning shell thus gives rise to the lower branch in Figure~\ref{fig:m4he} where the 15.78~\msun\ model is located.
At the time of collapse, in the 15.78~\msun\ model, oxygen is not completely exhausted in the region already dominated by Si, and as consequence $M_4$ is located at 1.47 \msun, inside the final Si-core (see Figure~\ref{fig:mf_profiles}) which is very extended, reaching out to 8,000~km. 
In the 15.79~\msun\ model, on the other hand, $M_4$ is at 1.78 \msun, about 3,000~km from the center and still aligned with the edge of the Si-core at the time of collapse.
Not surprisingly, having more mass concentrated in a less volume has a significant effect on the density.
As Figure~\ref{fig:init_den_s}, illustrates, above fairly similar iron cores, the density structures of the 15.78 and 15.79 models are quite different in the silicon and oxygen-rich regions.
Most notable is a sharp drop in density of the 15.78~\msun\ model at an enclosed mass of 1.47~\msun, where the base of the powerful convective oxygen burning episode was located just prior to collapse.
The most significant difference in the entropy profiles of these models (Figure~\ref{fig:init_den_s}) corresponds to this density drop, making clear the cause of the difference in $M_4$.

Of particular importance for the post-collapse evolution of these models is the difference in their compactness, defined as $\xi_M = (M/\msun) / (R(M)/1000$ km).  
Smaller compactness has been suggested as an indicator of easier explodability by \citet{OCOt11}, at least for progenitors of the same or very similar ZAMS \citep{VaLaRe21}.
For the 15.78 and 15.79~\msun\ models, $\xi_{2.5} = 0.136$ and 0.206, respectively, and 0.301 and 0.604 for $\xi_{1.75}$.
The parameterized simulations of \citet{OCOt11} indicated that a progenitor having a compactness parameter $\xi_{2.5} > 0.45$ would lead to black hole formation rather than an explosion. 
Other more elaborate criteria for the explodability of a given progenitor have been developed \citep[e.g.,][]{ErJaWo16, MuHeLi16, SuYaTa16, WaVaBu22,  BoRoLi22}, and they lead to similar conclusions regarding the explodability of 15.78~\msun\ model versus that of the 15.79~\msun\ model.

\section{Neutrino-Radiation Hydrodynamics}
\label{sec:methods}

The supernova simulations that follow were computed using the \chimera\ neutrino radiation hydrodynamics code \citep{BrBlHi20} and are part of the \chimera\ `F-series' of simulations.
\chimera\ uses PPMLR hydrodynamics, multi-pole self gravitation with a monopole GR potential correction, ray-by-ray 4-species neutrino transport by the flux-limited diffusion method, a dense nuclear equation of state (EoS) at high temperatures where nuclear statistical equilibrium (NSE) applies and a nuclear network where it doesn't.

The neutrino opacities used are those used since the B-series simulations \citep{BrMeHi13,BrLeHi16} with an improvement to the nucleon-nucleon bremsstrahlung computation.
For bremsstrahlung, we have modified the formulation in \citet{HaRa98} to treat nn-, pp-, and np-bremsstrahlung separately with the appropriate degeneracy factors.

The NSE EoS is the SFHo EoS of \citet{StHeFi13} incorporated into \chimera\ \citep{Land18}  using the WeakLib framework via the CompOSE\footnote{\url{https://compose.obspm.fr}} database.
As in the Series-E runs \citep{Land18,LaLeMe22}, we use the \cite{Coop85} electron EoS with SFHo and also in the non-NSE network region for continuity.

The nuclear network is the \texttt{anp56} network used in the D-series and E-series \citep{LaLeMe22} simulations.
It contains the same 14 $\alpha$-chain nuclei (\isotope{He}{4},\isotope{C}{12}--\isotope{Zn}{60}) as the \texttt{alpha} network used in the B-series \citep{BrMeHi13,BrLeHi16} and C-series \citep{LeBrHi15} with the neutron, proton, and \isotope{Fe}{56} added actively, but non-reactively, to the network.
The inclusion of these three `inert' species, which are not connected by reactions to the rest of the network, allows us to compute NSE compositions directly with the XNet nuclear network when needed at the boundary of the NSE region, rather than extracting them from the NSE EoS.

Together with internal code updates, this constitutes the default inputs and code for Series-F \chimera\ simulations.

Initial conditions for the core-collapse and supernova calculations with \chimera\ are taken from the 15.78 and 15.79 \msun\ pre-supernova progenitors described above.
In the following, we will refer the supernova simulations by their \chimera\ model designations F15.78 and F15.79.
The inner 3.56~\msun\ of both progenitors out to the \isotope{He}{4} shell, with radius 61,250~km for F15.78 and 48,960~km for F15.79, were remapped onto 720 radial shells of unequal widths that generally increase with radius. 
The simulations were carried out in full 2D with 240 zones of uniform width (0.75\degree) in $\theta$ from the onset of collapse.
The small roundoff errors supply the perturbations from which the fluid instabilities grow.
It is well known that imposing axisymmetry results in unipolar or bipolar deformations along the symmetry axis due to the boundary conditions imposed at the poles which deflects fluid motions parallel to the pole.
In addition, the inverse turbulent energy cascade distributes the energy in an unphysical way to the largest scales (see Kraichnan 1967; Hanke et al. 2012; Couch 2013; Radice et al. 2015)
Nevertheless, 2D simulations have manifested similarities in overall explosion outcomes \citep{Mull15, VaBuRa18b, BuRaVa19} are relatively fast now and are useful for obtaining insights concerning the interplay of essential physics and, if interesting, can motivate future 3D simulations.

\begin{figure}
	\includegraphics[width=\columnwidth,clip]{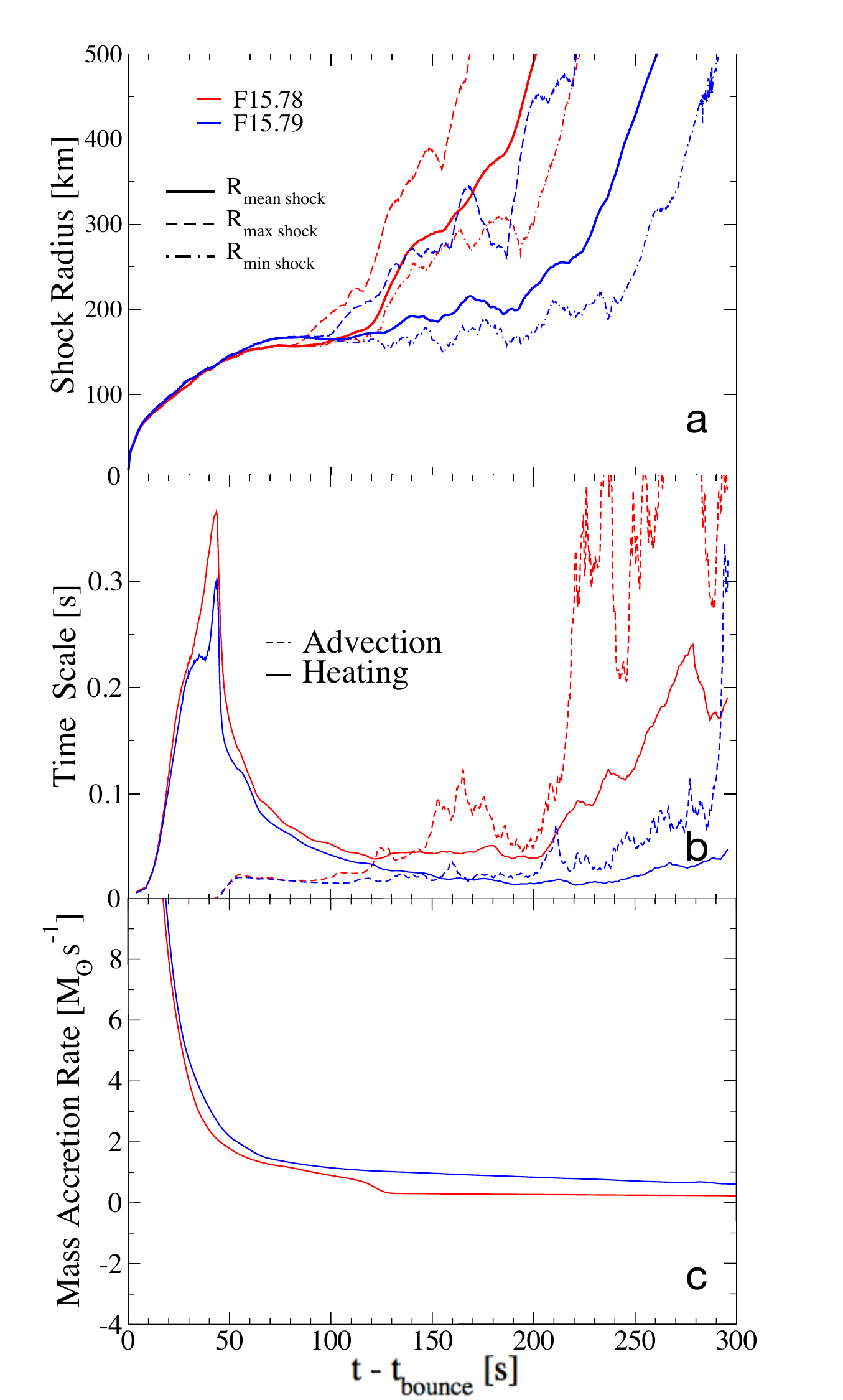}
	\caption{\label{fig:early_shock}
	As a function of post-bounce time, (a) the mean, minimum, and maximum shock positions as a function of post-bounce time; (b) the advection and heating timescales; (c) mass accretion rate at the shock location. The red and blue lines correspond respectively to F15.78 and F15.79.
	}
\end{figure}

\section{Onset of Explosion}
\label{sec:resullts}

\begin{figure}
	\includegraphics[width=\columnwidth,clip]{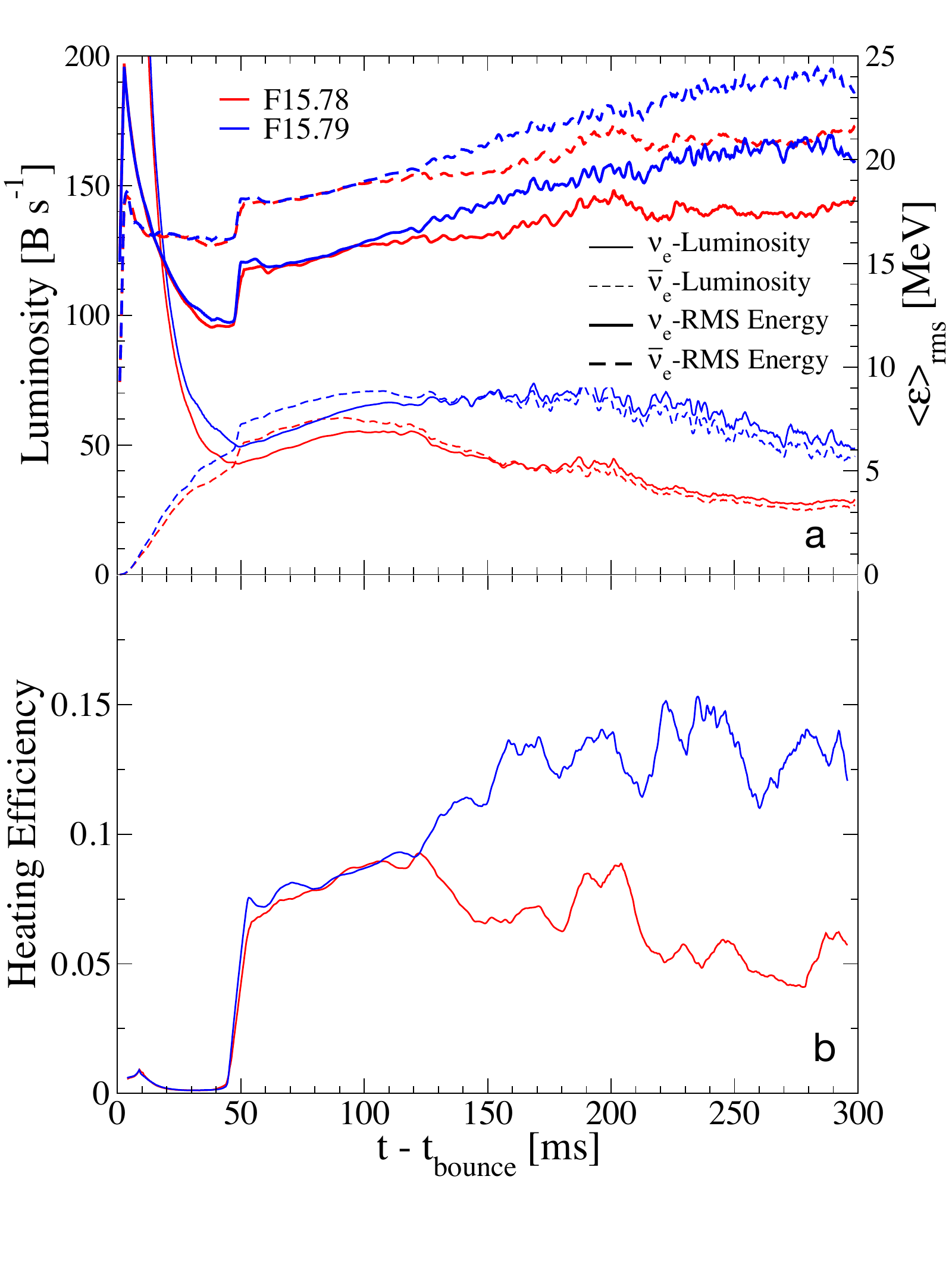}
	\caption{\label{Nu_Eff}
	As a function of post-bounce time, (a) \nue\ and \nuebar\ luminosities and rms energies at the gain radius; (b) neutrino heating efficiency
	}
\end{figure}

The shock trajectories of both models, Figure~\ref{fig:early_shock}(a), are quite similar from bounce through shock stagnation at a radius of $\sim$160~km, reflecting the similar structure of their iron cores. 
Following  stagnation, the shock location is set by the quasi-steady state balance between the accretion ram pressure at the shock and the total pressure, thermodynamic and turbulent, of the immediate post-shock material. 
In the neutrino driven mechanism for CCSNe, the post-shock pressure is generated by the deposition of energy into the heating region by neutrinos. 
The revival of the stationary (stalled) shock will occur when the total post-shock pressure (thermal and turbulent) begins to exceed  the pre-shock accretion ram pressure to the point where a stationary solution for the shock radius no longer exists \citep{BuGo93, JaMu96, Jank01, Jank12, SuHaJa16}. 
The different epochs and modes of shock revival for the two models are a consequence of the different density structures of their progenitors, shown in Figure~\ref{fig:init_den_s}.
For F15.78 (with the smaller compactness parameters), shock revival occurs at 120~ms (unless otherwise specified all times refer to post-bounce times) when the density decrement at the Si-Si/O interface at 1.47~\msun\ is advected through the shock, resulting in a sudden drop in the mass accretion rate at the shock front (Figure~\ref{fig:early_shock}(c)).
The sudden decline in the ram pressure ahead of the shock along with the still significant mass accretion luminosity from the surface of the proto-NS is the trigger for shock revival in this model.
Lacking a density decrement, there is no corresponding sudden decrease in the accretion ram to trigger shock revival in F15.79. 
In this model, shock revival occurs about 100~ms later than in F15.78, following a period during which the continuing core and accretion luminosities, coupled with a gradual hardening of the \nue\ and \nuebar\ spectra (Figure~\ref{Nu_Eff}(a)), continue to pump energy into the post-shock region.
Eventually an accretion/luminosity critical condition is satisfied (as formulated for 1D by \citet{BuGo93} and extended to multi-D by \citet{MuJa15, SuHaJa16}) and shock revival ensues.

This description of the onset of shock revival is reflected in the often used timescale inequality criterion. 
Shock revival is triggered when the the advection or dwell timescale, $\tau_{\rm adv}$, of a fluid element in the gain region, the region inside the shock where heating exceeds cooling, becomes longer than the heating timescale, $\tau_{\rm heat}$, the timescale for an e-fold increase in the total energy of a fluid element by neutrino heating \citep{JaKiRa01, ThQuBu05, BuJaRa06, Fern12}.
With $\tau_{\rm adv}$ approximated as $M_{\rm gain}/\dot{M}$, where steady-state conditions are assumed, and $\tau_{\rm heat} = E_{\rm gain}/\dot{Q}_{\rm gain}$, where $E_{\rm gain}$ and $\dot{Q}_{\rm gain}$ are the integrated total energy and the integrated net heating rate of material in the gain layer, respectively, these timescales are plotted for both models in Figure~\ref{fig:early_shock}(b).
The sudden switch from $\tau_{\rm adv} < \tau_{\rm heat}$ to $\tau_{\rm adv} > \tau_{\rm heat}$ at 120~ms for F15.78 is brought about by the abrupt increase in $\tau_{\rm adv}$ due to the advection of the density decrement through the shock and its immediate consequent expansion. 
Absent a density decrement trigger, no such sudden increase in $\tau_{\rm adv}$ is observed for F15.79 until 210~ms.
Rather, there is a sustained period from 120 to 210~ms during which $\tau_{\rm adv}$ slowly increases and $\tau_{\rm heat}$ slowly decreases, indicating a heating driven build-up of internal energy and pressure inside a slowly expanding heating region.
At 210~ms, $\tau_{\rm adv}$ exceeds $\tau_{\rm heat}$ sufficiently for shock revival to occur.

Another indicator of shock revival dynamics is the heating efficiency, $\eta_{f}$, defined by
\begin{equation}
\eta_{f} = \frac{\dot{Q}}{ L_{\nue} + L_{\nuebar}},
\label{h_eff}
\end{equation}
and plotted for the two models in Figure~\ref{Nu_Eff}(b).  As before, $\dot{Q}$ is the net neutrino energy deposition rate in the gain region, the region inside the shock where neutrino energy deposition exceeds emission, and $L_{\nue}$ and $L_{\nuebar}$ are the \nue\ and \nuebar\ luminosities at the base of the gain region.
Until the density decrement of F15.78 reaches the shock at 120 ms, the heating efficiencies for both models are quite similar, reflecting their similar core structures at bounce and the similar rates of mass advection through the shock during this time. 
The heating efficiencies rise during this time primarily because of the increase in the neutrino rms energies, as shown in Figure~\ref{Nu_Eff}(a).
The decline in $\eta_{f}$ for F15.78 after 120~ms is due to a combination of the leveling off of its neutrino rms energies (Figure~\ref{Nu_Eff}(a)) and the decline in the density of the heating region following the expansion of the shock.
The continuing rise in \nue\ and \nuebar\ rms energies in F15.79 past 120 ms leads to increasing heating efficiency (Figure~\ref{Nu_Eff}(b)) that persists until the occurrence of shock revival for this model at 210~ms, similar to earlier \chimera\ models \citep{BrMeHi13,LeBrHi15} where the shock also revived without a sudden drop in ram pressure.

Fluid instabilities have long been known to play a key role in enhancing the possibility of shock revival by increasing the dwell time of material in the gain region, thereby increasing the time and efficiency at which material can absorb energy. The addition of turbulent pressure via the radial Reynolds stress behind the shock, together with the thermal pressure, also facilitates pushing the shock outward \citep{BuHa96, CoOt15, NaBuRa19}. 
The neutrino heated layer develops a negative entropy gradient which can render it convectively unstable \citep{HeBeHi94, BuHa96, JaMu96} when the convective growth rate exceeds the rate at which accreting material is swept through the gain region, parameterized by \citet{FoScJa06}
\begin{equation}
\chi = \int_{\langle R_{\rm g} \rangle}^{\langle R_{\rm s} \rangle} \frac{ {\rm Im } \langle \omega_{\rm BV} \rangle }{ | \langle v_{\rm r} \rangle | } .
\end{equation}
\citet{FoScJa06} showed that when $\chi \gtrsim 3$ convective growth is expected to overcome the stabilizing effect of matter inflow, which we see at 69 ms for F15.78 and 75 ms for F15.79. 
These times correlate well with the first appearance of stable convective structures in the gain region of our models. 
Convection persists for only about 50 ms before runaway shock expansion sets in for F15.78, driven by the density decrement, and about 150~ms for  F15.79.

The standing accretion shock instability (SASI), first investigated in the context of shock revival in \ccsne\ by \citet{BlMeDe03} and further investigated \citep[][and many subsequent works]{OhKoYa06, FoScJa06, FoGaSc07, YaYa07, ScJaFo08}, can also be important in promoting conditions for shock revival by establishing large scale post shock flows that periodically push the shock outward \citep{BlMeDe03}. 
The dipole and quadrupole modes of the shock surface appear in both models and the periods match the periods computed using the formulation for SASI period in \citet{OhKoYa06} and \citet{ScJaFo08}, compensating for neutrino heating as \citet{YaYa07} suggest.
This suggests that the SASI is present in both of our models, but as is typical in CCSN simulations with strong neutrino-driven convection, we cannot assess the relative importance of the SASI in aiding explosion.

Both of our models follow the typical early pattern for post-bounce evolution --- shock stagnation, development of neutrino driven convection and SASI oscillations of the shock, a slow increase in neutrino heating efficiency, and post-shock thermal heating.
At this point, differences in the $\rho(M)$ vs $M$ profiles of the two models, illustrated in Figure~\ref{fig:init_den_s}, cause the subsequent evolutionary scenarios of the two models to diverge.
The large decrement in density at 1.47~\msun\ in the 15.78~\msun\ progenitor created by a late pre-collapse oxygen shell burning episode terminates the slow pre-revival build-up to explosion in F15.78 when that layer accretes through the shock. A rapid drop in external ram pressure results, followed by a sudden expansion of the shock, and triggering the explosion directly.
The 15.79~\msun\ progenitor did not have such a late burning episode, nor the resulting density decrement, so explosion in F15.79 is only triggered through a slower process of pressure build-up from neutrino heating and a more gradual decline in ram pressure at the shock until a critical condition is reached.
This provides the first major impact of the structural differences in these two progenitors, resulting in differences in the explosions generated. 
From this point, effects of the progenitor structure are combined with the building explosions to control the further development of the explosions.

Recent work has investigated the effect of progenitor asphericities due to pre-collapse convection in oxygen and silicon burning shells on the post-collapse evolution of CCSNe \citep{CoOt13, CoOt15, MuJa15, AbZhRa16, Mull16, MuMeHe17, VaBuRa18a}. While the exact mechanism still needs to be elaborated, this work indicates that vigorous large-scale convection in pre-collapse progenitors will shorten the time to shock revival for successful models and increase the possibility of shock revival for those models which would otherwise not explode. It is doubtful that incorporating convection in the oxygen or silicon burning shells in the F15.78 progenitor would have much effect on its shock revival time, as a robust shock revival already ensues shortly after bounce when the density decrement at the Si/Si-O interface encounters the shock. For F15.79, shock revival occurs while the shock is still within the extensive Si shell, so vigorous convective burning in this shell might reduce the time for shock revival.

\section{Explosion dynamics}
\label{sec:expl}

In this section we will give an overview of the evolutionary differences between the two models followed in subsequent sections by a detailed look at the origin of these differences, tracing them back to differences in the internal structures of their progenitors.

\subsection{Overview}
\label{sec:overview}

\begin{figure}[h]
	\includegraphics[width=\columnwidth,clip]{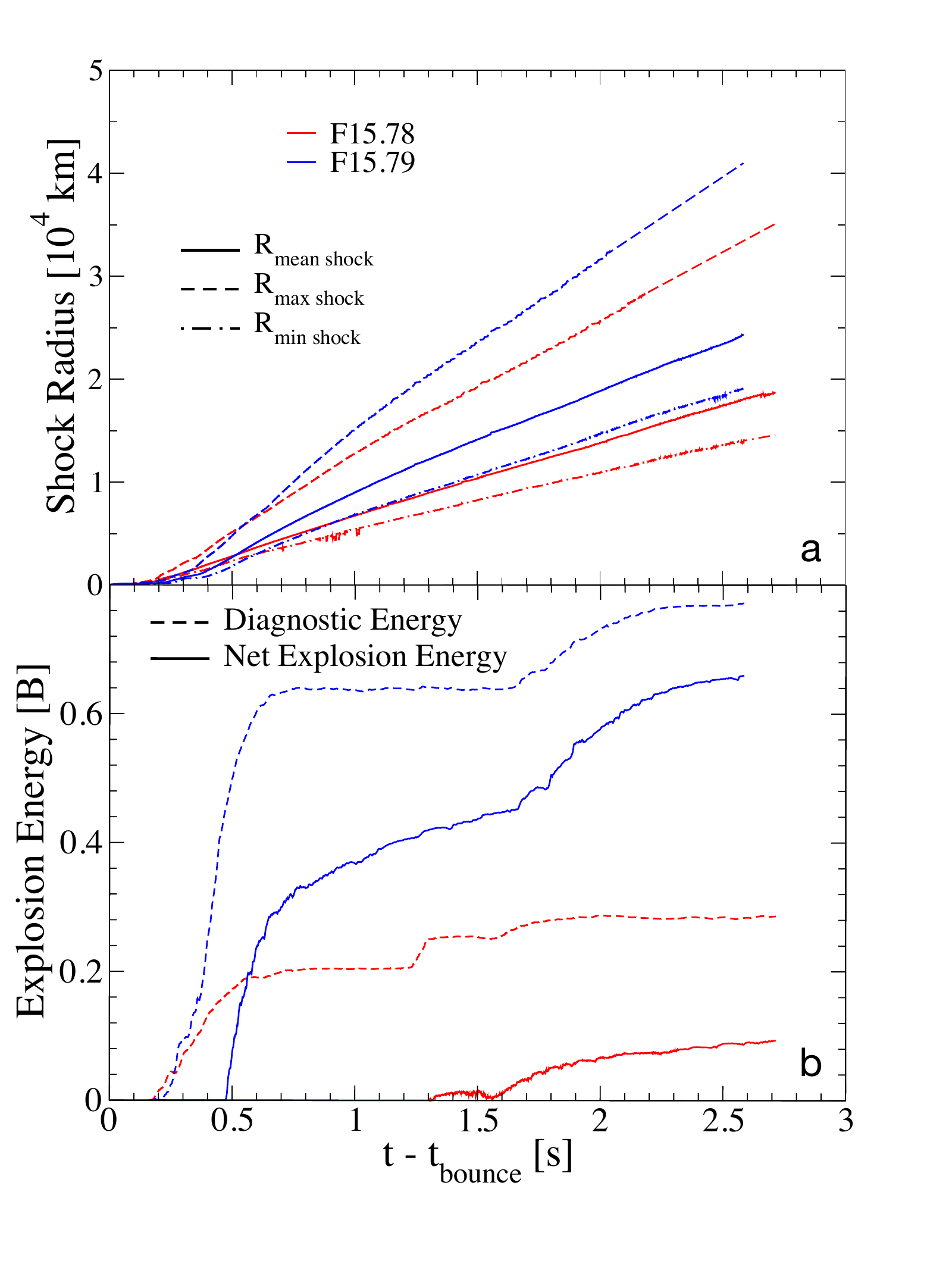}
	\caption{\label{Shock_Energy}
	Comparison of the (a) mean, minimum, and maximum radii of the shocks of both models, and (b) the diagnostic and net explosion energies, as a functions of post-bounce time.
	}
\end{figure}

Following shock revival, the dynamic evolution of the two models is considerably different, as exemplified by the difference in shock trajectories (Figure~\ref{Shock_Energy}(a)) and in the explosion energy histories (Figure~\ref{Shock_Energy}(b)). 
The shock grows faster in F15.79 after the explosion in that model is finally launched, catching up with the (mean) shock radius of F15.78 about 0.5~s after bounce.
The ``diagnostic energy" is the sum of the total energies (gravitational, internal, and kinetic) of all zones for which this energy is positive \citep{BuJaRa06, MuJaMa12, BrMeHi13}.
The net explosion energy is the diagnostic energy plus the binding energy of the on- and off-grid material ahead of the shock \citep{BrMeHi13}.
\citep[See][Appendix A for the details of the energies used]{BrLeHi16}.
While growth rates for the diagnostic energy are similar up to 0.3~s after bounce, the growth rate in F15.79 then accelerates while the growth rate in F15.78 declines.
Both measures of the explosion energy therefore become much larger in F15.79, with the diagnostic energy more than 3$\times$ larger at 1~s after bounce and $\approx$2.5$\times$ by the end of the simulations.
By the end, an even greater difference ($\approx$7$\times$) develops in their net energies. 
In neither model has the net energy leveled off by the end of the simulations, but if they do not change significantly during further evolution \citep[see][however]{Mull15} F15.79 is within the observed range of explosion energies for $M_{\rm MS} \sim 16 \msun$  \citep{MaAnBe22} while F15.78 falls below the observed range.

\begin{figure}
	\includegraphics[width=\columnwidth,clip]{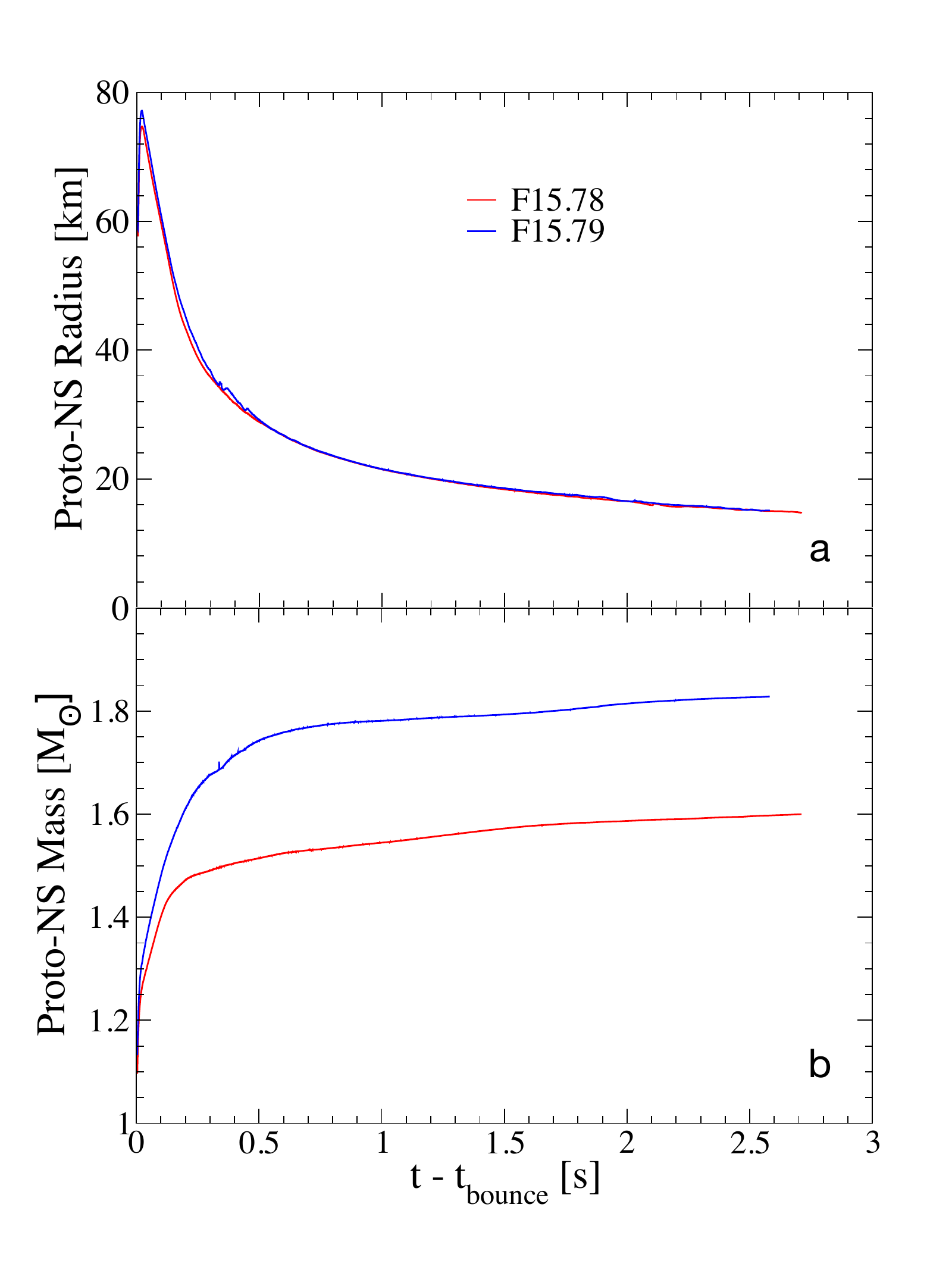}
	\caption{\label{ProtoNSMR}
	Comparison of the (a) proto-neutron radii, and (b) proto-neutron star masses of the two models as a function of post-bounce time.
	}
\end{figure}

The proto-NS radii (Figure~\ref{ProtoNSMR}(a)) of the two models are strikingly similar, while the baryonic masses (Figure~\ref{ProtoNSMR}(b)) are substantially different.
(Following common practice, we use the region with density $>10^{11}$~\gcc\ to define the proto-NS.)
The baryonic masses are still slowly increasing at the end of the simulations, at this time having attained values of 1.59 and 1.82~\msun\ for F15.78 and F15.79, respectively, which correspond to cold-NS gravitational masses of 1.46 and 1.62~\msun\ using the approximation formula of \citet{TiWoWe96}.
These masses are in the range of the observed distribution of slowly rotating pulsars \citep{ScPoRa10, OzPsNa12, OzFe16}.
While less closely constrained than the masses of binary neutron stars, slowly rotating pulsars seem a more likely outcome for single, non-rotating progenitor stars.

\begin{figure}
	\fig{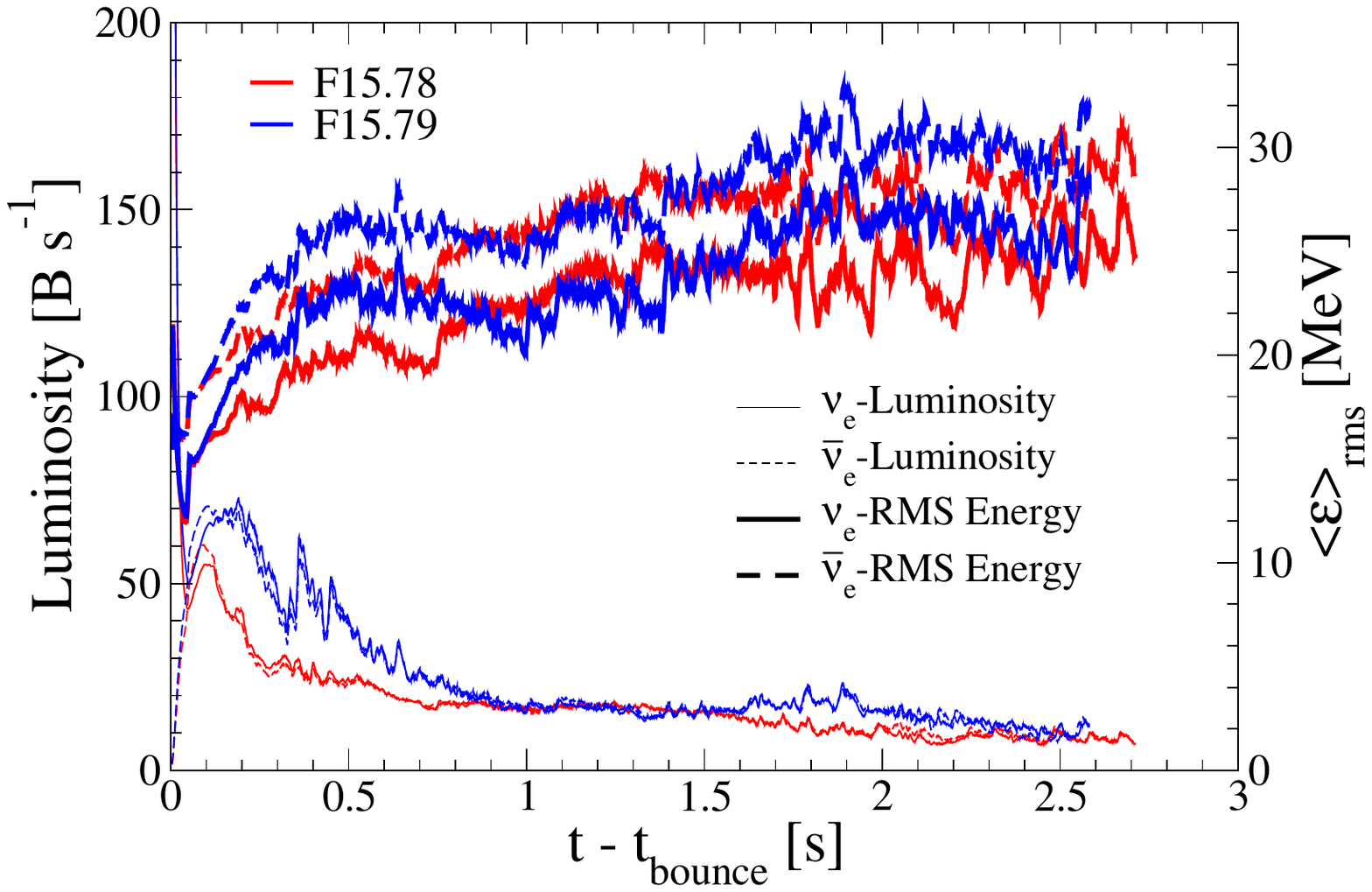}{\columnwidth}{}
	\caption{\label{LumRMS}
	The \nue\ and \nuebar\ luminosites and RMS energies at the gain layer.
	}
\end{figure}

The neutrino luminosities and rms energies for the duration of the simulation are shown in Figure~\ref{LumRMS} for the two models. There are clearly substantial differences between these quantities for the two models in the time period from 0.2 to 0.8~s and again from 1.6 to 2.3 s, with these quantities being larger for F15.79. These differences are important to the strength of the explosion, as they partially account for heating rate and heating efficiency differences between the two models that will be explored in detail below.

\begin{figure*}
	\includegraphics[scale = 0.50]{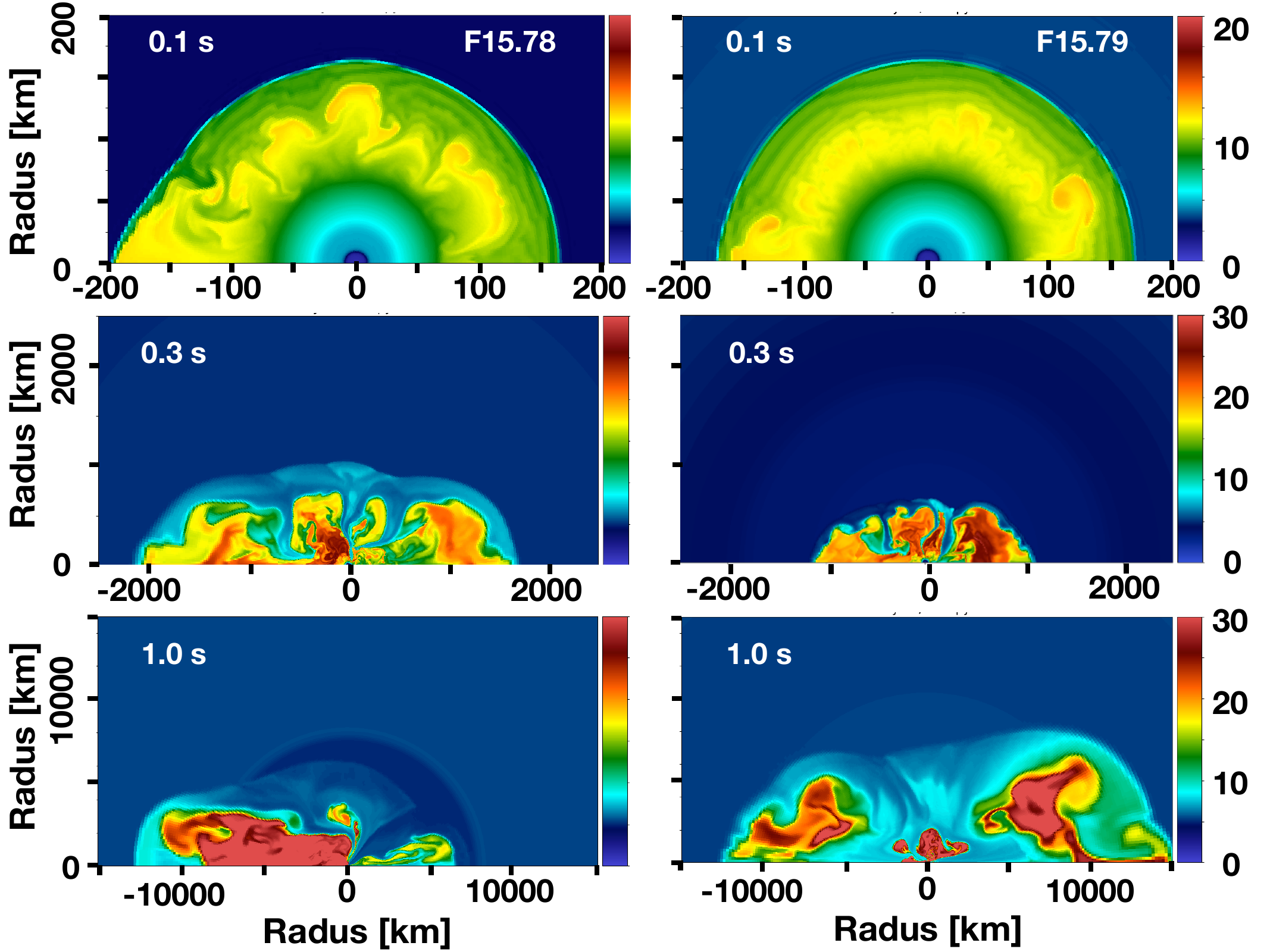}
	\caption{\label{F15_78_79_evol}
	Specific entropy  of F15.78 and F15.79 at post-bounce times of 0.1, 0.3, and 1.0~s post-bounce. The radial scales are the same for both models and are 0 to 200 km for the 0.1~s plot,  0--2500~km for the 0.3~s plot, and 0--15,000 km for the 1.0~s plot.
	}
\end{figure*}

Figure \ref{F15_78_79_evol} displays the morphological evolution of the two models by means of 2D entropy plots. 
At 0.1~s, both models exhibit neutrino driven convection, more highly developed in F15.78. 
By 0.3~s, both models have evolved into a highly prolate configuration with F15.78 having expanded farther at this time. 
At 1~s, both models have expanded much farther with F15.79 having overtaken F15.78. 
F15.78 has become almost unipolar at this time, the transition from bipolar to unipolar occurring roughly between 0.4 and 0.6~s, while F15.79 remains clearly bipolar.

Considering the diagnostic energies of the two models (Figure~\ref{Shock_Energy}(b)) again, it is clear that the major differences between the two models gets established during the first 0.6~s.
Our primary focus in Sections~\ref{sec:heating}--\ref{sec:Out} will therefore be to examine the origin of these evolutionary differences between the two models during this period of time, relating them to differences in the progenitors.
In Section~\ref{sec:Bursts}, we will take a look at the causes for the later jumps in the diagnostic energy at $\approx$1.2~s and $\approx$1.55~s in F15.78 and at $\approx$1.65~s in F15.79.

\subsection{Heating Rates}
\label{sec:heating}

To understand how two non-rotating progenitors with almost equal masses have such dissimilar explosion energy histories during the first 0.6~s, we begin with the basic driver of the explosion, the neutrino heating rate $\dot{Q}^{+}$.
Neutrino heating can be expressed as an integral over the volume specific heating rate $\dot{q}^{+}$, whose dominant contribution is given by
\begin{equation}
\dot{q}^{+} = \left\{ \frac{ X_{\rm n} }{ \lambda_{0}^{a} } \frac{ L_{\nu_{\rm e}} }{ 4 \pi r^{2} } \frac{ \langle E_{\nu_{\rm e}}^{2} \rangle }{ \langle \cal{F}_{\nu_{\rm e}} \rangle }
+ \frac{ X_{\rm p} }{ \bar{\lambda}_{0}^{a} } \frac{ L_{\bar{\nu}_{\rm e}} }{ 4 \pi r^{2} } \frac{ \langle E_{\bar{\nu}_{\rm e}}^{2} \rangle }{ \langle {\cal F}_{\bar{\nu}_{\rm e}} \rangle } \right\} \frac{\rho}{m_{\rm B}}
\label{eq:volspecheat}
\end{equation}
where $X_{\rm n/p}$  are the neutron and proton mass fractions, $L_{\nu_{\rm e}/{\bar{\nu}_{\rm e}} }$ and $E_{\nu_{\rm e}/{\bar{\nu}_{\rm e}} }$ are the \nue\ and \nuebar\ luminosities and energies, respectively, $\langle E_{\nue/\nuebar}^{2} \rangle \equiv \langle E_{\nue/\nuebar}^{3} \rangle/\langle E_{\nue/\nuebar} \rangle$, ${\cal F}_{\nu_{\rm e}/{\bar{\nu}_{\rm e}} }$ the \nue\ and \nuebar\ inverse flux factors (inverse ratios of the first to the zeroth angular moments of the neutrino distribution), $m_{\rm B}$ is the mean baryon mass, and $\lambda_{0}^{a}$, $\bar{\lambda}_{0}^{a}$ are weak interaction constants related to the absorption mean free path.
The quantities in angle brackets are spectral energy averages.

\begin{figure}
	\includegraphics[width=\columnwidth,clip]{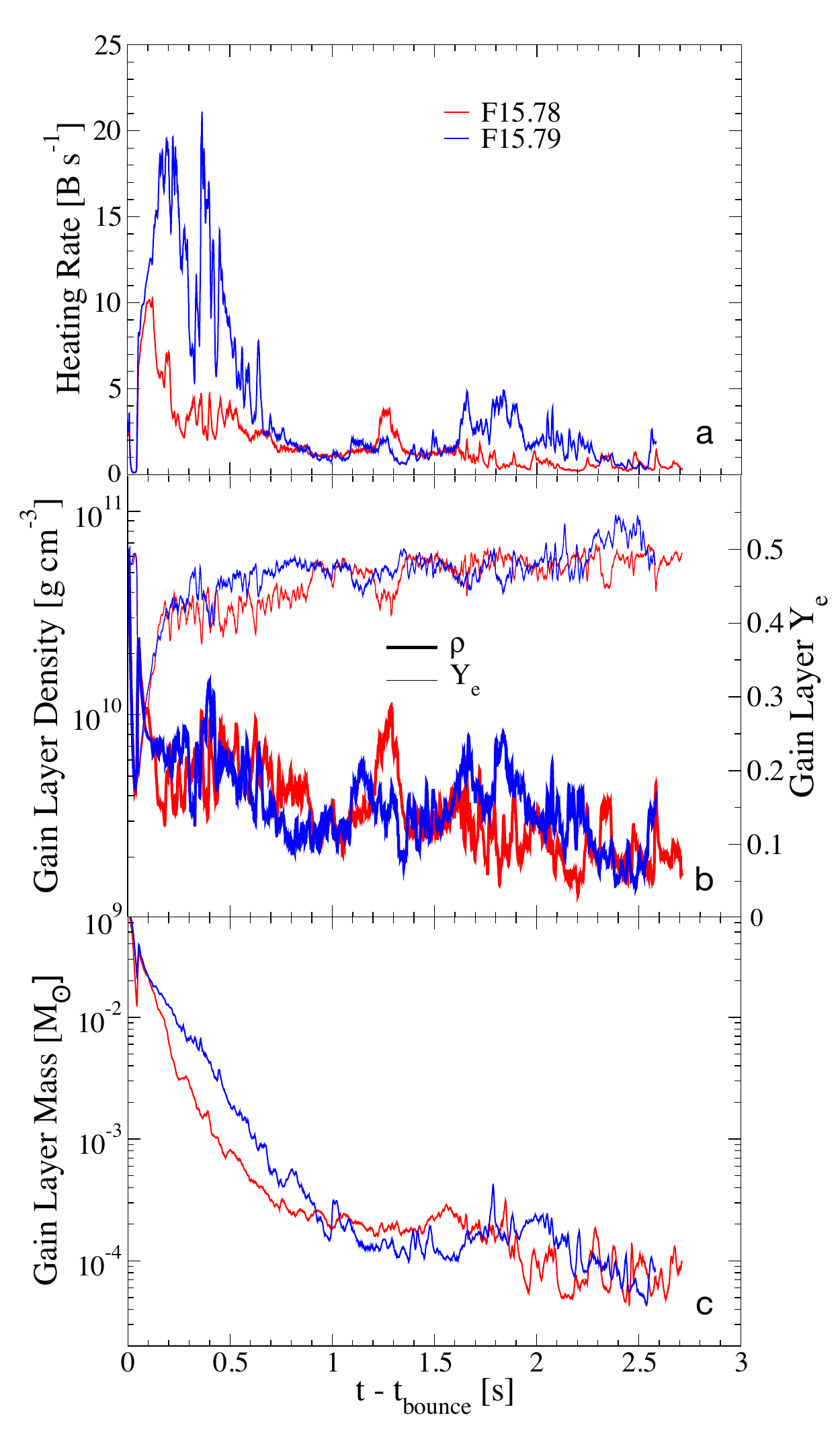}
	\caption{\label{H_Rates}
	(a) Neutrino heating rates; (b) density an electron fraction at the gain layer; (c) gain region mass
	}
\end{figure}

The total net neutrino heating rates for the two models are shown in Figure~\ref{H_Rates}(a). 
The rates for the two models are practically the same and essentially zero for the first 50 ms post-bounce, after which they are within 15--20\% of each other until 120~ms post-bounce.
At this point, the hearing rates for both models undergo a $\sim 0.5$~s duration peak with the heating rates for F15.79 exceeding those of F15.78 by a factor of 2--3 until $\approx$0.6~s post bounce. 
These heating rate peaks are coincident with the major rises in the diagnostic energies of the models (Figure~\ref{Shock_Energy}(b)). 
Both heating rates are relatively small beyond $\approx$0.6~s post bounce except for a brief rise in the rate for F15.78 at $\approx$1.2--1.3~s and a rise in the F15.79 rate from $\approx$1.65--2.3~s that will be discussed in Section~\ref{sec:Bursts}.

To examine the origin of the differences in the heating rates between the two models, we compare each of the factors that determine these rates as given by Equation~(\ref{eq:volspecheat}), namely, the radius r, the neutrino luminosities and rms energies, the neutron and proton mass fractions, the flux factors, and the gain region masses. 

The compositional mass fraction and neutrino flux factor differences do not contribute much to the differences in the heating rates.
The matter throughout much of the gain region consists of free neutrons and protons, so the sum of the nucleon mass fractions, $X_{\rm n}$ and $X_{\rm p}$ in Equation~(\ref{eq:volspecheat}) is unity. 
The partitioning of the nucleons between $X_{\rm n}$ and $X_{\rm p}$ depends on the the value of the electron fraction, \Ye\ at the gain radius (Figure~\ref{H_Rates}(b)), which is about 10\% larger for for F15.79 from 0.5 to 0.9~s and otherwise nearly the same in both models.
The factors multiplying $X_{\rm n}$ and $X_{\rm p}$ in Equation~(\ref{eq:volspecheat}) differ by as much as 16\%, due mainly to the luminosity and rms energy differences (Figures~\ref{Nu_Eff}(a) and \ref{LumRMS}), but when multiplied by a maximum 10\% difference in the mass fractions, the net contribution to the heating rate differences is rather minor.
The flux factors depend, to first order, on the ratio of the neutrinosphere radii to the gain layer radii, and these are approximately the same as a function of post-bounce time for the two models.

The primary cause for the heating rate differences is the differences in the \nue\ and \nuebar\ luminosities and rms energies of the two models, shown in Figures~\ref{Nu_Eff}(a) and \ref{LumRMS}. 
The neutrino quantities for F15.79 exceed those for F15.78 from $\approx$0.2~s to $\approx$0.8~s, and again from $\approx$1.6~s to $\approx$2.3~s reflecting approximately the same difference pattern in the neutrino heating rates shown in Figure~\ref{H_Rates}(a).
The \nue\ and \nuebar\ luminosity and rms energy differences collectively account for about a factor of 2 difference in the heating rates in the first 0.6 s. 

\begin{figure}
	\includegraphics[width=\columnwidth,clip]{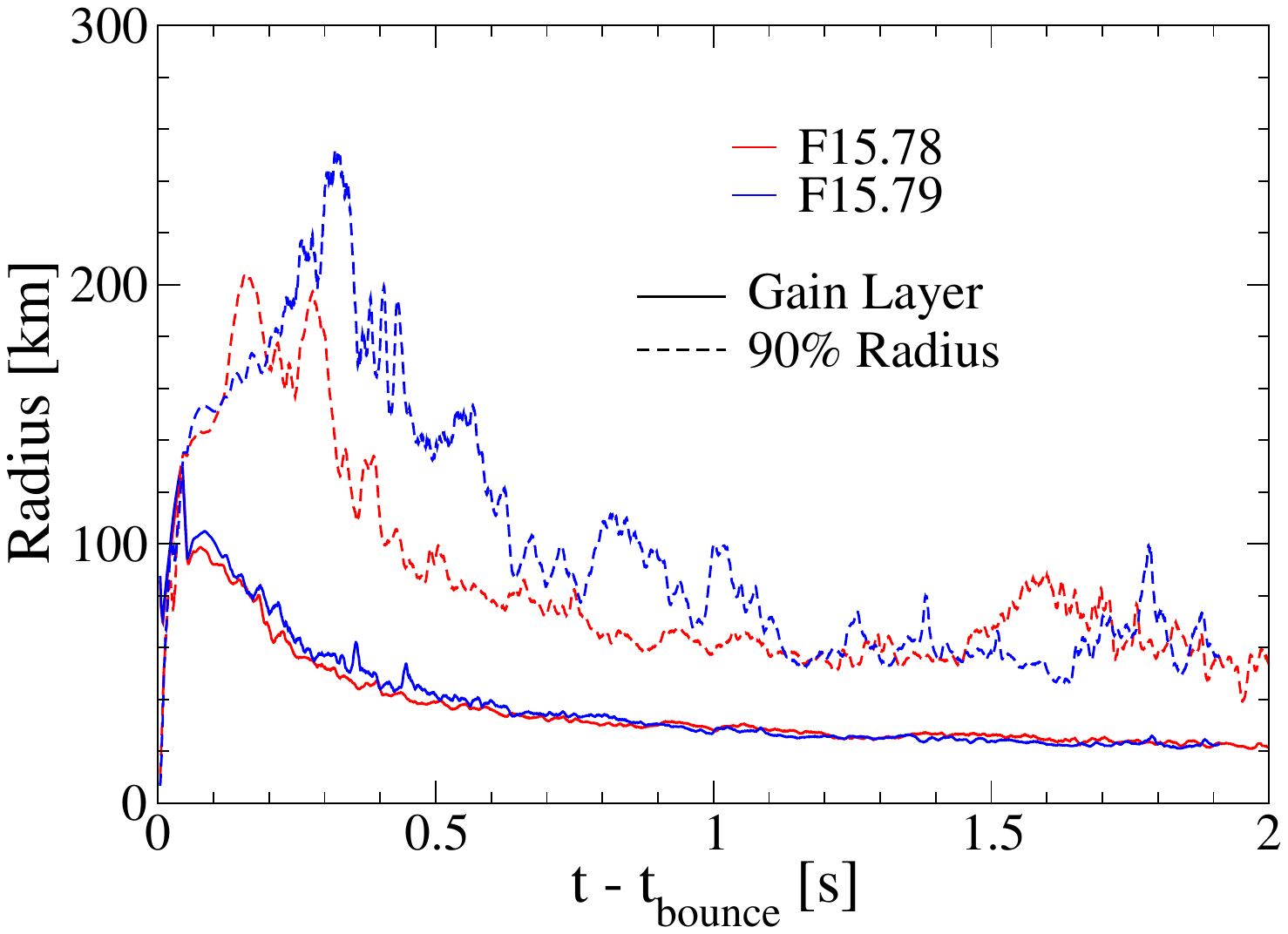}
	\caption{\label{H_Radii}
	Gain layer radii and radii enclosing 90\% of the heating rate (smoothed).
	}
\end{figure}

The other primary contribution to the heating rate differences between the two models is the difference in their gain region masses (Figure~\ref{H_Rates}(c)), defined here as the mass lying between the gain layer and the surface enclosing 90\% of the heating rate (Figure~\ref{H_Radii}).
Like the \nue\ and \nuebar\ luminosities and rms energies, the mass of the gain region for F15.79 exceeds that of F15.78 by as much as a factor of 2 until about 0.9 s. 
This can be traced to the larger densities in the matter surrounding the inner core of the 15.79~\msun\ pre-SN progenitor and its consequently larger accretion rates during the 0.2--0.9~s time period, as will be discussed in Section~\ref{sec:Lum} with our discussion of the accretion luminosity.
After 1.1~s, and particularly after 1.5~s, the gain region mass of F15.78 undergoes a surge and tends to exceed that of F15.79 for $\approx$0.6~s, while that of F15.79 undergoes a surge after 1.65~s and tends to exceed that of F15.78 again for $\approx$0.3~s, though the masses involved have become very small, $\sim 10^{-4}$~\msun.
These late surges in the gain region masses are associated with uptakes in the diagnostic and total energies and will be explored in Section~\ref{sec:Bursts}.

\begin{figure}
	\includegraphics[width=\columnwidth,clip]{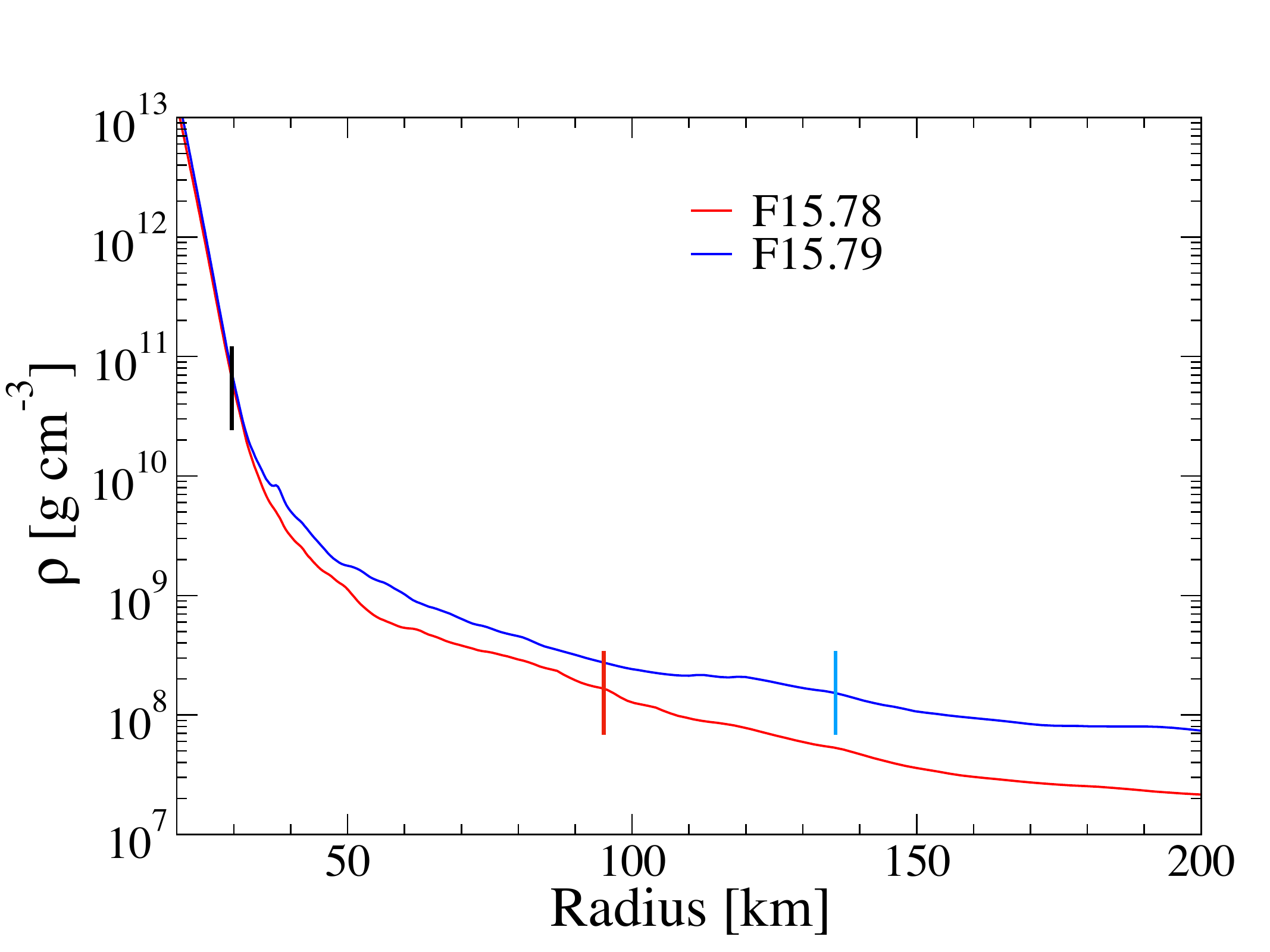}
	\caption{\label{rho_vs_r}
	Density of the models against radius at 0.5~s post-bounce. The black line segment locates the approximate radii of the \nue\ and \nuebar\ neutrinospheres. The red and blue line segments mark the radius within which 90\% of the heating rate is contained for F15.78 and F15.79, respectively. 
	}
\end{figure}

The greater gain region mass, and therefore heating rate contribution in F15.79 from $\approx$0.2--0.9~s arises from two factors: the larger (on average) density of the F15.79 gain region and its greater radial extent.
Figure~\ref{rho_vs_r} illustrates the differences in the density profiles of the two models at 0.5~s post-bounce. 
While the density is approximately the same at the base of the gain region for the two models, the figure shows that the density of F15.79 exceeds that of F15.78 as we move outward through the rest of the gain region and that the gain region of F15.79 extends well beyond that of F15.78. 
This is also evident in Figure~\ref{H_Radii}, showing the gain radius and the radius enclosing 90\% of the neutrino heating rate. 
The density profile at 0.5~s is considerably flatter for F15.79 than for F15.78 with the density in F15.78 decreasing more rapidly with radius.
This leads to a smaller heating region for F15.78 as follows from the radial dependence of the heating rate per unit radius, $\partial \dot{\cal{Q}}^{+}(r)/\partial r$ given by
\begin{equation}
\frac{ \partial \dot{\cal{Q}}^{+}(r)}{\partial r} = 4 \pi r^{2} \dot{q}^{+}
\label{eq:6}
\end{equation}
where $\dot{q}^{+}$ is the volume specific heating rate given by Equation~(\ref{eq:volspecheat}).
The mean free paths in the heating region tend to be large and the neutrino luminosities in Equation~(\ref{eq:volspecheat}) are therefore roughly constant across this region, which leads to $\dot{q}^{+} \propto \rho/r^{2}$, where the $1/r^{2}$ factor is due to the radial dilution of the neutrino flux. 
Thus, the radial dependence cancels out and  $\partial \dot{\cal{Q}}^{+}(r)/\partial r$  scales with $\rho$, and the more rapid decrease of $\rho$ with $r$ in F15.78 results in its smaller thickness of the heating region. 

\begin{figure}[h]
	\includegraphics[width=\columnwidth,clip]{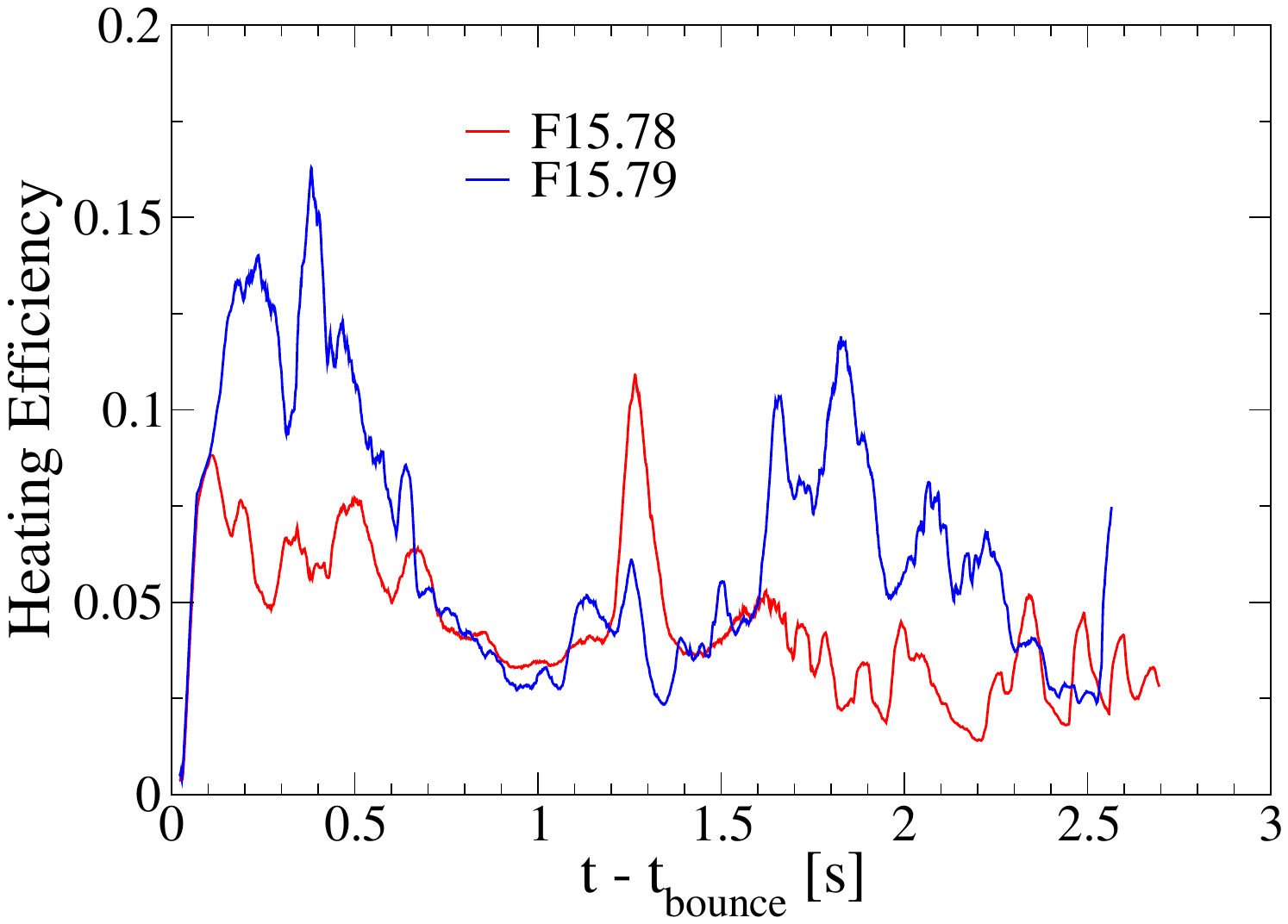}
	\caption{\label{heat_eff_all}
	Neutrino heating efficiencies as defined by Equation~\ref{h_eff}. 
	}
\end{figure}

A comparison that encapsulates most of the factors accounting for the differences in the heating rates between the two models is a comparison of their heating efficiencies, as defined by Equation~(\ref{h_eff}) and plotted for the first 0.3~s in Figure~\ref{Nu_Eff} and for the full simulations in Figure~\ref{heat_eff_all}.
To the extent that the factors multiplying $L_{\nue}$ and $L_{\nuebar}$ in Equation~(\ref{eq:volspecheat}) are equal, the heating efficiency is just given by that equation without the factors $L_{\nue}$ and $L_{\nuebar}$.
Referring to Figure~\ref{heat_eff_all} in particular, it is clear that the heating efficiencies of F15.79 are a factor 2--3 times those of F15.78 in the critical 0.2--0.6~s period due to its larger rms \nue/\nuebar\ energies and heating region mass. 
This reflects the corresponding differences in the heating rates, shown in Figure~\ref{H_Rates}(a).
Figure~\ref{heat_eff_all} also shows succeeding uptakes in the heating efficiencies of F15.78 at 1.2~s and 1.55~s and of F15.79 at 1.65~s.
These are also associated with uptakes in the explosion energies and are discussed in Section~\ref{sec:Bursts}.

We conclude that the primary causes for the differences in the heating rates for the two models are the differences in the \nue\ and \nuebar\ luminosities and rms energies, and the differences in the masses and density profiles of the gain regions. These in turn will be related below to the differences in the internal structures of the progenitors.

\subsection{Neutrinospheres}
\label{sec:muspheres}

The differences between the neutrino luminosities and rms energies of the two models are related to differences in the geometric and thermodynamic conditions of their mean neutrinospheres, specifically, to corresponding differences in their radii and temperatures.
The use of mean neutrinospheres is an approximation, of course, but useful for a first order understanding of the differences in the neutrino luminosities and rms energies between the two models.

\begin{figure}
	\includegraphics[width=\columnwidth,clip]{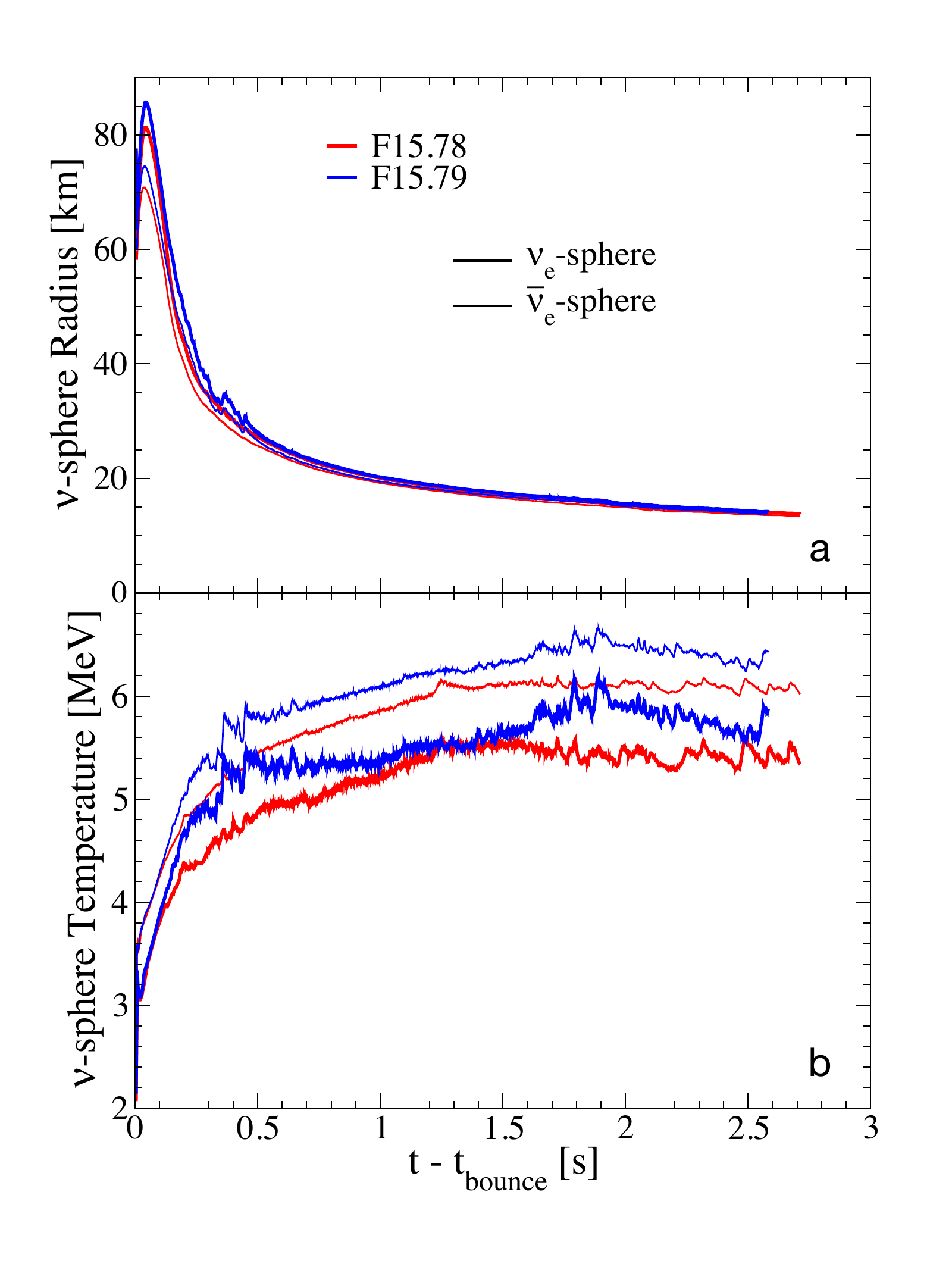}
	\caption{\label{nu_sphere}
	\nue/\nuebar neutrinopheric radii (a) and temperatures (b) as a function of post-bounce time.
	}
\end{figure}

Figure~\ref{nu_sphere}(a) plots the \nue/\nuebar-sphere radii.
The \nue/\nuebar-sphere radii of F15.79 exceeds those of F15.78 by as much as 5 km shortly after bounce, but thereafter diminishes to a couple of km at 0.25 s, and less than 1 km after 0.5 s.
The nearly equal \nue/\nuebar-sphere radii, together with the nearly equal proto-NS radii noted earlier, reflect, in part, the nearly constant mass-radius relation of the SFHo EoS for cold matter \citep{StHeFi13}.
These initial differences in radii may explain the differences in the luminosities of the two models during the first several hundred ms, during which times their $\nue/\nuebar$-sphere temperatures (Figure~\ref{nu_sphere}b) are almost equal, but \nue/\nuebar-sphere radius differences do not play a direct role in the $\nue/\nuebar$ luminosity and rms energy differences after that.

The near equality of the $\nue/\nuebar$-sphere radii of the two models does imply, however, that their luminosity and rms energy differences are simply related. 
To the extent that Stefan's and Wien's laws (applied to fermions) are applicable here, the $\nue/\nuebar$ luminosities are $\propto T^{4}$ while the rms energies are $\propto T$.
To first order, we therefore expect that
\begin{equation}
\frac{\Delta L_{\nue/\nuebar} }{ L_{\nue/\nuebar} } = 4 \frac{\Delta E_{\nue/\nuebar} }{ E_{\nue/\nuebar} }
\label{eq:7}
\end{equation}
where $\Delta L_{\nue/\nuebar}$ and $\Delta E_{\nue/\nuebar}$ are the $\nue/\nuebar$ luminosity and rms energy differences, respectively, between the two models.
Referring to Figure~\ref{LumRMS}, it is evident, despite the noise, that in the time period from 0.2~s to 0.8~s, during which the differences in neutrino luminosity and rms energy between the two models are significant, Equation~(\ref{eq:7}) is approximately satisfied; likewise in the time period from 1.65~s to 2.3~s.
At other times, the luminosities and rms energies are roughly similar.
Our focus will now turn to the differences in the neutrino luminosities between the two models, keeping in mind the fact that the corresponding neutrino rms energy differences are related to the neutrino luminosities differences as given approximately by Equation~(\ref{eq:7}). 

\subsection{Distinguishing Core and Accretion Luminosities}
\label{sec:Lum}

To relate the differences in the \nue\ + \nuebar\ luminosities and rms energies of the models to differences in their progenitor structures, it is necessary, at least approximately, to separate from the total \nue\ + \nuebar\ luminosities, the component that emanates from the core (the core luminosity) and the component that arises from the energy released by recent mass accretion onto the proto-NS surface (the accretion luminosity), as these components arise from different features of the progenitors.
The accretion luminosity mirrors in time the radial density structure of the progenitor, and structural changes in the accretion streams as they affect the mass accretion rate onto the proto-NS, while the core luminosity is governed more by the time integrated assembly history of the proto-NS.

\begin{figure}
	\includegraphics[width=\columnwidth,clip]{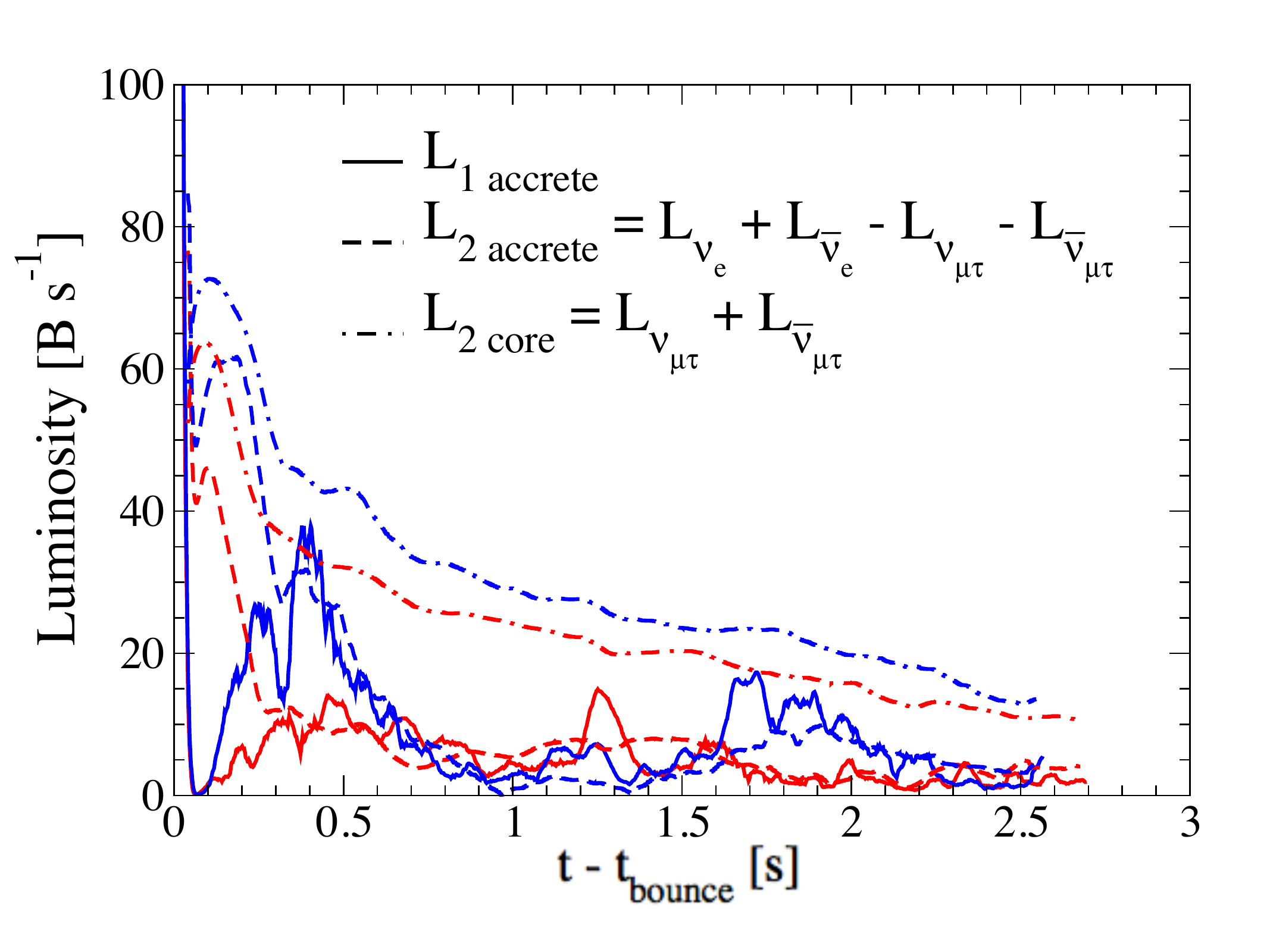}
	\caption{\label{L_accrete}
	Accretion luminosity as defined by Equation~(\ref{eq:7}) (solid lines), the accretion luminosity as defined by Equation~(\ref{eq:8}) (dashed lines), and the core luminosity as defined by the sum of $L_{\numt} + L_{\numtbar}$ (dot-dashed lines) assuming that the core luminosities are the same for each neutrino flavor. 
	}
\end{figure}

One measure of the \nue\ + \nuebar\ accretion luminosity, which we will denote by $L_{1 \, \rm accrete}$, is given by the rate of kinetic energy accretion, namely,
\begin{equation}
L_{1 \, \rm accrete} = \sum_{j, v_{r} < 0} \frac{1}{2} \left( v_{r}^{2} + v_{\theta}^{2} \right)_{j, \rm r\_gain} \times \dot{M}_{j, \rm r\_gain}
\label{eq:8}
\end{equation}
where, for each radial ray $j$, $\dot{M}_{j, \rm r\_gain}$ is the rate of mass flow through the gain radius, and $v_{r}$ and $v_{\theta}$ are the radial and angular velocities at the gain radius.
The sum is over all radial rays for which the fluid has negative radial velocities at the gain radius. 
We choose the gain radius to compute the rate of kinetic energy accretion as it is close to the surface of the proto-NS and is most likely above the radius at which much of the flow is disrupted and thermalized.
$L_{1 \, \rm accrete}$ for the two models is plotted by the solid lines in Figure~\ref{L_accrete}.

An alternative scheme for distinguishing between the core and accretion \nue\ + \nuebar\ luminosities is to use the fact that, after the initial \nue\ burst at shock breakout, the core luminosity of each neutrino flavor is approximately the same \citep{Jank95}. 
Given this, the \nue\ + \nuebar\ accretion and core luminosities, denoted respectively by $L_{2 \, \rm accrete}$ and $L_{2 \, \rm core}$, are computed by
\begin{eqnarray}
L_{2 \, \rm accrete} &=&  L_{\nue} + L_{\nuebar} - L_{\numt} - L_{\numtbar} , \nonumber \\
L_{2 \, \rm core} &=& L_{\numt} + L_{\numtbar},
\label{eq:9} 
\end{eqnarray}
where, according to the above assumption, the sum of the \nue\ and \nuebar\ core luminosities is given approximately by the sum of the \numt\ and \numtbar\ luminosities, and the accretion luminosities are given by the difference between the total and the core luminosities \citep[e.g.,][]{MuJa14}. 
Here $L_{\numt}$ ($L_{\numtbar}$) refers to one of $L_{\nu}$ or $L_{\tau}$ ($L_{\numubar}$ or $L_{\nutaubar}$). 
$L_{2 \, \rm accrete}$ as defined by Equation~\ref{eq:9} is plotted in Figure~\ref{L_accrete} by the dashed lines, and the core luminosities as represented by the sum of the \numt\ and \numtbar\ luminosities are plotted by the dot-dashed lines.

Both methods of distinguishing accretion from core luminosities are only as good as the underling assumptions in their definitions. 
$L_{1 \, \rm accrete}$ assumes that the \nue\ + \nuebar\ accretion luminosity is given by the rate of kinetic energy accretion through the gain layer, and $L_{2 \, \rm accrete}$ suffers from its dependence on the assumption that the core luminosities of each neutrino flavor are the same, and from the numerically precarious procedure of obtaining  the accretion luminosities by the subtraction of two large, comparably sized numbers.
Despite their limitations, both methods of defining $L_{\rm accrete}$ agree rather well, with a few exceptions.
Before 0.3~s, the \numt\ + \numtbar\ luminosity may be considerably smaller than the \nue\ + \nuebar\ core luminosity, invalidating the underlying the assumption of their equality, and causing $L_{2 \, \rm accrete}$ to be much larger than $L_{1 \, \rm accrete}$.
Conversely, $r_{\rm gain}$ is large during this period (Figure~\ref{H_Radii}), and Equation~(\ref{eq:8}) may be underestimating the kinetic energy inflow.
Despite these discrepancies, both indicate that the core and accretion luminosities of F15.79 exceed those of F15.78 in the interval from 0.1 to 0.8 s, during which time most of the diagnostic energies of the models are acquired, and the differences between the models established.
The differences between $L_{1 \, \rm accrete}$ and $L_{2 \, \rm accrete}$ after 0.3~s arise from 1.2 to 1.6~s in F15.78, where $L_{1 \, \rm accrete}$ exhibits several large fluctuations about $L_{2 \, \rm accrete}$, not seen in Figure~(\ref{LumRMS}), and from 1.65 to 2.0~s in F15.79 where $L_{1 \, \rm accrete}$ exhibits several large peaks relative to $L_{2 \, \rm accrete}$, again not seen in Figure~(\ref{LumRMS}).
These disagreements occur at the onset of accretion events (cf. Figure~\ref{Mass_Accrete}) where quantities are changing rapidly, and it may be that the direct relation between the inflow of kinetic energy and the accretion luminosity breaks down at these events. 
Despite their differences, both measures of the accretion luminosity increase during the above intervals, and are associated with uptakes in the explosion energy of the models, as can be seen in Figure~\ref{Shock_Energy}(b). 
These accretion events and their effect on the explosion energy will be discussed in more detail in Section~\ref{sec:Bursts}. 

\begin{deluxetable}{cccccc}
\tablecaption{Cooling time scales ($\Delta t_{\rm cool}$) for the indicated layers in the proto-NS\label{tab:Cool} for averaged model.}
\tablecolumns{5}
\tablewidth{0pt}
\tablehead{
\colhead{} & \colhead{$t - t_{\rm b} = 0.3$~s} & \colhead{$t - t_{\rm b} = 0.5$~s} & \colhead{$t - t_{\rm b} = 0.7$~s} \\
\cline{2-4}
\colhead{Shell [\gcc]}  & \multicolumn{3}{c}{$\Delta t_{\rm cool}$ [s]} 
}
\startdata
$10^{9...10}$    & $3.9 \times 10^{-3}$ & $2.0 \times 10^{-3}$ & $8.5 \times 10^{-4}$ \\
$10^{10...11}$  & $1.0 \times 10^{-2}$ & $3.4 \times 10^{-3}$ & $1.6 \times 10^{-3}$ \\	
$10^{11...12}$  & $4.8 \times 10^{-2}$ & $1.3 \times 10^{-2}$ & $1.0 \times 10^{-2}$ \\
$10^{12...13}$  & $2.9 \times 10^{-1}$ & $1.4 \times 10^{-1}$ & $1.1 \times 10^{-1}$
\enddata
\end{deluxetable}  

A further approach at distinguishing accretion from core luminosities, which we introduce below as an added corroboration of our above methods, utilizes the fact that the accretion luminosity is generated near the proto-NS's surface where down flows are shock-terminated and partially thermalized while the core luminosity arises from the neutrino production and diffusion from deeper layers.
We use the additional fact that the cooling time scales, which we define below, increase for successively deeper spherical regions in the proto-NS when subjected to a given outflow of energy.
These time scales are listed in Table~\ref{tab:Cool}, and are computed from
\begin{equation}
\Delta t_{\rm cool} = \left( \frac{3}{2} k \frac{ \Delta M }{ m_{\rm B} } \right)  \left( L_{\rm in} - L_{\rm out} \right)^{-1} \Delta T_{\rm cool},
\label{eq:10}
\end{equation}
where $\Delta M$ is the mass of a given spherical region in the proto-NS, $m_{\rm B}$ is the average neutron-proton mass, $L_{\rm in} - L_{\rm out}$ is the neutrino luminosity into minus the neutrino luminosity out of $\Delta M$, and $\Delta T_{\rm cool}$ is a given change in temperature.
The volume is assumed constant and the heat capacity assumes the nucleons behave like an ideal gas with negligible contribution from the partially degenerate electrons.
In computing the cooling time scale entries of Table~\ref{tab:Cool} from Equation (\ref{eq:10}), we have taken $\Delta M$ to be the mass of the spherical region enclosed between the indicated density limits, the net luminosity $( L_{\rm in} - L_{\rm out} )$ to be -1~\Bethes, and $\Delta T_{\rm cool}$ to be -1~MeV.
Both models provide similar values for $\Delta M$ in corresponding regions, and we have used the average of the two in the computation of the table entries.
The fiducial times from bounce chosen to cover the period during which the bulk of the diagnostic energy of both models is acquired are 0.3, 0.5, and 0.7~s.

Given the cooling time scales listed in Table~\ref{tab:Cool} for the conditions described above, distinguishing core from accretion luminosities at a given post-bounce time, $t_{0}$, can be accomplished, at least approximately, by taking as initial conditions the configuration of the model at the time $t_{0}$ and continuing its simulation with the hydrodynamics and $Y_{\rm e}$ frozen, thereby eliminating the effects of accretion. 
With an appropriately chosen time interval $t - t_{0}$ for continuing the frozen hydro/$Y_{\rm e}$ simulation, neutrino transport will quickly cool the accretion hot spots near the proto-NS surface with their small cooling time scales, thereby eliminating the accretion luminosity, while transport from the deeper layers, with their much larger cooling time scales, will be largely unaffected.
The luminosity remaining at the end of the above time interval should therefore be the core luminosity. 
Clearly this mode of separating core from accretion luminosities is not completely clean as there must be some overlap in their sources. 
It should provide semi-quantitative results, however, which can be compared with the previous methods.

\begin{figure*}
	\includegraphics[scale = 0.50]{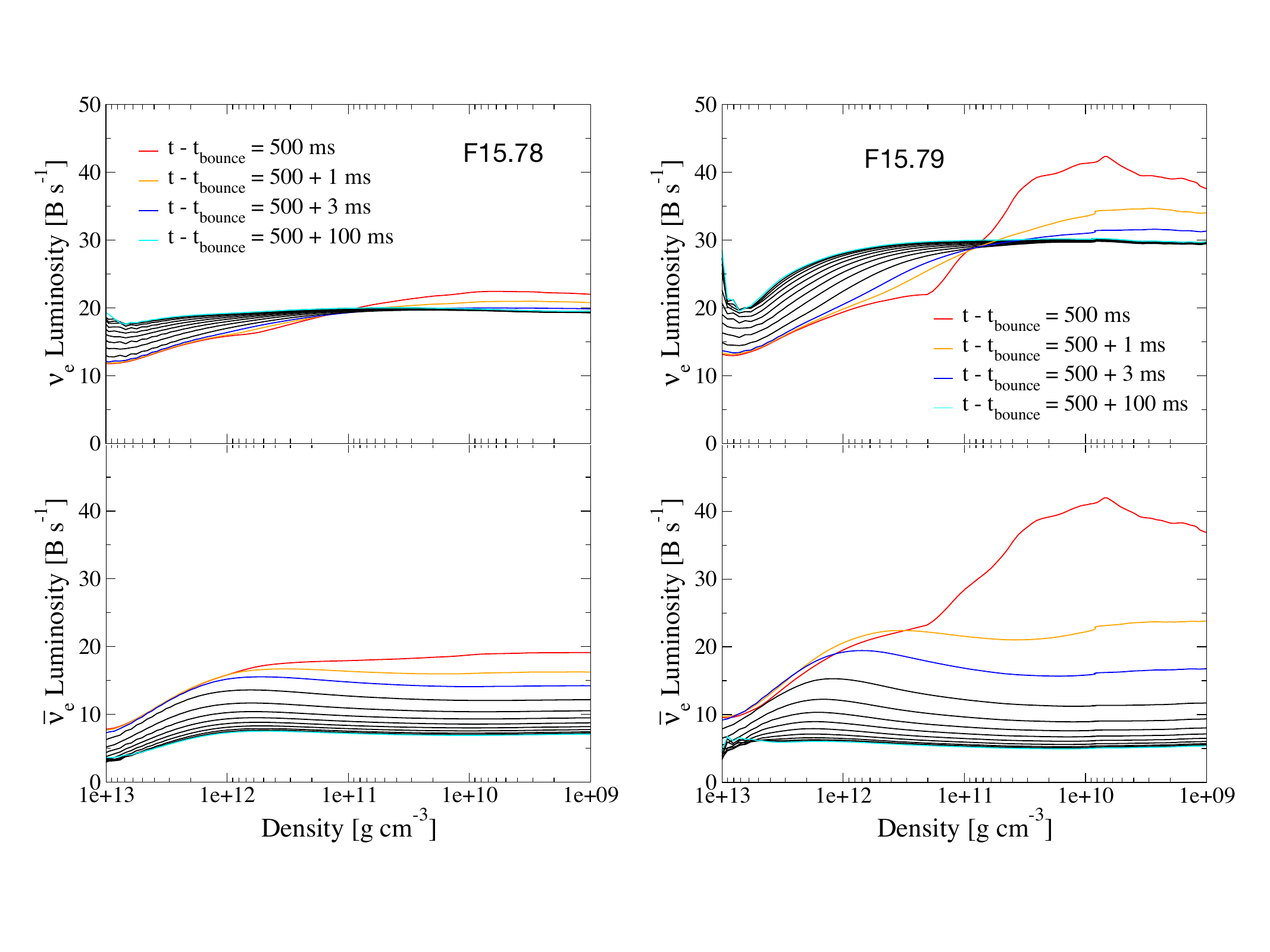}
	\caption{\label{F15_78_79_500}
	For F15.78 (left panels) and F15.79 (right panels), the upper and lower panel show the \nue\ and \nuebar\ luminosity, respectively, of the model at 500 ms by the red line. 
The orange, blue, and cyan lines show the luminosities at the indicated times after 500 ms during which the hydro and $Y_{\rm e}$ were frozen. 
The black lines show the frozen hydro and $Y_{\rm e}$ luminosities from 10 ms to 90 ms in 10 ms intervals generated from the 500 ms initial configuration.
	}
\end{figure*}

To determine values for $t - t_{0}$ that will be just sufficient to eliminate the effects of accretion, we have taken the spherical region in the proto-NS between $10^{10}$ and $10^{11}$ g cm$^{-3}$ as representative of the region where most of the accretion luminosity is generated. 
Values of $( L_{\rm in} - L_{\rm out} )$ at time $t_{0}$ from this region turn out to be typically about $- 4$ B s$^{-1}$ for F15.78 and about 3 times that in magnitude for F15.79. 
With these values of $( L_{\rm in} - L_{\rm out} )$ together with the time scale values of $\Delta t_{\rm cool}$ listed in Table~ \ref{tab:Cool} and computed with $( L_{\rm in} - L_{\rm out} )$ = $- 1$ B s$^{-1}$, we have chosen $t - t_{0} = 3$ ms for both F15.78 and F15.79.
This choice is further suggested by Figure~\ref{F15_78_79_500}, which plots snapshots of the \nue\ and \nuebar\ luminosities at 0.5~s of the frozen hydro/$Y_{\rm e}$ simulations, and the corresponding figures for the other fiducial times. 
These figures show that the \nue\ luminosity ceases to change rapidly after 3 ms, indicating that the source of accretion luminosity has been radiated away, and furthermore that the \nuebar\ luminosity begins to decrease at densities above $10^{12}$ g cm$^{-3}$ at 3 ms, indicating that the effects of the core luminosity are beginning to appear.
From the results of these simulations at the fiducial times, the core and accretion luminosities as given by this procedure, and denoted by $L_{3 \, \rm core}$ and $L_{3 \, \rm accrete}$, respectively, are listed for both F15.78 and F15.79 in Table~\ref{tab:f1578_9}.

\begin{deluxetable}{ccccc}
\tablecaption{Core and accretion luminosities\label{tab:f1578_9}}
\tablecolumns{5}
\tablewidth{0pt}
\tablehead{
\colhead{$t - t_{\rm b}$} & \colhead{$L_{3 \, \nue {\rm accrete}}$} & \colhead{$L_{3 \, \nue {\rm core}}$} & \colhead{$L_{3 \, \nuebar {\rm accrete}}$} & \colhead{$L_{3 \, \nuebar {\rm core}}$}\\
\colhead{[s]} & \colhead{[B s$^{-1}$]} & \colhead{[B s$^{-1}$]} & \colhead{[B s$^{-1}$]} & \colhead{[B s$^{-1}$]} 
}
\startdata
\multicolumn{5}{c}{F15.78}\\
\hline
0.3 & 3 & 25 & 6 & 18\\
0.5 & 3 & 20 & 5 & 14\\	
0.7 & 4 & 15 & 6 & 15\\
 \hline   
 \multicolumn{5}{c}{F15.79}\\   
 \hline                     
0.3 & 8 & 44 & 28 & 20\\
0.5 & 10 & 32 & 25 & 17\\	
0.7 & 5 & 22 & 11 & 15 
\enddata
\end{deluxetable}

A comparison of the core and accretion luminosities, $L_{2 \, \rm core}$ and $L_{2 \, \rm accrete}$, given by Equation~(\ref{eq:9}), with $L_{3 \, \rm core}$ and $L_{3 \, \rm accrete}$ is presented in Table~\ref{tab:L2_L3} for the post-bounce fiducial times of 0.3, 0.5, and 0.7 s. 
As in the definition of $L_{2 \, \rm core}$ and $L_{2 \, \rm accrete}$, each of the quantities $L_{3 \, \rm core}$ and $L_{3 \, \rm accrete}$ is the sum of the \nue\ and \nuebar\ core and accretion luminosities, respectively. 
Several trends are evident from this table. 
Firstly, for both F15.78 and F15.79, the core and accretion luminosities given by Equation (\ref{eq:9}) and by the frozen hydro/$Y_{\rm e}$ simulations agree rather closely for $t - t_{0} =$ 0.5 and 0.7~s.
On the other hand, at $t - t_{0} =$ 0.3~s the core and accretion luminosities given by Equation (\ref{eq:9}) are respectively smaller and greater than those given by the frozen hydro/$Y_{\rm e}$ simulations. 
The origin of these differences is not clear, but a similar pathology was previously apparent in the comparison of $L_{1 \, \rm accrete}$ and $L_{2 \, \rm accrete}$ at early times, as shown in Figure~\ref{L_accrete}, indicating the difficulty of separating core and accretion luminosities at these early times.
Secondly, common to both methods of computing the core and accretion luminosities is the fact that the accretion luminosities are considerably smaller than the core luminosities during this time, and that both luminosities are larger for F15.79 than F15.78.
A takeaway from a comparison of the $L_{2}$ and $L_{3}$ core and accretion luminosities is that the $L_{2}$ (and therefore the $L_{1}$) luminosities plotted in Figure~\ref{L_accrete} likely represents semiquantitatively the actual core and accretion luminosities for the models at times greater than 0.3 s. 
As noted above, for times less than 0.3~s it is likely that the assumption underlying Equation~(\ref{eq:9}) is not valid and/or the accuracy of Equation~(\ref{eq:8}) in accounting for the rate of kinetic energy infall at the radius where it is thermalized may be off.
In the next two subsections we will attempt to account for these differences in the core and accretion luminosities between the two models, particularly for the period 0.12--0.8~s during which most of the diagnostic energy is acquired.

\begin{deluxetable}{ccccc}
\tablecaption{Comparison of core and accretion luminosities\label{tab:L2_L3}}
\tablecolumns{5}
\tablewidth{0pt}
\tablehead{
\colhead{$t - t_{\rm b}$} & \colhead{$L_{2 \, {\rm accrete}}$} & \colhead{$L_{3 \, {\rm accrete}}$} & \colhead{$L_{2 \, {\rm core}}$} & \colhead{$L_{{3 \, \rm core}}$}\\
\colhead{[s]} & \colhead{[B s$^{-1}$]} & \colhead{[B s$^{-1}$]} & \colhead{[B s$^{-1}$]} & \colhead{[B s$^{-1}$]} 
}
\startdata  
\multicolumn{5}{c}{F15.78}\\
\hline                      
0.3 & 13 & 9 & 39 & 43\\
0.5 &   8 & 7 & 33 & 34\\	
0.7 &   7 &   7 & 30 & 30\\
 \hline   
 \multicolumn{5}{c}{F15.79}\\   
 \hline                          
0.3 & 48 & 36 & 51 & 64\\
0.5 & 34 & 35 & 47 & 47\\	
0.7 & 15 & 16 & 38 & 37 
\enddata
\end{deluxetable}

\subsection{Accretion Luminosity Differences}
\label{sec:acc_lum}

\begin{figure}
	\includegraphics[width=\columnwidth,clip]{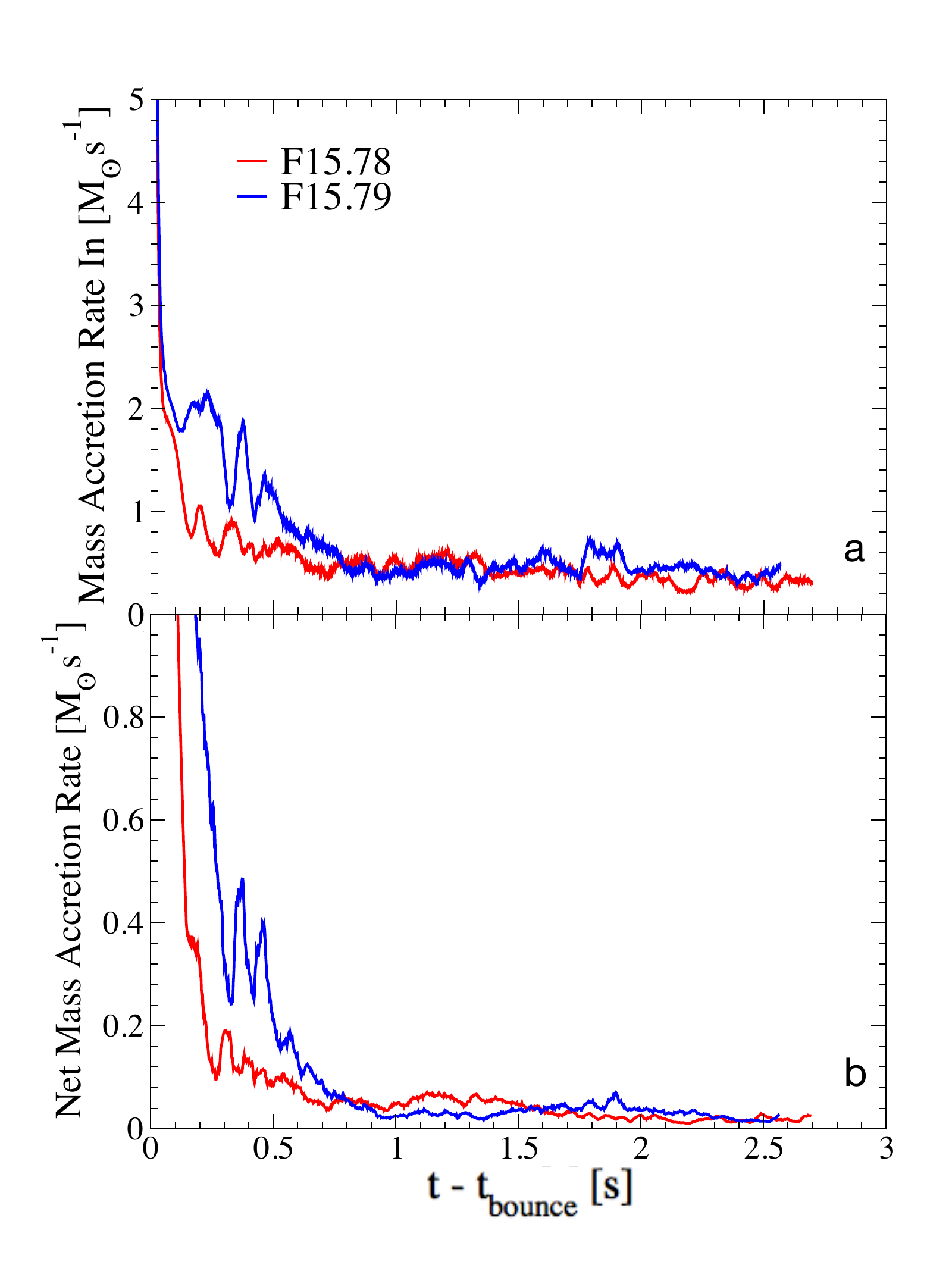}
	\caption{\label{Mass_Accrete}
	(a) Inflowing mass accretion rate through through the gain layer $r_{\rm gain}$; (b) net mass accretion rate through the same surface defined as the inward minus the outward mass accretion rate.
	}
\end{figure}

A counterpart to the measure of the accretion luminosity $L_{1 \, \rm accrete}$, defined by Equation~(\ref{eq:8}), is the rate of mass accretion itself, which is directly related to the density structure of the progenitor, mediated at later times by the progress of the shock. 
Figure~\ref{Mass_Accrete}(a) plots the in-flowing mass accretion rate, computed by summing over the mass accretion rates of all radial rays with negative radial velocities at the gain layer $r_{\rm gain}$.
Figure~\ref{Mass_Accrete}(b) plots the net mass accretion rate through the same layer, defined as the in-flowing minus the out-flowing mass accretion rates.
Both measures of the mass accretion rates show that the accretion rate for F15.79 substantially exceeds that of F15.78 for the time period 0.12 to 0.8 s.
This behavior of the accretion rates with time follows from the density profiles of the progenitors (Figure~\ref{fig:init_den_s}). 
The inflowing rates for the two models (Figure~\ref{Mass_Accrete}(a)) are initially similar, but diverge at $\approx$0.12~s, when the density decrement of F15.78 reaches the vicinity of $r_{\rm gain}$.
The earlier revival of the shock due to the density decrement further impedes accretion after this time in F15.78. 
The faster infall accretion velocities of F15.79 thereafter, due to the stronger gravity of its larger developing enclosed mass, causes material at $r_{\rm gain}$ to originate from farther out, and therefore lower density, regions in the 15.79-\msun\ progenitor. 
This, in turn, causes its mass accretion rate to decline faster than that of F15.78 until the mass accretion rates of the two models become similar again at $\approx$0.8~s.
We regard the net mass accretion rate (Figure~\ref{Mass_Accrete}(b)), which excludes matter that flows inwardly then outwardly, as the source of the accretion luminosity, as this excluded matter is presumed to deposit momentum hydrodynamically but none of its kinetic energy is assumed to be thermalized.
Assuming nearly equal accretion efficiencies for the two models, the neutrino accretion luminosity originating from the mass accretion rate should roughly scale with these latter rates.
We will refer to the accretion luminosity computed from the net mass accretion rate by $L_{\rm m \, accrete}$.

A comparison of $L_{\rm m \, accrete}$ (Figure~\ref{Mass_Accrete}(b)) with $L_{1 \, \rm accrete}$ and $L_{2 \, \rm accrete}$ (Figure~\ref{L_accrete}) shows that each measure gives similar qualitative descriptions of the accretion luminosities of the models when compared to each other.
These relative accretion rates are correlated with the net \nue/\nuebar\ luminosities profiles (Figure~\ref{LumRMS}).
Specifically, all accretion luminosity measures give values for F15.79 that are larger than F15.78 by a factor of 2--3, from about 0.12 to 0.6--0.8~s, during which time most of the diagnostic energy of the models is acquired.
After 0.8~s, $L_{\rm m \, accrete}$ (Figure~\ref{Mass_Accrete}(b)) for F15.78 exceeds that for F15.79 from about 0.9 to 1.55~s, with the reverse being true from about 1.6 to 2.3~s.
This is mirrored in the relative values of F15.78 versus F15.79 for both $L_{1 \, \rm accrete}$ and $L_{2 \, \rm accrete}$ (Figure~\ref{L_accrete}), except that the values for F15.79 are higher relative to F15.78.
The reason is that, the gain radii of the two models being nearly equal, (Figure~\ref{H_Radii}) the larger mass of F15.79 interior to $r_{\rm gain}$ causes the kinetic energy of the in-falling matter to be larger at $r_{\rm gain}$, thereby boosting $L_{1 \, \rm accrete}$, which is related to the rate of kinetic energy infall, and $L_{2 \, \rm accrete}$ which is based on luminosities.
In both cases, more energy is accreted and thermalized in F15.79 for the same accreted mass.

All measures of the accretion luminosity indicate that until about 0.8~s, the accretion luminosity is responsible for a significant fraction of the luminosity differences between the two models. 
This, in turn, contributes a significant fraction to the heating rate differences between the two models (Figure~\ref{H_Rates}(a)).
The accretion luminosity measures also exhibit broad accretion luminosity upturns initiated at 1.2~s for F15.78 and 1.65~s for F15.79, and these, along with the accretion luminosity peaks also exhibited in Figure~\ref{L_accrete}, will be discussed in Section~\ref{sec:Bursts}.

\subsection{Core Luminosity Differences}
\label{sec:core_lum}

Consider now the core luminosities.
We first note that the differences in the core \nue\ and \nuebar\ luminosities of the models may be inferred from those of  \numt\ and \numtbar\ as the latter species decouple from the matter deeper in the core and are less affected by the accreted material at the proto-NS surface.
This, and the fact that the core luminosities of all neutrino flavors are approximately the same, led us to the core luminosity definition $L_{2 \, \rm core}$ given by Equation~(\ref{eq:9}) and plotted in Figure~\ref{L_accrete}.
Both the plots of $L_{2 \, \rm core}$ for the models and the entries in Table~\ref{tab:L2_L3} indicate that the core luminosities of F15.79 exceed those of F15.78.

\begin{figure}
	\includegraphics[width=\columnwidth,clip]{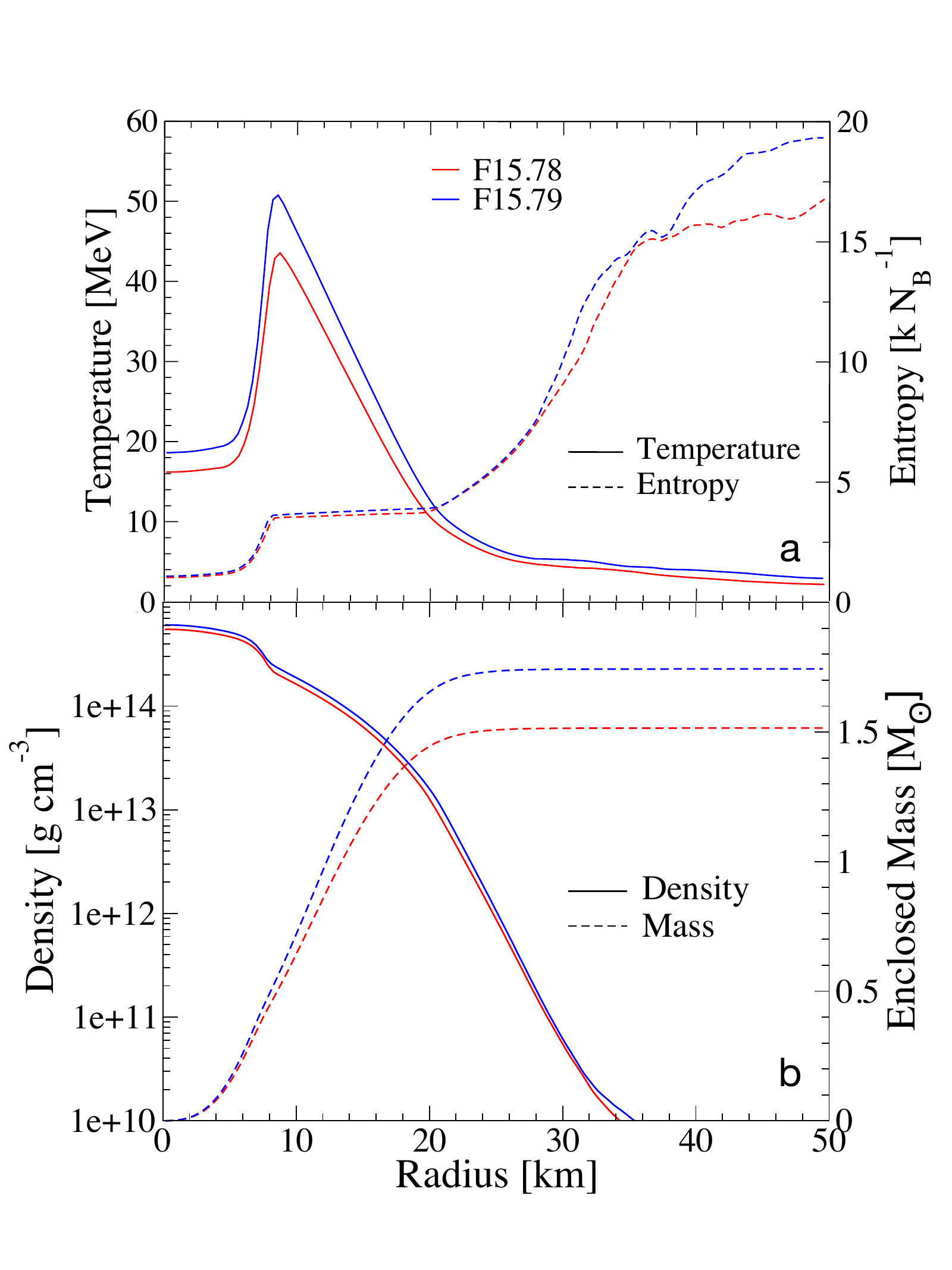}
	\caption{\label{Profiles}
	Structure of the model cores at 0.5~s post-bounce; (a) the temperature and entropy of the modes, (b) the density and enclosed nass of the two models.
	}
\end{figure}

These differences in core luminosities arise from the differences in the structures of the cores established $\sim$~0.1~s after bounce.
Figures~\ref{Profiles} plots the structures of the cores at 0.5~s post-bounce.
The low entropy feature from the center to about 8~km is the cold inner core, resulting from the homologous, nearly adiabatic collapse of the inner part of the initial Chandrasekhar-mass iron cores of the progenitors, the parts inside their respective sonic points, which have the same radius in both models. 
The inner core is surrounded by a warmer outer core that extends to $\approx$20~km, very nearly the same radius in both models, and is smoothed by proto-NS convection.
The mass of F15.79 core at this time, however, is about 16\% greater than the core of F15.78 as a result of the initial rapid mass accretion rate in the former, therefore the core densities are about 16\% greater in F15.79.
The additional compression of the F15.79 core due to the rapid mass accretion at its surface being isentropic, the core entropies are nearly the same in both models so that $T \propto \rho^{\gamma}$, with $\gamma \sim 1 \ldots 3/2$ for a mixture of degenerate leptons and partially degenerate nucleons at constant entropy.
(The equation of state gives $T \propto \rho$ in this region).
The temperature in the two models exhibits a negative radial gradient extending from the inner--outer core boundary at $\approx$10~km (Figure~\ref{Profiles}(a)) for both models, to the neutrinospheres at $\approx$30~km (Figure~\ref{nu_sphere}(a)) for both models.
The neutrino degeneracy parameter ($\mu_{\nu}/kT$) slowly varies along this interval, so neutrino transport should be dominated by energy rather than lepton transport, which can be crudely approximated by equilibrium diffusion:
\begin{equation}
L = - 4\pi r^{2} \frac{7}{8} \frac{a c}{3}  \ell \frac{d T^{4}}{dr} \qquad \ell \propto T^{-2} \rho^{-1} ,
\label{eq:12}
\end{equation}
where $\ell$ is the mean free path and the factor $7/8$ reflects Fermion rather than photon transport.
With the radial intervals the same for both models, and with $T \propto \rho$, and $\rho \propto M_{\rm core}$, we get that
\begin{equation}
L_{\rm core} \propto M_{\rm core} .
\label{eq:13}
\end{equation}
Thus, core luminosity scales approximately as $M_{\rm core}$ for our models and is therefore greater for F15.79 by about 16\%. 
Integrating over the life of the proto-NS, the core luminosity clearly dominates over the accretion luminosity (Figure~\ref{L_accrete}). 
However, the periods of rapid growth in the explosion energy (0.12--0.8~s) coincide with the periods when the accretion luminosity most aggressively augments the core luminosity, especially in the F15.79 model. 
Mass accretion peaks and upturns (Figure~\ref{Mass_Accrete}) have disproportionately large effects on the explosion energy, as they not only increase the accretion luminosity, but also the density (Figure~\ref{H_Rates}(b)) and mass (Figure~\ref{H_Rates}(c)) of the gain region, and thereby the net neutrino energy deposition rate from both the core and accretion luminosities. 
One can therefore not ignore the contributions of recent accretion to the heating rate (Figure~\ref{H_Rates}(a)) and the explosion energy (Figure~\ref{Shock_Energy}) differences between these two models.

\subsection{Outflows}
\label{sec:Out}

\begin{figure}
	\includegraphics[width=\columnwidth,clip]{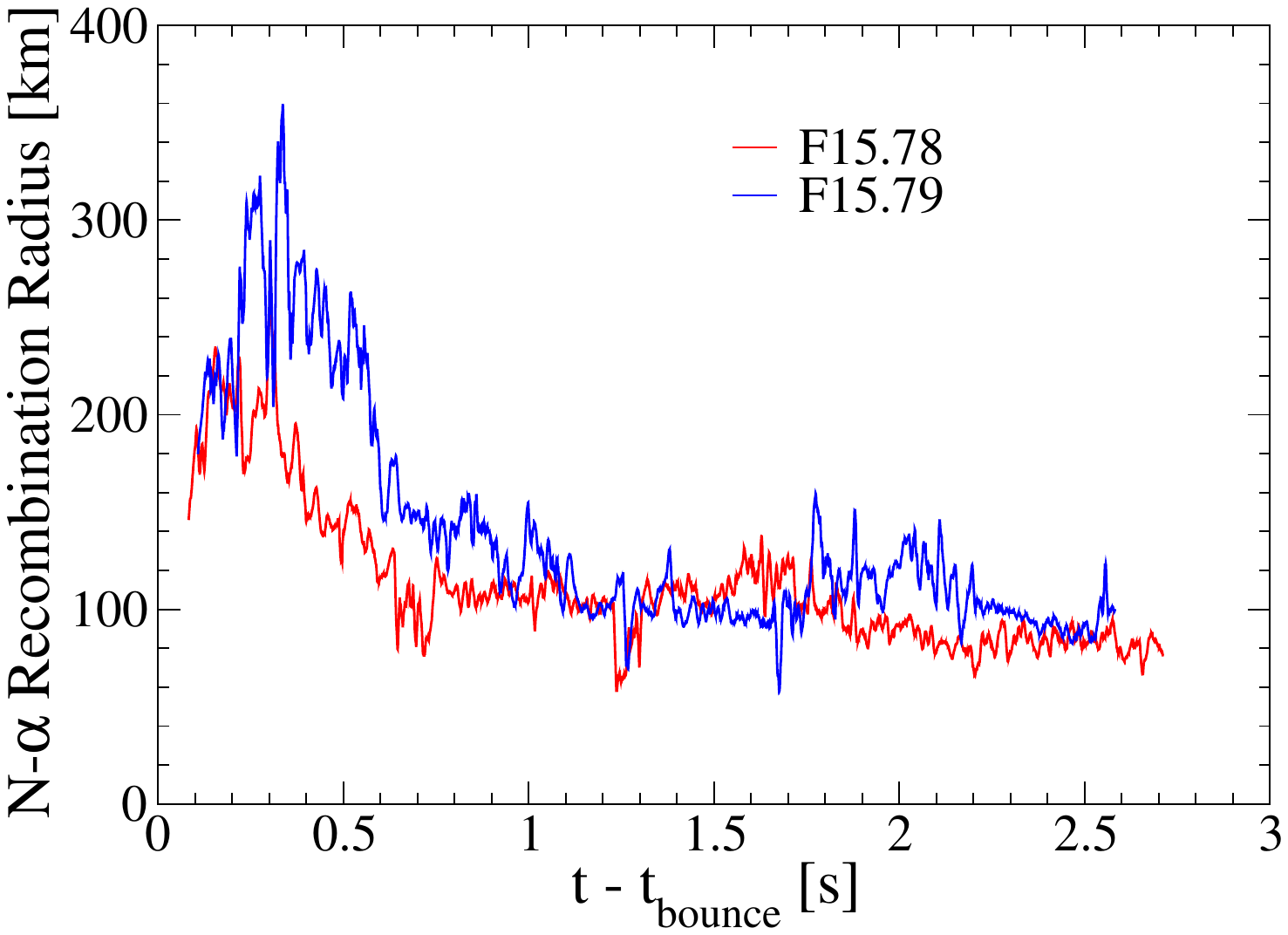}
	\caption{\label{recom}
	Nucleon--alpha recombination radius in outflowing material
	}
\end{figure}

We conclude our discussion of the important time period 0.12--0.8~s where most of the diagnostic energy of the models was acquired with a brief look at the delivery of energy from the heating region to the shock.
The larger neutrino heating rate for F15.79 relative to F15.78 (Figure~\ref{H_Rates}(a)) leads to greater mass and energy ejected during the course of the explosion, as is evident from Figures~\ref{Shock_Energy}(a) and (b). 
To examine this transfer of neutrino deposited energy to the shock, we set up \citep[analogous to the approachs of][]{Mull15, BrLeHi16} an ejecta volume, $V_{\rm ejecta}$, with an inner radius, $R_{\rm ejecta}$, at 400~km and an outer radius at the position of the shock, and consider matter with positive and negative radial velocities at $R_{\rm ejecta}$, i.e., matter flowing into and out of $V_{\rm ejecta}$, respectively, through the surface at $R_{\rm ejecta}$.

The value of 400 km is chosen for $R_{\rm ejecta}$ because it lies outside the gain layer and most of the region where significant neutrino energy deposition is occurring (Figure~\ref{H_Radii}), and outside the radius at which recombination of nucleons to $\alpha$-particles takes place (Figure~\ref{recom}).
Partial recombination of alpha-particles into heavier nuclei does occur in $V_{\rm ejecta}$ for some of the matter, and that can add  $\sim 1$~MeV~baryon$^{-1}$ ($\sim 10^{18} \; {\rm ergs \; g}^{-1}$) to the energy of the ejecta.
Nuclear burning of shock heated material adds little energy.
The principle source of the diagnostic energy in $V_{\rm ejecta}$ is the enthalpy flow into $V_{\rm ejecta}$ through the $R_{\rm ejecta}$ surface.

In the following discussion, thermodynamic and other quantities associated with the up-flows at $R_{\rm ejecta}$ are taken as averages over the up-flow mass fluxes. 
That is, for a given quantity $X$, the flow rate of $X$ into $V_{\rm ejecta}$ is given by
\begin{equation}
X = \sum_{j, u_{j} > 0} X_{j} \dot{M}_{j} \; / \sum_{j, u_{j} > 0} \dot{M}_{j}
\label{eq:14}
\end{equation}
where $X_{j}$ and $\dot{M}_{j}$ are the values of $X$ and the mass flux of radial ray $j$ at $R_{\rm ejecta}$.
Down-flow quantities are similarly defined.

\begin{figure}
	\includegraphics[width=\columnwidth,clip]{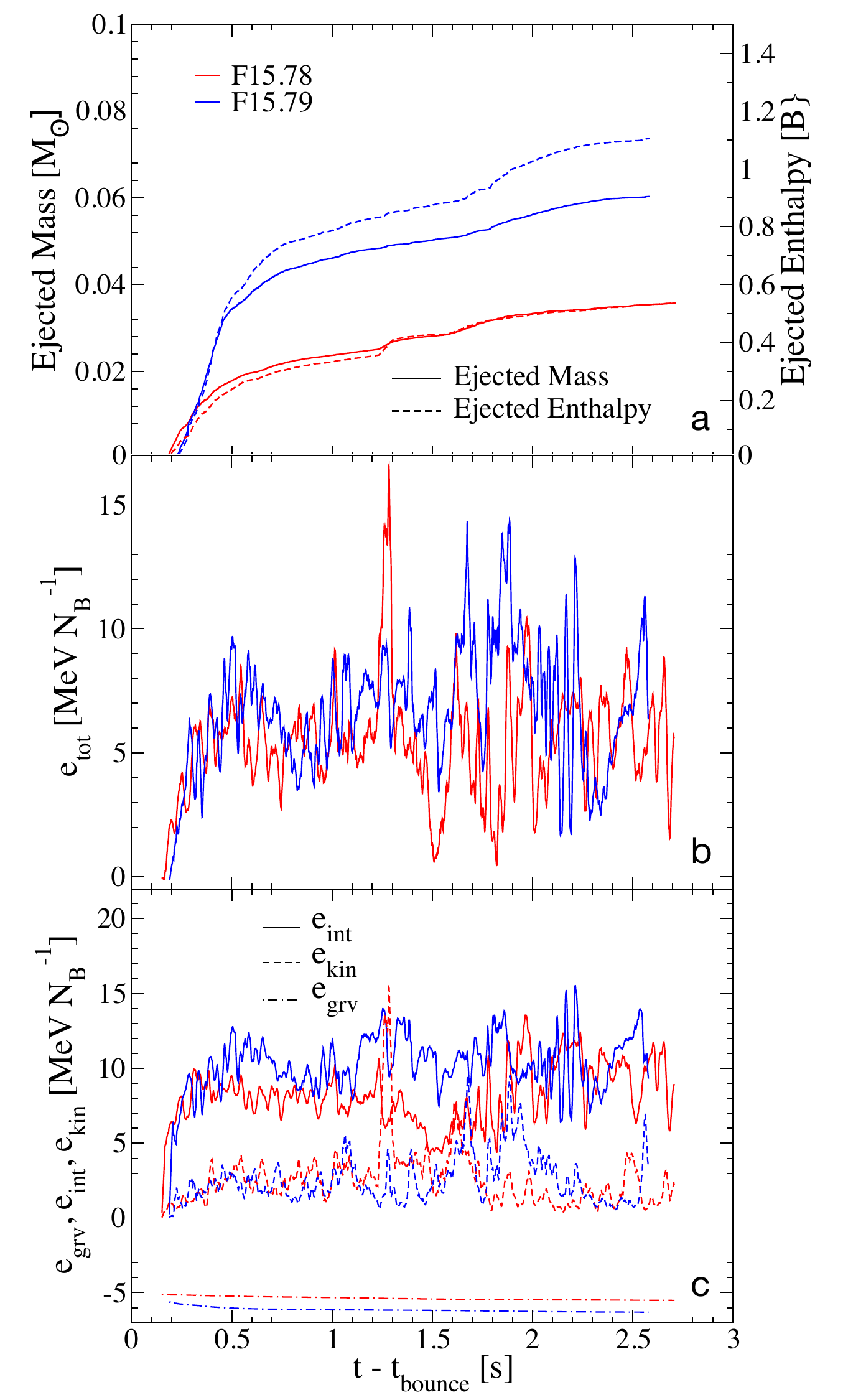}
	\caption{\label{Outflow}
	(a) Cumulative mass and enthalpy flows into $V_{\rm ejecta}$ through $R_{\rm ejecta}$; (b) total energy per baryon at $R_{\rm ejecta}$; (c) components of the total energy per baryon at $R_{\rm ejecta}$.
	}
\end{figure}

Figure~\ref{Outflow}(a) shows the accumulated mass and enthalpy flowing into $R_{\rm ejecta}$, where the enthalpy, $h$, is defined as
\begin{equation}
h = e_{\rm tot} + p/\rho, \qquad e_{\rm tot} = e_{\rm int} + e_{\rm kin} + e_{\rm grv},
\label{eq:15}
\end{equation}
where $e_{\rm int}$, $e_{\rm kin}$, and $e_{\rm grv}$ are the baryon specific internal (not including rest mass), kinetic, and gravitational energies, and $p$ is the pressure.
The cumulative enthalpy input into $V_{\rm ejecta}$ is a measure of the diagnostic energy in that volume, modulo recombination of $\alpha$-particles into heavy nuclei, unbound material entrained in downflows, and the upper moving boundary condition at the shock. 
Figure~\ref{Outflow}(a) shows that the cumulative mass flux into the ejecta region is larger for F15.79 than for F15.78, as to be expected, given its larger heating rate, but the difference in the cumulative enthalpy flux into the ejecta region is not proportional to the cumulative mass flux.
To illustrate this, the enthalpy ordinate in Figure~\ref{Outflow}(a) is scaled so that the plot of cumulative ejected enthalpy lies on top of that for the cumulative ejected mass for F15.78. 
The cumulative enthalpy plot for F15.79, while initially coincident with its plot of cumulative ejected mass, begins at 0.5~s to deviate such that the cumulative total enthalpy to mass ratio is about 20\% larger than in F15.78.
This can be traced to the larger baryon specific total energy in F15.79 of the material at $R_{\rm ejecta}$ at these times, shown despite the noise in Figure~\ref{Outflow}(b).
Examining the components of the total energy at $R_{\rm ejecta}$ shown in Figure~\ref{Outflow}(c), we note first that the gravitational potential, $e_{\rm grv}$, is more negative for F15.79 than F15.78 as a consequence of its larger enclosed mass (Figure~\ref{ProtoNSMR}(b)).
The baryon specific kinetic energies are about the same for the two models, but the internal energies are clearly different, being significantly larger for F15.79, over compensating for its smaller gravitational energy, and accounting for its larger value of $h$ at $R_{\rm ejecta}$.
Much of the internal energy acquired by rising fluid elements is from the recombination of free nuclei to \isotope{He}{4}, and is approximately the same for both models.
The greater internal energy acquired by a rising fluid element in F15.79 is due to its having been subjected to a more intense rate of neutrino heating and for a longer period of time, the former because of greater \nue/\nuebar\ luminosities and rms energies for F15.79, and the latter because of the larger width of its gain region (Figure~\ref{H_Radii}).
The larger outflow of matter in F15.79 and the greater baryon specific enthalpy of this matter, both a result of its larger heating rate, results in a significantly greater rise in its explosion energy in comparison with F15.78.

\subsection{Accretion Transients}
\label{sec:Bursts}

We now briefly examine the diagnostic and total energy uptakes, initiated at 1.2 and 1.55~s for F15.78, and at 1.65~s for F15.79, as is manifested in Figure~\ref{Shock_Energy}(b). 
These energy uptakes are caused by corresponding mass accretion uptakes, with their proportionately large effects on the explosion energy (Section~\ref{sec:core_lum}), together with accretion stream rearrangement transients.
These are likely stochastic in nature and therefore not immediately related to differences in the internal structures of the progenitor models, which is the focus of this paper. 
Given that, we will provide only a brief account of these accretion transients.

\subsubsection{F15.78}
\label{sec:Bursts78}

\begin{figure}
	\includegraphics[width=\columnwidth,clip]{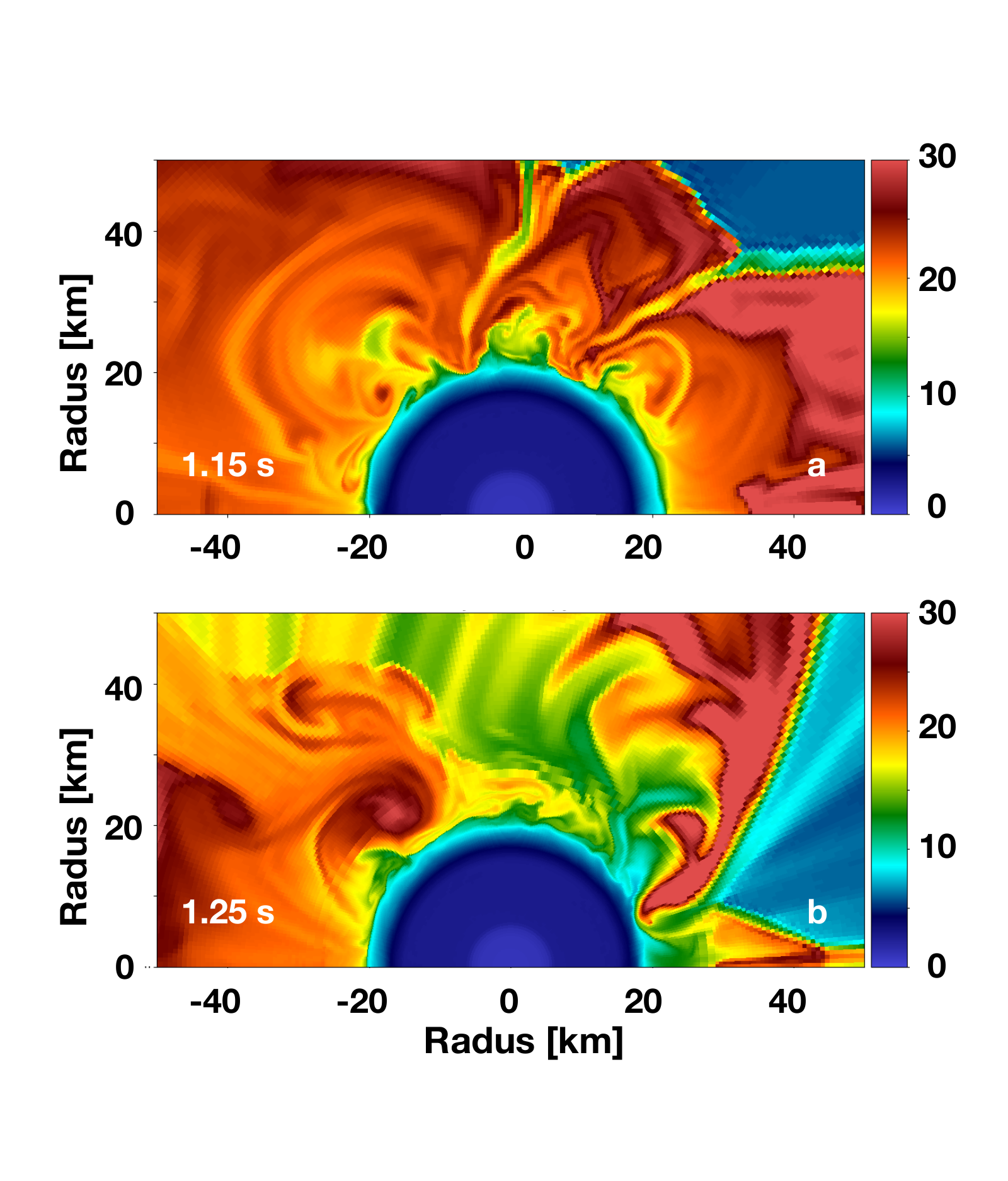}
	\caption{\label{nu_burst}
	(a) Specific entropy of F15.78 at post-bounce times of (a) 1.15~s, and (b) 1.25~s.
	}
\end{figure}

At 1.2 s, the overall shock geometry of F15.78 is similar to that at 1~s and is nearly unipolar, as shown in the bottom left-hand panel of Figure~\ref{F15_78_79_evol}. 
At this time, the inward mass flow through $r_{\rm gain}$ increases (Figure~\ref{Mass_Accrete}), and a prominent accretion stream develops in the equatorial sector just to the north (right) of the equator with matter down-flows ultimately directed toward the vicinity of the proto-NS from the northward direction.
Immediately prior to 1.2~s, this accretion stream is encountering a standing accretion shock at a position approximately 20 km above the 20-km radius surface of the proto-NS.
This flow geometry changes after 1.2~s such that the standing accretion shock develops a cusp, allowing some of the pre-shocked material to reach the surface of the proto-NS. 
This change in the mass accretion geometry is illustrated in Figure~\ref{nu_burst}, where the proto-NS is the blue (low-entropy), semi-circular figure at the bottom center, and the low-entropy mass accretion stream is the blue 'tongue' at the upper right (a) and the center-right (b).
Note how the accretion stream has penetrated close to the surface of the proto-NS in the interval from 1.15 to 1.25~s.
This results in an increase in the accretion luminosity as computed by Equation~(\ref{eq:8}) and displayed in Figure~\ref{L_accrete}, as the standing accretion shock, originally above $r_{\rm gain}$ now falls below it.
The accretion luminosity as computed by Equation~(\ref{eq:9}) and shown in Figure~\ref{L_accrete} also responds to the change in the mass infall rate accretion geometry, but more gently.
The altered mass infall rate and geometry also increases the gain layer density (Figure~\ref{H_Rates}(b)), and ultimately the \nue/\nuebar\  luminosities and rms energies (Figure~\ref{LumRMS}), resulting, in turn, in a subsequent increase in the heating rate (Figure~\ref{H_Rates}(a)), and in the rate outflowing mass and enthalpy (Figure~\ref{Outflow}(a)).

The energy uptake at 1.55~s also arises from the penetration of an accretion stream to the vicinity of the proto-NS surface. 
In this case, the penetration is immediately preceded by a change in the direction of the accretion stream near the proto-NS surface from being directed from the north to being directed from the south. 
The scenario producing an increase in the rate outflowing mass and enthalpy as a result of this change is similar to that described above for the energy uptake at 1.2~s, but more mild.

\subsubsection{F15.79}
\label{sec:Bursts79}

The overall configuration of F15.79 at 1.65~s is similar to its configuration at 1 s, shown at the bottom right panel of Figure~\ref{F15_78_79_evol}.
At 1.65~s, its configuration is highly prolate with several strong equatorial accretion channels, both of which swerve to the south of the proto-NS, and are ultimately directed towards the vicinity of proto-NS from the south polar direction.
Prior to 1.65~s, the accretion flow near the vicinity of the proto-NS encounters an accretion shock located approximately 30 km above the 20-km radius surface of the proto-NS, with a cusp near the southern edge of the flow through which a small amount of unshocked  material penetrates to the vicinity of the proto-NS surface.
At 1.65~s, both accretion flows are joined and the accretion shock is pushed to within a couple of km of the proto-NS surface.
This results in a near doubling of the mass accretion rate through $r_{\rm gain}$ (Figure~\ref{Mass_Accrete}(a) and(b)) that, when combined with the increased release of gravitational potential energy, results in a dramatic increase in the release of neutrino accretion luminosity, as reflected in the accretion luminosity displayed in Figure~\ref{L_accrete}.
The concomitent rise in the \nue/\nuebar\ rms energies (Equation~(\ref{eq:7}) and Figure~\ref{LumRMS}) and the increase in the density of the gain region (Figure~\ref{H_Rates}(b)) combine to at least double the neutrino heating efficiency (Figure~\ref{heat_eff_all}).
The increase in the heating efficiency and the neutrino luminosity and rms energy increases the heating rate (Figure~\ref{H_Rates}(a)) by several, and the rate of mass and enthalpy ejected (Figure~\ref{Outflow}(a)), resulting in the uptake in the explosion energy. 

\section{Nucleosynthesis}
\label{sec:nucleo}

\begin{figure}
	\fig{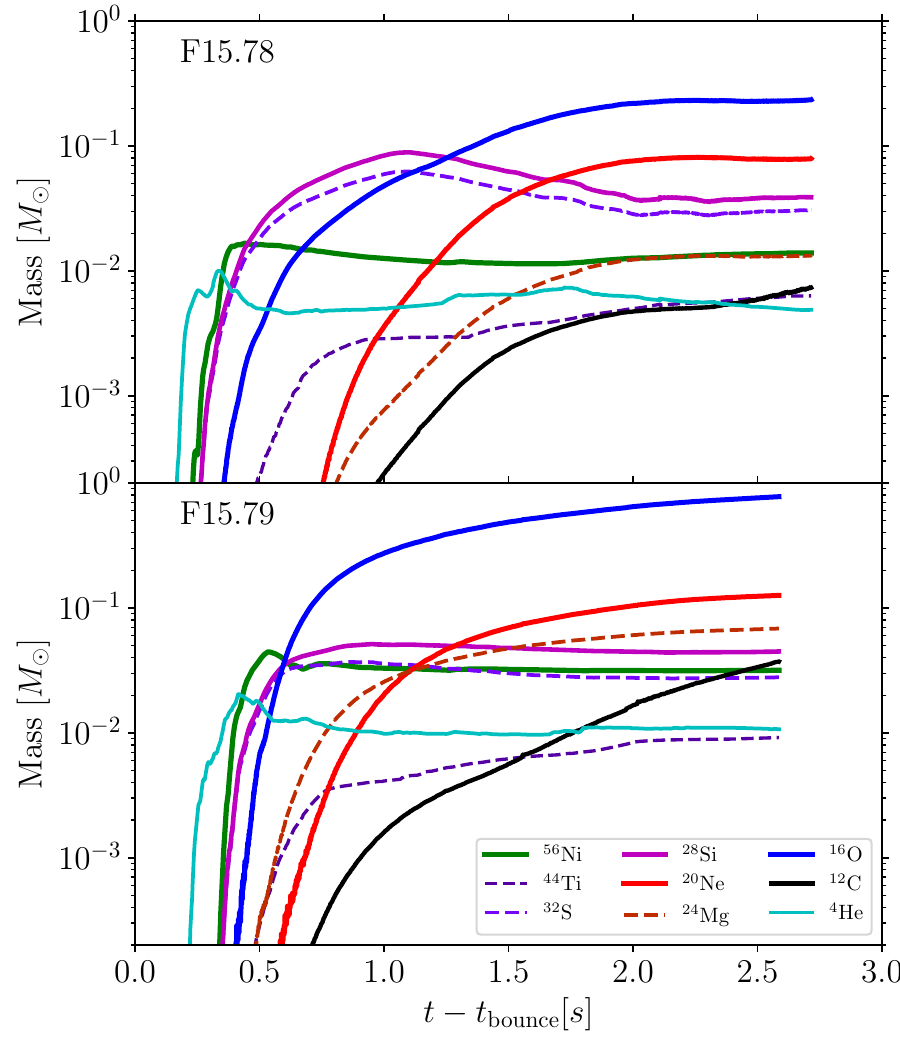}{\columnwidth}{}
	\caption{\label{fig:abu_evol}
	Time evolution of the unbound mass of the most important nuclear species in the $\alpha$-network for F15.78 (top) and F15.78 (bottom).
	}
\end{figure}

The small \texttt{anp56} nuclear reaction network included within these simulations is not suitable to give accurate nucleosynthesis predictions but 
 we can use the approximate abundances it provides 
to illustrate the impact of the differences in the stellar structure of the two models has on the composition of the ejecta.
\citet{HaHiCh17} have discussed the caveats of small reaction networks for nucleosynthesis predictions, but found that the yields of the most common isotopes, including \isotope{Ni}{56}, \isotope{Si}{28}, and \isotope{O}{16} are relatively well reproduced by an $\alpha$-network, while deviations can be large for \isotope{Ti}{44}. 
What is important for this analysis is the relative outcomes between the models that parallel some of the features of the explosion dynamics discussed in Section~\ref{sec:expl} and differences in the progenitor composition.

\subsection{Isotopic Yields}
 
In Figure \ref{fig:abu_evol} we show the evolution of the composition of our estimated ejecta defined as the material contained in all zones with positive total energy (thermal + kinetic + gravitational).
This mirrors the diagnostic energy (Figure 7).
In both simulations, the first material to become unbound is very hot and consists mainly of free nucleons and $\alpha$ particles. 
Afterwards, \isotope{Ni}{56} appears from both the freeze out of NSE and explosive Si-burning as the shock progresses through the Si-shell. Unburnt \isotope{S}{32} and \isotope{Si}{28} becomes unbound afterwards, as incomplete Si-burning, followed by O-burning occur in the shock.
\isotope{Ti}{44} appears with some delay from the $\alpha$-rich freeze out of high-entropy ejecta. 
When the shock leaves the Si-shell, \isotope{O}{16}, \isotope{Ne}{20} and \isotope{Mg}{24} and eventually \isotope{C}{12} join the ejecta, reflecting mostly the ejection of progenitor composition. 
The formation of accretion downstreams, which can also entrain part of the previously unbound material, causing it to fall back, is reflected in the decrease of the unbound  \isotope{Ni}{56} and especially \isotope{Si}{28}/\isotope{S}{32} mass. 
Explosive Si-burning is completed in both models at the end of the simulation and the unbound mass of \isotope{Si}{28}/\isotope{S}{32} and \isotope{Ni}{56} have stabilized.
The ongoing increase of \isotope{O}{16} and \isotope{C}{12} shows the shock ejecting the O/Ne- and C-shells without inducing significant nuclear reactions.
Qualitatively, these features are common to both models discussed here, and other CCSN simulations \citep[see, e.g.,][]{HaHiCh17,BrLeHi16,EiNaTa18,SiMuQi20}, but the details reflect important differences between the models.
 
 \paragraph{\isotope{Ni}{56} yield}
Due to the rather weak and very asymmetric explosion of F15.78, most of the \isotope{Ni}{56} is produced in the direction where the explosion is strongest in a relatively narrow cone along the axis.
Most of this \isotope{Ni}{56}, created by explosive Si burning in F15.78, is ejected unhindered with only a gradual decrease in the unbound mass visible in Figure~\ref{fig:abu_evol} reflecting ongoing accretion.
A slight boost in \isotope{Ni}{56} accompanies the boosts in explosion energy after 1.2~s (see Figure~\ref{Shock_Energy}) as more matter joins the ejecta.
In the other directions, where the explosion is weaker, there is little Si-burning.
In F15.78, the unbound \isotope{Ni}{56} mass reaches its peak at about 400~ms, which is earlier than the peak in \isotope{Ni}{56} for F15.79 at around 550~ms. This also reflects the slightly delayed shock expansion in the F15.79 (see Figure \ref{Shock_Energy}).
F15.79 exhibits a higher overall accretion rate (see Figure~\ref{fig:early_shock}) and associated accretion luminosity early on that causes a significant amount of  the initially formed \isotope{Ni}{56} to fall back, leading to the more pronounced peak at 550~ms.

At the end of the simulation, there is $1.4 \times 10^{-2} \msun$ of \isotope{Ni}{56} unbound for F15.78 compared to $3.2 \times 10^{-2} \msun$ in F15.79.
Observationally, this also implies that the explosion of a star similar to the progenitor of F15.78 would appear much dimmer than a star similar to F15.79.
The higher \isotope{Ni}{56} yield and 2.5 times higher explosion energy in F15.79 follows the strong correlation between the explosion energy and the production of \isotope{Ni}{56} seen in observations \citep[see, e.g.,][]{MaAnBe22}.
As discussed in Section~\ref{sec:preSN}, the 15.79-\msun\ pre-SN model has higher compactness than the 15.78-\msun\ pre-SN model, correlating with the higher \isotope{Ni}{56} yields in F15.78 and F15.79, respectively, in agreement with the finding of \citet{EbCuFr19}, for neutrino-driven, spherically-symmetric explosions, that higher compactness correlates with higher \isotope{Ni}{56} yields.

\paragraph{\isotope{Ti}{44} yield}
The amount of \isotope{Ti}{44} is overestimated by the \texttt{anp56} reaction network by roughly an order of magnitude \citep{HaHiCh17} compared to the results of more realistic reaction network, therefore its value should be considered to represent a broader composition of species resulting from $\alpha$-rich freeze-out from NSE.
The numbers quoted here should therefore be compared directly to observations with care.
In both simulations,  \isotope{Ti}{44} is first produced at the edge of the initial, high-velocity blast that ejects hot material from the Fe-core as free nucleons and $\alpha$ particles.
When this material expands into the O-shell and rapidly cools, \isotope{Ti}{44} forms.
This leads to the increase in unbound \isotope{Ti}{44} between 400~ms and 700~ms visible in Figure~\ref{fig:abu_evol} for both simulations.
Note this is $\sim200$~ms after the rise of \isotope{Ni}{56}.
About 45\% of the final  \isotope{Ti}{44} mass becomes unbound during this phase, less than 1~s after bounce. 
Additional \isotope{Ti}{44} is produced later in the high-entropy, neutrino-heated material.
For example, Figure~\ref{fig:abu_evol} shows an increase in the unbound $\alpha$ particle mass in F15.78 at  $\sim$1.3~s after bounce that is associated with the increase of the heating rate (see Section~\ref{sec:Bursts78} \& Figure~\ref{H_Rates}(a)).
This outburst of neutrino heating drives an outflow from the region near the proto-NS that then undergoes an $\alpha$-rich freeze-out as it expands.
More than half of the final \isotope{Ti}{44} yield in F15.78 only forms after this event, in the freeze-out of the neutrino heated material. 
The evolution of the \isotope{Ti}{44} mass for F15.79 (lower panel of Figure~\ref{fig:abu_evol}) is more smooth, but is also shows episodes of increased \isotope{Ti}{44} formation at about 1.3~s and 1.8~s, which are associated with the episodes of increased heating shown in Figure~\ref{H_Rates} (see Section~\ref{sec:Bursts79}). Since  \isotope{Ti}{44} results from an  $\alpha$-rich freeze-out from NSE, the unbound mass of \isotope{He}{4} shows similar features.

At the end of the simulations, we find ejected masses of $6.3 \times 10^{-3}$~\msun\ of \isotope{Ti}{44} for F15.78 and $9.2 \times 10^{-3}$~\msun\ for F15.79. 
It is interesting to note that the unbound mass of  \isotope{Ti}{44} does not scale in the same way with the explosion energy as \isotope{Ni}{56}. As a consequence, the  \isotope{Ti}{44}/ \isotope{Ni}{56} mass ratio is higher in F15.78 (0.45) than in F15.79 (0.29). 
Due to the very low explosion energy of F15.78, it is likely that a significant fraction of the \isotope{Ti}{44} produced at late times may eventually fall back onto the neutron star. 
A clear assessment of the nucleosynthesis yields of this model requires following the explosion to much later times.

\paragraph{\isotope{Si}{28} and \isotope{O}{16} yield}
The lower compactness and smaller core of the 15.78~\msun\ progenitor model and the stronger explosion have a significant impact on the evolution of \isotope{Si}{28} in the ejecta. 
Overall, the 15.78~\msun\ pre-SN model contains 0.86~\msun\ of \isotope{Si}{28} which is slightly more than the 0.79~\msun\ of \isotope{Si}{28} in  the 15.79~\msun\ pre-SN model.
In the 15.79~\msun\ pre-SN model, the Si-core is much more compact and only 0.18~\msun\ of \isotope{Si}{28} is found outside of the NSE region at collapse, compared to 0.28~\msun\ of  \isotope{Si}{28} in the 15.78~\msun\ pre-SN model. 
In the 15.78~\msun\ pre-SN model, the Si-rich layer is also very extended in radius, reaching out to almost 8,000 km (Figure~\ref{fig:mf_profiles}). 
It takes almost 1~s for the average shock radius in F15.78 to reach the O-shell.
 A large amount of \isotope{Si}{28} becomes unbound without being burnt reaching more than 0.1~\msun\ at around 800~ms after bounce.  At its peak, the amount of unbound  \isotope{Si}{28} is one order of magnitude larger than the unbound  \isotope{Ni}{56} mass.
As the unipolar morphology of F15.78 is established, the strong and persistent accretion stream captures more than half of the temporarily unbound \isotope{Si}{28} from the progenitors Si-shell.
For the 15.79~\msun\ progenitor, on the other hand, the Si-shell is more compact, extending only to about 3,000~km and the shock leaves the Si-shell only 600~ms after bounce. As a consequence, a larger fraction of
 \isotope{Si}{28} is burnt to \isotope{Ni}{56}, such that the difference between unbound  \isotope{Si}{28} and  \isotope{Ni}{56} remains much smaller than in the F15.78 model.

While the \isotope{Ni}{56} mass is very different between the two simulations, the ejected \isotope{Si}{28} mass is more similar, with $3.9\times 10^{-2} \msun$ for F15.78 and $4.5 \times 10^{-2} \msun$ of \isotope{Si}{28} for F15.79.
This leads to a much larger \isotope{Si}{28}/ \isotope{Ni}{56} mass ratio in F15.78. The larger final yield of \isotope{Si}{28} for F15.79, which is based on the progenitor model with less \isotope{Si}{28} at collapse, is a sign of stronger explosive O-burning in the F15.79 model due to the more energetic shock compared to the F15.78 model.
Figure~\ref{fig:mf_profiles} also shows that most of the Si-shell of the 15.78~\msun\ pre-SN model contains at least 10\%~\isotope{O}{16}. Consequently the unbound mass of \isotope{O}{16} also increases gradually as \isotope{Si}{28} is swept up by the shock in F15.78. This also reflects the relatively weak shock, which does not allow for a significant amount of explosive O-burning. In the 15.79~\msun\ progenitor, on the other hand, the Si-shell is free of  \isotope{O}{16} and it becomes unbound more abruptly once the shock has reached the O-shell.
The large difference in the amount of unbound \isotope{O}{16} at the end of the simulation is partly due to the morphology of the explosion in F15.78, where most of the unbound  \isotope{O}{16} is ejected in the narrow cone where the explosion is strongest, while a large fraction of the O-shell and C-shell remain undisturbed and thus are not unbound at the end of the simulation. In order to determine the final yields of \isotope{O}{16} and \isotope{C}{12}, the simulation needs to be continued to even later times, since the usual assumption that the shock will successfully eject the remainders of these layers seems suspect in this highly aspherical, weakly explosing case.

\paragraph{\isotope{Ne}{20} and \isotope{Mg}{24} yields}
The ejection of  \isotope{Ne}{20} and \isotope{Mg}{24} also reflects the progenitor structure.
Figure~\ref{fig:mf_profiles} shows that there is \isotope{Mg}{24} at the bottom of the O-burning shell between about 3--4,000~km in the 15.79~\msun\ pre-SN progenitor, where \isotope{Ne}{20} has been burnt.
As a consequence, in F15.79, \isotope{Mg}{24} appears in the unbound material before \isotope{Ne}{20}.
For F15.78, both \isotope{Ne}{20} and \isotope{Mg}{24} start to be unbound at about the same time when the maximum shock radius reaches 10,000~km --- the edge of the O-shell containing  \isotope{Ne}{20} and \isotope{Mg}{24}.
Since \isotope{Mg}{24} is about one order of magnitude less abundant than \isotope{Ne}{20} in the pre-SN model, the \isotope{Mg}{24} mass in the ejecta increases much more slowly. 

\paragraph{Comparison to other models}
While the focus of this study is on the differences between two models based on stellar progenitors with almost identical mass, we briefly discuss our results in the context of other models from the available literature. We focus on \isotope{Ni}{56} because it is most commonly reported and we expect a reasonable estimate from our
small $\alpha$-network.
Post-processing parametrized explosions in spherical symmetry with a large reaction network, \citet{CuEbFr19} find much higher yields for \isotope{Ni}{56} for stellar models of comparable initial mass, e.g., $9.86\times 10^{-2}$~\msun\ for a 15.8~\msun\ progenitor model and $7.54\times 10^{-2}$~\msun\ for a 16.0~\msun\ progenitor model, both from \citet{WoHeWe02}. They also find $6.64\times 10^{-2}$~\msun\ of \isotope{Ni}{56} for a 16~\msun\ progenitor model from \citet{WoHe07}. These higher yields are consistent with the higher explosion energies larger than $10^{51}$~erg obtained with their spherically symmetric parameterization \citep{EbCuFr19}. The differences in the yields from models with similar initial mass and from slightly different versions of the same stellar evolution code illustrate a general sensitivity to the progenitor structure that is, in relative terms, comparable to the differences we find here. It remains, however, unclear whether the origin of these differences are the same.   
The \isotope{Ni}{56} yield of F15.79 is close to the value of 0.038~\msun\ estimated by \citet{BoYaKr21} based on a
3D simulation of a 19~\msun\ progenitor that also employs an $\alpha$~network.
\citet{EiNaTa18} have also noted very low \isotope{Ni}{56} yields based on long-term axisymmetric simulations extended to more than 5~s after bounce and  a strong sensitivity to the
ejection criterion. These authors find a range from $9.5 \times 10^{-3}$ to $6.6 \times 10^{-2}$ \msun\ of  \isotope{Ni}{56} for their 17~\msun\ model depending on the ejection criterion and 
argue that a large amount of \isotope{Ni}{56} is eventually accreted. While we find yields that are higher than the values reported by \citet{EiNaTa18}, the simulations need to be run longer to clarify which material is eventually ejected.

\begin{figure}
	\fig{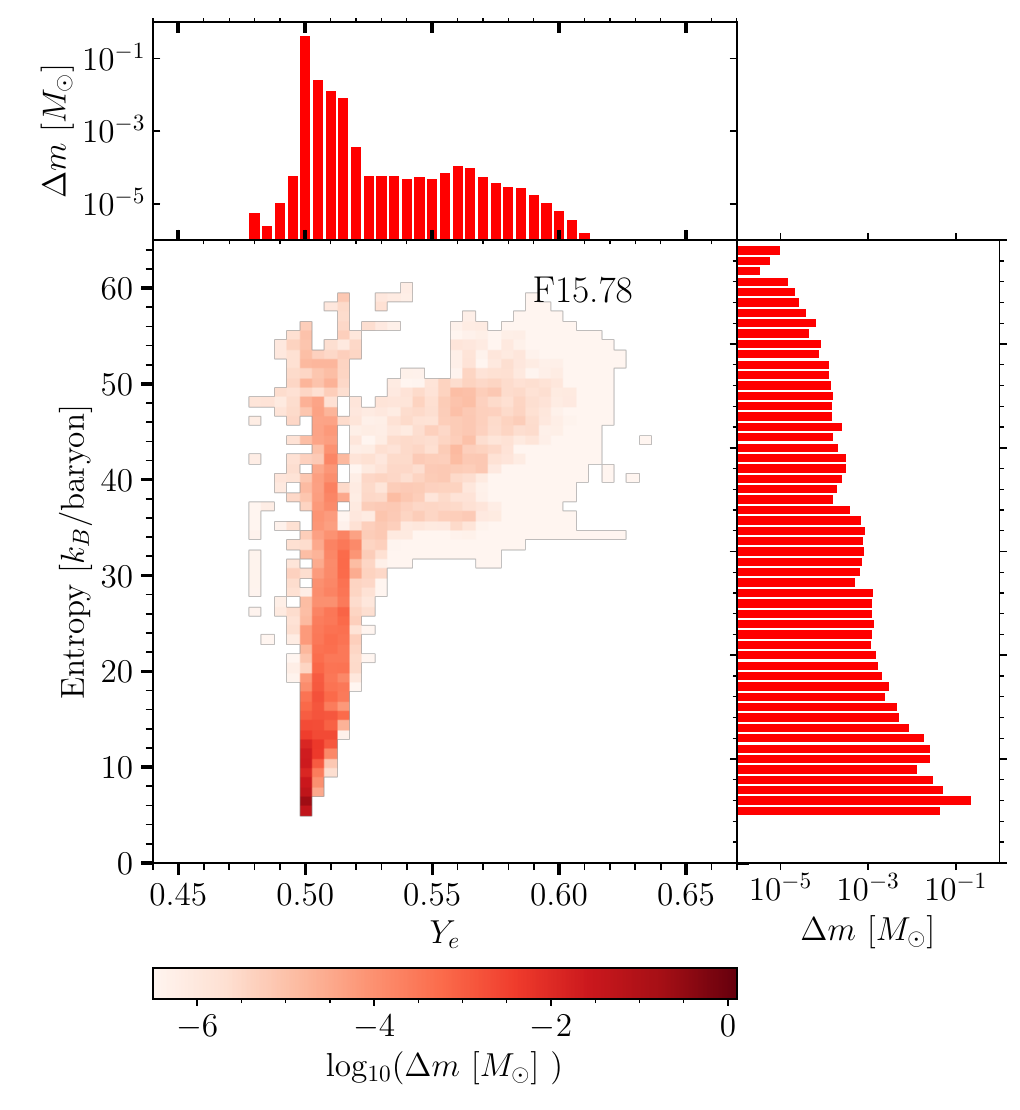}{\columnwidth}{(a)}
	\fig{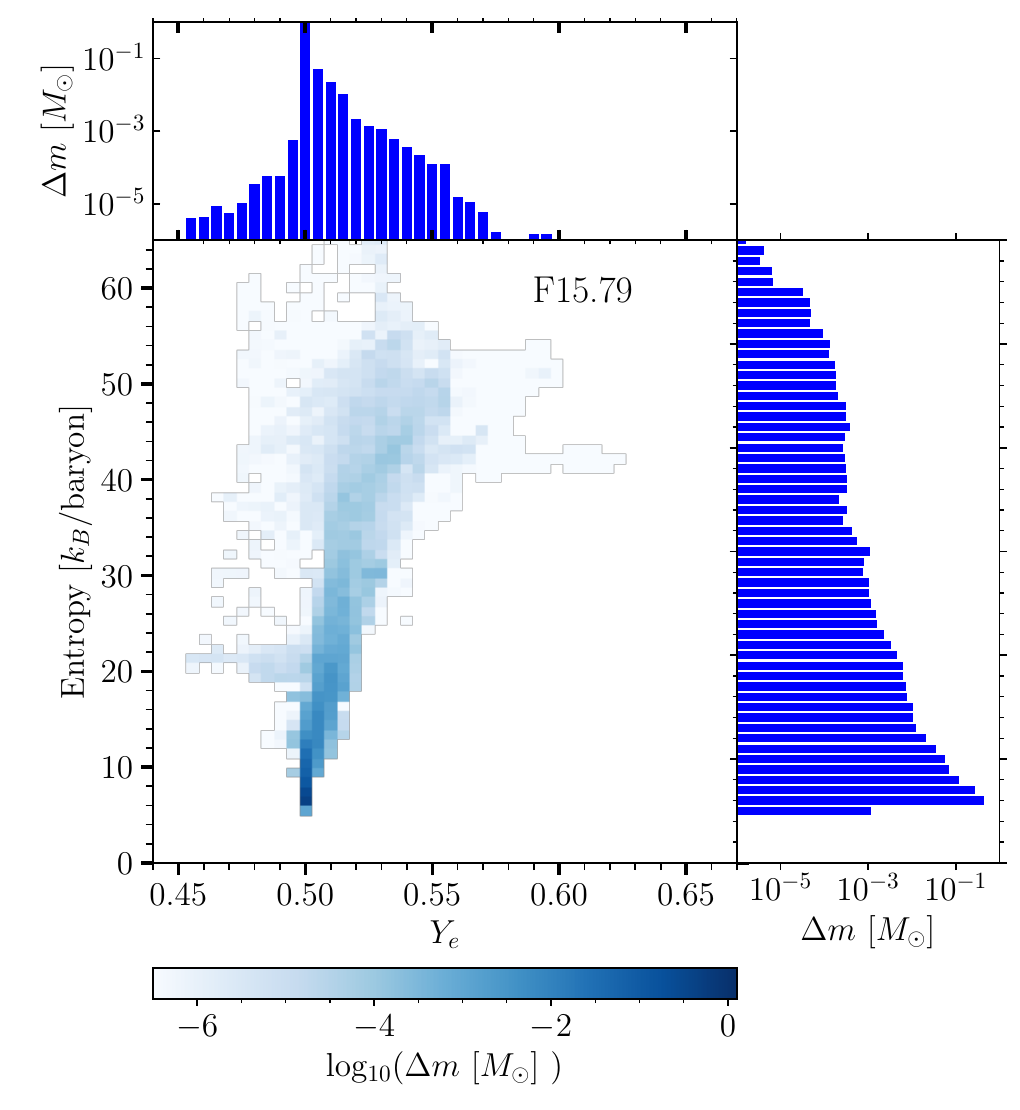}{\columnwidth}{(b)}
	\caption{\label{fig:ye_histo}
	Distribution of  $Y_e$ and entropy of unbound matter at the end of the simulations for (a) F15.78 and (b) F15.79. 
	}
\end{figure}

\subsection{Entropy and $Y_e$ Distributions}

More than 15,000 passive Lagrangian tracer particles have been included in each simulation that will be post-processed with
a large nuclear reaction network for detailed nucleosynthesis yields in future work.
To assess the possibilities for neutron- and proton-rich nucleosynthesis of heavier elements, Figure~\ref{fig:ye_histo} shows the distributions of $Y_e$ and entropy for the unbound material at the end of the simulation for each simulation.

In both simulations, the unbound material is mostly proton rich ($Y_e>0.5$), with $Y_e$ ranging from 0.48 to 0.61 for F15.78 and from 0.46 to close to 0.60 for F15.79. 
Most of the material that is just swept up by the shock stays at values of $Y_e\approx 0.5$ and entropy below 15~\kbbar.
The distribution for F15.78 is particularly narrow, with the vast majority of the ejecta centered around $Y_e\approx0.5$ but a relatively long tail of high $Y_e$.
This can be understood as a consequence of the early, but relatively weak explosion, with low neutrino heating rates described in Section~\ref{sec:expl}.
The explosion of F15.79 experiences more neutrino heating (Figure~\ref{H_Rates}) and consequently shows more material with $Y_e>0.5$.
As noted by \citet{WaMuJa18}, there is a general correlation between high entropy and high $Y_e$ because neutrino exposure tends to increase both quantities. 
The distributions in $Y_e$ shown in Figure \ref{fig:ye_histo} are much narrower than the results presented by \citet{WaMuJa18} but they report the value of $Y_e$ at freeze-out, rather than at the end of the simulation, which does not allow a direct comparison.
 
There are two noticeable features that merit further discussion.
First, it is unexpected to find the most proton-rich material in F15.78, which exhibits a lower explosion energy and, on average, less energetic neutrino emission.
From the distribution in $Y_e$--entropy space in Figure~\ref{fig:ye_histo}(a), it is clear that the highest $Y_e$ material branches off from the bulk of the ejecta. 
Closer inspection shows that the high-$Y_e$ material is part of a plume that develops at around 1.7~s after bounce in the direction opposite to the unipolar explosion, i.e., to the right in Figure~\ref{F15_78_79_evol}.
The  material expands rapidly along the symmetry axis  as it is pushing against the weakening downstream that moves away from the axis.
Strong neutrino heating results in high $Y_e$ and significant entropy.
Comparing these conditions to the results presented by \citet{XiSiSe20}, we anticipate a small contribution to the lighter $p$-nuclei up to \isotope{Kr}{78}. 
Because of the higher neutrino luminosity in F15.79 relative to F15.78, it is surprising that F15.78 reaches higher values of $Y_e$. 
This is connected to the unusual unipolar morphology of the F15.78 explosion and probably affected by the assumption of axial symmetry, which favors the development of the outflow along the axis.
Due to the marginal explosion in this model, it is not clear whether a significant fraction of this material would eventually become part of the ejecta.
While F15.78 reaches unexpectedly high values of $Y_e$, very little mass is ejected under such conditions.
The total mass of material with $Y_e>0.5$ is larger for  F15.79, almost 0.1~\msun, compared to only 0.02~\msun\ of proton-rich ejecta in F15.78.

The second noticeable feature is the neutron-rich material in F15.79 with a distribution extending down to $Y_e$ of nearly 0.45 and an entropy of around 20~\kbbar.
This material is part of an outflow that forms at around 1.9~s after bounce, when the dominant accretion streams start to shift from a position close to the equator to one side of the proto-NS.
The formation of the downstream and the resulting increase of the density make the neutrinosphere on that side of the proto-NS move outward, reducing the average neutrino energies.
 Most of the material in this downstream is accreted onto the proto-NS, but a small fraction of the material gets compressed against the symmetry axis and is ejected in a cone along the axis without getting close to the proto-NS and without strong neutrino exposure. 
Since the neutrino energies on that side of the proto-NS are relatively low due to the increased density and the material never gets very close to the proto-NS, the $Y_e$ remains low due to electron captures when compressed against the axis.
This outflow is also probably exaggerated by the 2D nature of the simulation that allows the downflow to be deflected by the axis.
While the conditions of this material are interesting for the nucleosynthesis of elements heavier than Fe, the total mass of unbound material with $Y_e<0.47$ is less than $3.5\times10^{-5}$~\msun, satisfying the constraint that CCSNe should not eject more than $10^{-4}$~\msun\ of material with such low $Y_e$ in order to prevent the overproduction of certain heavy elements \citep{HoWoFu96}. 
 A detailed analysis with tracer particles is necessary to determine the potential contribution of this type of material to chemical evolution. We note that similar ejecta with $Y_e$ as low as $0.4$ and $S\approx 15$~\kbbar\ have been found previously in 2D CCSN simulation presented by \citet{WaMuJa18} for a 27~\msun\ progenitor model. 

The appearance of individual outflows with conditions that are interesting for the nucleosynthesis of heavy elements very late in the simulation emphasizes the need to run long-term supernova simulation that capture the conditions of these low-mass outflows. 
The emergence of these features is the result of chaotic fluid flows, rather than a direct effect of differences in these progenitor models.
However, the overall accretion rate, which is governed by the interplay of the progenitor structure and the developing explosion, does impact the appearance and evolution of outflows like these.

\section{Summary and Conclusions}
\label{sec:summary}

Recent stellar evolution modeling of massive stars \citep{SuWoHe18} has indicated a bimodal variability, related to specific patterns of shell burning interactions, which can result in two nearly equal zero mass progenitors ending up with significantly different internal structures at the time of core collapse.
To explore the implications of this bimodal variability for core collapse and subsequent evolution, we have followed the core-collapse evolution in 2D for almost 3~s of two nearly equal mass progenitors, 15.78 and 15.79~\msun, with different internal structures as detailed in Section~\ref{sec:preSN}.
Shock stagnation after bounce occurs in both models at approximate 160 km.
The 15.78~\msun\ progenitor exhibits a substantial density drop at the S/Si-O composition interface resulting in significant drop in the accretion ram pressure for F15.78 when that interface arrives at the shock front, approximately 120~ms after bounce.
This caused a strong shock expansion which, in turn, triggered runaway conditions for an explosion.
With no strong density decrement characterized in its progenitor, the explosion for F15.79 sets in 100 ms later than in  F15.78 due to a gradual decrease in the accretion ram pressure at the shock front together with hardening of the neutrino energy spectra and a corresponding increase in the neutrino heating efficiencies.
Despite the early shock revival of F15.78, more neutrino energy is deposited into the heating region of F15.79 and the shock of the latter rapidly overtakes that of F15.78.
By the end of the simulations, the diagnostic energy of F15.79 is $\approx$0.74~B and that of F15.78 is $\approx$0.26~B.

The more powerful shock and greater energetics of F15.79 that arise during the 0.2--0.8~s post-bounce period can be traced ultimately to the more massive ($M_{4} = 1.78$~\msun\ versus 1.47~\msun) and denser core of its progenitor. 
The immediate consequence of this is a larger mass accretion rate during the 0.2--0.8~s epoch, giving rise to a greater \nue/\nuebar\ accretion luminosity and a denser and more extended gain region. 
At the same time, the added accumulated mass on the F15.79 inner core supplied by the larger mass accretion rate forces the core to undergo a greater adiabatic compression and a resulting higher peak temperature, leading to a greater \nue/\nuebar\ core luminosity.
Since the \nue/\nuebar\ luminosities of both models emanate from neutrinospheres of approximately the same radii, the greater \nue/\nuebar\ accretion and core luminosities in F15.79 result in greater \nue/\nuebar\ rms energies.
Finally, the greater \nue/\nuebar luminosities and rms energies in F15.79 and the more massive and flatter density profile of its gain region leads to a substantially larger \nue/\nuebar\ heating rate and efficiency.
This produces a more intense outflow of unbound matter with higher total per baryon energies accounting for its greater energetics in F15.79.

The late energy uptakes exhibited by F15.78 at 1.2 and 1.55~s and by F15.79 at 1.65~s are caused by transient rearrangements in the accretion streams which are stochastic in nature and seem unrelated directly to differences in progenitor structure.
However, they provide well-isolated examples of the connection between accretion streams and the growth of the explosion energy. 

The differences in the unbound mass of \isotope{Ni}{56} based on the limited $\alpha$-reaction network are consistent with the different explosion energies resulting in $1.4\times 10^{-2} \msun$ of \isotope{Ni}{56} for F15.78 and $3.2 \times 10^{-2} \msun$ in F15.79 with the more energetic explosion.
The structure of the Si shell in the 15.78~\msun\ progenitor leads to a relatively large amount of unburnt  \isotope{Si}{28} to become unbound in F15.78 and possibly a larger \isotope{Ti}{44}/\isotope{Ni}{56} mass ratio than F15.79.
In both models the ejecta is mostly moderately proton-rich, but small amounts of material with more extreme values of $Y_e$ ranging from 0.46 to 0.61 becomes unbound at late times, with F15.78 exhibiting higher values of $Y_e$ in the unbound material than F15.79.

We speculate on what differences we might expect had our two simulations been performed in 3D rather than 2D. 2D and 3D simulations performed from the same progenitor indicate that shock revival, in most cases, is a little less robust in 3D \citep{HaMuWo13,TaHaJa14, LeBrHi15, MeJaMa15}. An exception is \citet{NaBuRa19} who found that their 19 \msun\ progenitor exploded slightly earlier in 3D than in 2D. The shocks, once revived, however, tend to be more energetic in 3D, at least in the few cases where the comparison has extended far enough in time. A detailed analysis of the 2D versus 3D dynamics of shock propagation was provided by \citet{Mull15}. Without repeating his analysis here, he found, on the basis of 2D and 3D simulations initiated from a 11.5~\msun progenitor, that the mass outflow rate was greater in 3D despite similar heating rates, due to the smaller binding energy of matter at the gain radius. Along with the greater mass outflow, more enthalpy (though not mass specific enthalpy) was transported to the shock in 3D relative to its 2D counterpart, strengthening the former relative to the latter. These 2D--3D difference, extrapolated to our models, would suggest that shock revival for F15.78, being very robust, would probably be unaffected while in F15.79 it might be a little more delayed. The shocks, assuming the results of \citet{Mull15} hold generally, would be more energetic and the remanent masses would likely be smaller.

\begin{acknowledgements}

This research was supported by the U.S. Department of Energy, Offices of Nuclear Physics and Advanced Scientific Computing Research;
the NASA Astrophysics Theory Program (grant NNH11AQ72I); 
and the National Science Foundation PetaApps Program (grants  OCI-0749242, OCI-0749204, and OCI-0749248), Nuclear Theory Program (PHY-1913531, PHY-1516197) and Stellar Astronomy and Astrophysics program (AST-0653376).
This research used resources of the National Energy Research Scientific Computing Center (NERSC), a U.S. Department of Energy Office of Science User Facility located at Lawrence Berkeley National Laboratory, operated under Contract No. DE-AC02-05CH11231.
This research used resources of the Oak Ridge Leadership Computing Facility at Oak Ridge National Laboratory.
Research at Oak Ridge National Laboratory is supported under contract DE-AC05-00OR22725  from the Office of Science of the U.S. Department of Energy to UT-Battelle, LLC. 

\end{acknowledgements}

\facility{NERSC, OLCF}

\software{\chimera \citep{BrBlHi20}}


\begin{thebibliography}{}
\expandafter\ifx\csname natexlab\endcsname\relax\def\natexlab#1{#1}\fi
\providecommand{\url}[1]{\href{#1}{#1}}

\bibitem[{{Abdikamalov} {et~al.}(2016){Abdikamalov}, {Zhaksylykov}, {Radice},
  \& {Berdibek}}]{AbZhRa16}
{Abdikamalov}, E., {Zhaksylykov}, A., {Radice}, D., \& {Berdibek}, S. 2016,
  \mnras, 461, 3864

\bibitem[{Blondin {et~al.}(2003)Blondin, Mezzacappa, \& DeMarino}]{BlMeDe03}
Blondin, J., Mezzacappa, A., \& DeMarino, C. 2003, \apj, 584, 971

\bibitem[{{Boccioli} {et~al.}(2022){Boccioli}, {Roberti}, {Limongi}, {Mathews},
  \& {Chieffi}}]{BoRoLi22}
{Boccioli}, L., {Roberti}, L., {Limongi}, M., {Mathews}, G.~J., \& {Chieffi},
  A. 2022, arXiv e-prints, arXiv:2207.08361

\bibitem[{{Bollig} {et~al.}(2021){Bollig}, {Yadav}, {Kresse}, {Janka},
  {M{\"u}ller}, \& {Heger}}]{BoYaKr21}
{Bollig}, R., {Yadav}, N., {Kresse}, D., {et~al.} 2021, \apj, 915, 28

\bibitem[{Bruenn {et~al.}(2013)Bruenn, Mezzacappa, Hix, Lentz, Messer,
  Lingerfelt, Blondin, Endeve, Marronetti, \& Yakunin}]{BrMeHi13}
Bruenn, S.~W., Mezzacappa, A., Hix, W.~R., {et~al.} 2013, \apj, 767, L6

\bibitem[{{Bruenn} {et~al.}(2016){Bruenn}, {Lentz}, {Hix}, {Mezzacappa},
  {Harris}, {Messer}, {Endeve}, {Blondin}, {Chertkow}, {Lingerfelt},
  {Marronetti}, \& {Yakunin}}]{BrLeHi16}
{Bruenn}, S.~W., {Lentz}, E.~J., {Hix}, W.~R., {et~al.} 2016, \apj, 818, 123

\bibitem[{{Bruenn} {et~al.}(2020){Bruenn}, {Blondin}, {Hix}, {Lentz}, {Messer},
  {Mezzacappa}, {Endeve}, {Harris}, {Marronetti}, {Budiardja}, {Chertkow}, \&
  {Lee}}]{BrBlHi20}
{Bruenn}, S.~W., {Blondin}, J.~M., {Hix}, W.~R., {et~al.} 2020, \apjs, 248, 11

\bibitem[{{Buras} {et~al.}(2006){Buras}, {Janka}, {Rampp}, \&
  {Kifonidis}}]{BuJaRa06}
{Buras}, R., {Janka}, H.-T., {Rampp}, M., \& {Kifonidis}, K. 2006, A\&A, 457,
  281

\bibitem[{Burrows \& Goshy(1993)}]{BuGo93}
Burrows, A., \& Goshy, J. 1993, \apjl, 416, L75

\bibitem[{{Burrows} \& {Hayes}(1996)}]{BuHa96}
{Burrows}, A., \& {Hayes}, J. 1996, \prl, 76, 352

\bibitem[{{Burrows} {et~al.}(2019){Burrows}, {Radice}, \&
  {Vartanyan}}]{BuRaVa19}
{Burrows}, A., {Radice}, D., \& {Vartanyan}, D. 2019, \mnras, 485, 3153

\bibitem[{{Burrows} {et~al.}(2020){Burrows}, {Radice}, {Vartanyan}, {Nagakura},
  {Skinner}, \& {Dolence}}]{BuRaVa20}
{Burrows}, A., {Radice}, D., {Vartanyan}, D., {et~al.} 2020, \mnras, 491, 2715

\bibitem[{{Cooperstein}(1985)}]{Coop85}
{Cooperstein}, J. 1985, \nphysa, 438, 722

\bibitem[{{Couch} \& {Ott}(2013)}]{CoOt13}
{Couch}, S.~M., \& {Ott}, C.~D. 2013, \apjl, 778, L7

\bibitem[{{Couch} \& {Ott}(2015)}]{CoOt15}
---. 2015, \apj, 799, 5

\bibitem[{{Curtis} {et~al.}(2019){Curtis}, {Ebinger}, {Fr{\"o}hlich}, {Hempel},
  {Perego}, {Liebend{\"o}rfer}, \& {Thielemann}}]{CuEbFr19}
{Curtis}, S., {Ebinger}, K., {Fr{\"o}hlich}, C., {et~al.} 2019, \apj, 870, 2

\bibitem[{{Dolence} {et~al.}(2015){Dolence}, {Burrows}, \& {Zhang}}]{DoBuZh15}
{Dolence}, J.~C., {Burrows}, A., \& {Zhang}, W. 2015, \apj, 800, 10

\bibitem[{{Ebinger} {et~al.}(2019){Ebinger}, {Curtis}, {Fr{\"o}hlich},
  {Hempel}, {Perego}, {Liebend{\"o}rfer}, \& {Thielemann}}]{EbCuFr19}
{Ebinger}, K., {Curtis}, S., {Fr{\"o}hlich}, C., {et~al.} 2019, \apj, 870, 1

\bibitem[{{Eichler} {et~al.}(2018){Eichler}, {Nakamura}, {Takiwaki}, {Kuroda},
  {Kotake}, {Hempel}, {Cabez{\'o}n}, {Liebend{\"o}rfer}, \&
  {Thielemann}}]{EiNaTa18}
{Eichler}, M., {Nakamura}, K., {Takiwaki}, T., {et~al.} 2018, J. Phys. G: Nucl.
  Phys., 45, 014001

\bibitem[{{Ertl} {et~al.}(2016){Ertl}, {Janka}, {Woosley}, {Sukhbold}, \&
  {Ugliano}}]{ErJaWo16}
{Ertl}, T., {Janka}, H.-T., {Woosley}, S.~E., {Sukhbold}, T., \& {Ugliano}, M.
  2016, \apj, 818, 124

\bibitem[{{Fern{\'a}ndez}(2012)}]{Fern12}
{Fern{\'a}ndez}, R. 2012, \apj, 749, 142

\bibitem[{{Foglizzo} {et~al.}(2007){Foglizzo}, {Galletti}, {Scheck}, \&
  {Janka}}]{FoGaSc07}
{Foglizzo}, T., {Galletti}, P., {Scheck}, L., \& {Janka}, H.-T. 2007, \apj,
  654, 1006

\bibitem[{{Foglizzo} {et~al.}(2006){Foglizzo}, {Scheck}, \& {Janka}}]{FoScJa06}
{Foglizzo}, T., {Scheck}, L., \& {Janka}, H.-T. 2006, \apj, 652, 1436

\bibitem[{{Hamuy}(2003)}]{Hamu03}
{Hamuy}, M. 2003, \apj, 582, 905

\bibitem[{{Hanke} {et~al.}(2013){Hanke}, {M{\"u}ller}, {Wongwathanarat},
  {Marek}, \& {Janka}}]{HaMuWo13}
{Hanke}, F., {M{\"u}ller}, B., {Wongwathanarat}, A., {Marek}, A., \& {Janka},
  H.-T. 2013, \apj, 770, 66

\bibitem[{{Hannestad} \& {Raffelt}(1998)}]{HaRa98}
{Hannestad}, S., \& {Raffelt}, G. 1998, \apj, 507, 339

\bibitem[{{Harris} {et~al.}(2017){Harris}, {Hix}, {Chertkow}, {Lee}, {Lentz},
  \& {Messer}}]{HaHiCh17}
{Harris}, J.~A., {Hix}, W.~R., {Chertkow}, M.~A., {et~al.} 2017, \apj, 843, 2

\bibitem[{Herant {et~al.}(1994)Herant, Benz, Hix, Fryer, \& Colgate}]{HeBeHi94}
Herant, M., Benz, W., Hix, W.~R., Fryer, C.~L., \& Colgate, S.~A. 1994, \apj,
  435, 339

\bibitem[{{Hoffman} {et~al.}(1996){Hoffman}, {Woosley}, {Fuller}, \&
  {Meyer}}]{HoWoFu96}
{Hoffman}, R.~D., {Woosley}, S.~E., {Fuller}, G.~M., \& {Meyer}, B.~S. 1996,
  \apj, 460, 478

\bibitem[{{Janka}(1995)}]{Jank95}
{Janka}, H.~T. 1995, Astroparticle Physics, 3, 377

\bibitem[{{Janka}(2001)}]{Jank01}
{Janka}, H.-T. 2001, \aap, 368, 527

\bibitem[{{Janka}(2012)}]{Jank12}
---. 2012, Annu. Rev. Nucl. Part. Sci., 62, 407

\bibitem[{{Janka} {et~al.}(2001){Janka}, {Kifonidis}, \& {Rampp}}]{JaKiRa01}
{Janka}, H.-T., {Kifonidis}, K., \& {Rampp}, M. 2001, in Physics of Neutron
  Star Interiors, ed. D.~{Blaschke}, N.~K. {Glendenning}, \& A.~{Sedrakian},
  Vol. 578, 333

\bibitem[{Janka \& M{\"u}ller(1996)}]{JaMu96}
Janka, H.-T., \& M{\"u}ller, E. 1996, \aap, 306, 167

\bibitem[{{Landfield}(2018)}]{Land18}
{Landfield}, R.~E. 2018, PhD thesis, University of Tennessee, Knoxville, United
  States

\bibitem[{{Landfield} {et~al.}(2022){Landfield}, {Lentz}, {Mezzacappa}, {Hix},
  {et~al.}}]{LaLeMe22}
{Landfield}, R.~E., {Lentz}, E.~J., {Mezzacappa}, A., {Hix}, W.~R., {et~al.}
  2022, \apj, in prep

\bibitem[{{Lentz} {et~al.}(2015){Lentz}, {Bruenn}, {Hix}, {Mezzacappa},
  {Messer}, {Endeve}, {Blondin}, {Harris}, {Marronetti}, \&
  {Yakunin}}]{LeBrHi15}
{Lentz}, E.~J., {Bruenn}, S.~W., {Hix}, W.~R., {et~al.} 2015, \apjl, 807, L31

\bibitem[{{Martinez} {et~al.}(2022){Martinez}, {Anderson}, {Bersten}, {Hamuy},
  {Gonz{\'a}lez-Gait{\'a}n}, {Orellana}, {Stritzinger}, {Phillips},
  {Guti{\'e}rrez}, {Burns}, {de Jaeger}, {Ertini}, {Folatelli}, {F{\"o}rster},
  {Galbany}, {Hoeflich}, {Hsiao}, {Morrell}, {Pessi}, \& {Suntzeff}}]{MaAnBe22}
{Martinez}, L., {Anderson}, J.~P., {Bersten}, M.~C., {et~al.} 2022, \aap, 660,
  A42

\bibitem[{{Melson} {et~al.}(2015){Melson}, {Janka}, \& {Marek}}]{MeJaMa15}
{Melson}, T., {Janka}, H.-T., \& {Marek}, A. 2015, \apjl, 801, L24

\bibitem[{{M{\"u}ller}(2015)}]{Mull15}
{M{\"u}ller}, B. 2015, \mnras, 453, 287

\bibitem[{{M{\"u}ller}(2016)}]{Mull16}
---. 2016, Proc. Astron. Soc. Aust., 33, e048

\bibitem[{{M{\"u}ller} {et~al.}(2016){M{\"u}ller}, {Heger}, {Liptai}, \&
  {Cameron}}]{MuHeLi16}
{M{\"u}ller}, B., {Heger}, A., {Liptai}, D., \& {Cameron}, J.~B. 2016, \mnras,
  460, 742

\bibitem[{{M{\"u}ller} \& {Janka}(2014)}]{MuJa14}
{M{\"u}ller}, B., \& {Janka}, H.-T. 2014, \apj, 788, 82

\bibitem[{{M{\"u}ller} \& {Janka}(2015)}]{MuJa15}
---. 2015, \mnras, 448, 2141

\bibitem[{{M{\"u}ller} {et~al.}(2012){M{\"u}ller}, {Janka}, \&
  {Marek}}]{MuJaMa12}
{M{\"u}ller}, B., {Janka}, H.-T., \& {Marek}, A. 2012, \apj, 756, 84

\bibitem[{{M{\"u}ller} {et~al.}(2017){M{\"u}ller}, {Melson}, {Heger}, \&
  {Janka}}]{MuMeHe17}
{M{\"u}ller}, B., {Melson}, T., {Heger}, A., \& {Janka}, H.-T. 2017, \mnras,
  472, 491

\bibitem[{{Nagakura} {et~al.}(2019){Nagakura}, {Burrows}, {Radice}, \&
  {Vartanyan}}]{NaBuRa19}
{Nagakura}, H., {Burrows}, A., {Radice}, D., \& {Vartanyan}, D. 2019, \mnras,
  490, 4622

\bibitem[{{O'Connor} \& {Ott}(2011)}]{OCOt11}
{O'Connor}, E., \& {Ott}, C.~D. 2011, \apj, 730, 70

\bibitem[{{O'Connor} \& {Ott}(2013)}]{OCOt13}
---. 2013, \apj, 762, 126

\bibitem[{{Ohnishi} {et~al.}(2006){Ohnishi}, {Kotake}, \& {Yamada}}]{OhKoYa06}
{Ohnishi}, N., {Kotake}, K., \& {Yamada}, S. 2006, \apj, 641, 1018

\bibitem[{{Ott} {et~al.}(2018){Ott}, {Roberts}, {da Silva Schneider}, {Fedrow},
  {Haas}, \& {Schnetter}}]{OtRodSS18}
{Ott}, C.~D., {Roberts}, L.~F., {da Silva Schneider}, A., {et~al.} 2018, \apjl,
  855, L3

\bibitem[{{{\"O}zel} \& {Freire}(2016)}]{OzFe16}
{{\"O}zel}, F., \& {Freire}, P. 2016, \araa, 54, 401

\bibitem[{{{\"O}zel} {et~al.}(2012){{\"O}zel}, {Psaltis}, {Narayan}, \& {Santos
  Villarreal}}]{OzPsNa12}
{{\"O}zel}, F., {Psaltis}, D., {Narayan}, R., \& {Santos Villarreal}, A. 2012,
  \apj, 757, 55

\bibitem[{{Scheck} {et~al.}(2008){Scheck}, {Janka}, {Foglizzo}, \&
  {Kifonidis}}]{ScJaFo08}
{Scheck}, L., {Janka}, H.-T., {Foglizzo}, T., \& {Kifonidis}, K. 2008, A\&A,
  477, 931

\bibitem[{{Schwab} {et~al.}(2010){Schwab}, {Podsiadlowski}, \&
  {Rappaport}}]{ScPoRa10}
{Schwab}, J., {Podsiadlowski}, P., \& {Rappaport}, S. 2010, \apj, 719, 722

\bibitem[{{Sieverding} {et~al.}(2020){Sieverding}, {M{\"u}ller}, \&
  {Qian}}]{SiMuQi20}
{Sieverding}, A., {M{\"u}ller}, B., \& {Qian}, Y.~Z. 2020, \apj, 904, 163

\bibitem[{{Smartt}(2009)}]{Smar09}
{Smartt}, S.~J. 2009, \araa, 47, 63

\bibitem[{{Smartt}(2015)}]{Smar15}
---. 2015, Proc. Astron. Soc. Aust., 32, 16

\bibitem[{{Steiner} {et~al.}(2013){Steiner}, {Hempel}, \& {Fischer}}]{StHeFi13}
{Steiner}, A.~W., {Hempel}, M., \& {Fischer}, T. 2013, \apj, 774, 17

\bibitem[{{Sukhbold} {et~al.}(2018){Sukhbold}, {Woosley}, \&
  {Heger}}]{SuWoHe18}
{Sukhbold}, T., {Woosley}, S.~E., \& {Heger}, A. 2018, \apj, 860, 93

\bibitem[{{Summa} {et~al.}(2016){Summa}, {Hanke}, {Janka}, {Melson}, {Marek},
  \& {M{\"u}ller}}]{SuHaJa16}
{Summa}, A., {Hanke}, F., {Janka}, H.-T., {et~al.} 2016, \apj, 825, 6

\bibitem[{{Suwa} {et~al.}(2016){Suwa}, {Yamada}, {Takiwaki}, \&
  {Kotake}}]{SuYaTa16}
{Suwa}, Y., {Yamada}, S., {Takiwaki}, T., \& {Kotake}, K. 2016, \apj, 816, 43

\bibitem[{{Tamborra} {et~al.}(2014){Tamborra}, {Hanke}, {Janka}, {M{\"u}ller},
  {Raffelt}, \& {Marek}}]{TaHaJa14}
{Tamborra}, I., {Hanke}, F., {Janka}, H.-T., {et~al.} 2014, \apj, 792, 96

\bibitem[{{Thompson} {et~al.}(2005){Thompson}, {Quataert}, \&
  {Burrows}}]{ThQuBu05}
{Thompson}, T.~A., {Quataert}, E., \& {Burrows}, A. 2005, \apj, 620, 861

\bibitem[{{Timmes} {et~al.}(1996){Timmes}, {Woosley}, \& {Weaver}}]{TiWoWe96}
{Timmes}, F.~X., {Woosley}, S.~E., \& {Weaver}, T.~A. 1996, \apj, 457, 834

\bibitem[{{Ugliano} {et~al.}(2012){Ugliano}, {Janka}, {Marek}, \&
  {Arcones}}]{UgJaMa12}
{Ugliano}, M., {Janka}, H.-T., {Marek}, A., \& {Arcones}, A. 2012, \apj, 757,
  69

\bibitem[{{Vartanyan} {et~al.}(2018){Vartanyan}, {Burrows}, {Radice},
  {Skinner}, \& {Dolence}}]{VaBuRa18a}
{Vartanyan}, D., {Burrows}, A., {Radice}, D., {Skinner}, M.~A., \& {Dolence},
  J. 2018, \mnras, 477, 3091

\bibitem[{{Vartanyan} {et~al.}(2019){Vartanyan}, {Burrows}, {Radice},
  {Skinner}, \& {Dolence}}]{VaBuRa18b}
---. 2019, \mnras, 482, 351

\bibitem[{{Vartanyan} {et~al.}(2021){Vartanyan}, {Laplace}, {Renzo},
  {G{\"o}tberg}, {Burrows}, \& {de Mink}}]{VaLaRe21}
{Vartanyan}, D., {Laplace}, E., {Renzo}, M., {et~al.} 2021, \apjl, 916, L5

\bibitem[{{Wanajo} {et~al.}(2018){Wanajo}, {M{\"u}ller}, {Janka}, \&
  {Heger}}]{WaMuJa18}
{Wanajo}, S., {M{\"u}ller}, B., {Janka}, H.-T., \& {Heger}, A. 2018, \apj, 852,
  40

\bibitem[{{Wang} {et~al.}(2022){Wang}, {Vartanyan}, {Burrows}, \&
  {Coleman}}]{WaVaBu22}
{Wang}, T., {Vartanyan}, D., {Burrows}, A., \& {Coleman}, M. S.~B. 2022,
  \mnras, 517, 543

\bibitem[{{Woosley} \& {Heger}(2007)}]{WoHe07}
{Woosley}, S.~E., \& {Heger}, A. 2007, \physrep, 442, 269

\bibitem[{{Woosley} {et~al.}(2002){Woosley}, {Heger}, \& {Weaver}}]{WoHeWe02}
{Woosley}, S.~E., {Heger}, A., \& {Weaver}, T.~A. 2002, Rev. Mod. Phys., 74,
  1015

\bibitem[{{Xiong} {et~al.}(2020){Xiong}, {Sieverding}, {Sen}, \&
  {Qian}}]{XiSiSe20}
{Xiong}, Z., {Sieverding}, A., {Sen}, M., \& {Qian}, Y.-Z. 2020, \apj, 900, 144

\bibitem[{{Yamasaki} \& {Yamada}(2007)}]{YaYa07}
{Yamasaki}, T., \& {Yamada}, S. 2007, \apj, 656, 1019

\end{thebibliography}
\end{document}